\documentclass[14pt]{extarticle}

\usepackage[utf8]{inputenc}
\usepackage[T2A]{fontenc}
\usepackage[english]{babel}

\usepackage{setspace}
\usepackage{lineno}
\usepackage{subfigure}
\usepackage{csquotes}
\usepackage{amsmath}
\usepackage{amssymb}
\usepackage{amsxtra}
\usepackage{float}
\usepackage{graphicx}

\usepackage{cmap} 
\usepackage[unicode, pdftex]{hyperref} 
\newcommand{\snn}{\sqrt{s_{\mathrm{NN}}}}
\usepackage[
	style = gost-numeric,
	bibencoding=auto,
	backend=biber,
	sorting=none,
	language=auto,
	autolang=other,
	doi=false,
	url=false,
	movenames=false,
	maxbibnames=3
	]{biblatex}
	
\usepackage{indentfirst}
\usepackage{geometry}
\begin{document}

    \begin{center} 
        \textbf{\Large \underline{Analysis Note}}\\[.6cm] 
        
       \textbf{ \large{Directed flow $v_1$ of protons  in the Xe+Cs(I) collisions at 3.8 AGeV}}\\[.6cm] 
        
        \normalsize 
            Mikhail Mamamev\footnote{E-mail: mam.mih.val@gmail.com}, Arkadiy Taranenko\footnote{E-mail: AVTaranenko@mephi.ru}, Alexander Demanov, Petr Parfenov, Valery Troshin.
             
            \setcounter{footnote}{0}

            National Research Nuclear University MEPhI, Moscow, Russia \\
            Joint Institute for Nuclear Research, Dubna, Russia \\
        
         \noindent
         \underline{\hspace{17.5cm}}
  \begin{center}

  \end{center}

            \begin{center}

In this note, we present the directed flow $v_1$  measurements of protons from Xe+Cs(I)  collisions at 3.8 AGeV (BM@N run8). We
show the datasets, event and track selection cuts, centrality definition,
event plane reconstruction and resolution.
The $v_1$ results are presented as function of transverse momentum ($p_T$) and
rapidity ($y_{cm}$) for 10-30\% central Xe+Cs(I)  collisions.
The systematic uncertainty study will also be presented and discussed.
The $v_1$ measurements are compared with
results of JAM transport model calculations and published data from other experiments.\\

            \end{center}

            \normalsize
            \noindent
    \end{center}
    
      \begin{center}
    
    {\small The work has been supported by the Ministry of Science and Higher Education of the Russian Federation, Project “Fundamental and applied research at the NICA megascience experimental complex” № FSWU-2024-0024. }

       \end{center}

    \normalsize
    \noindent
    \underline{\hspace{17.6cm}}
\newpage
    \tableofcontents

\section{Introduction}

\begin{figure}[h]
    \centering
    \includegraphics[width=0.37\textwidth]{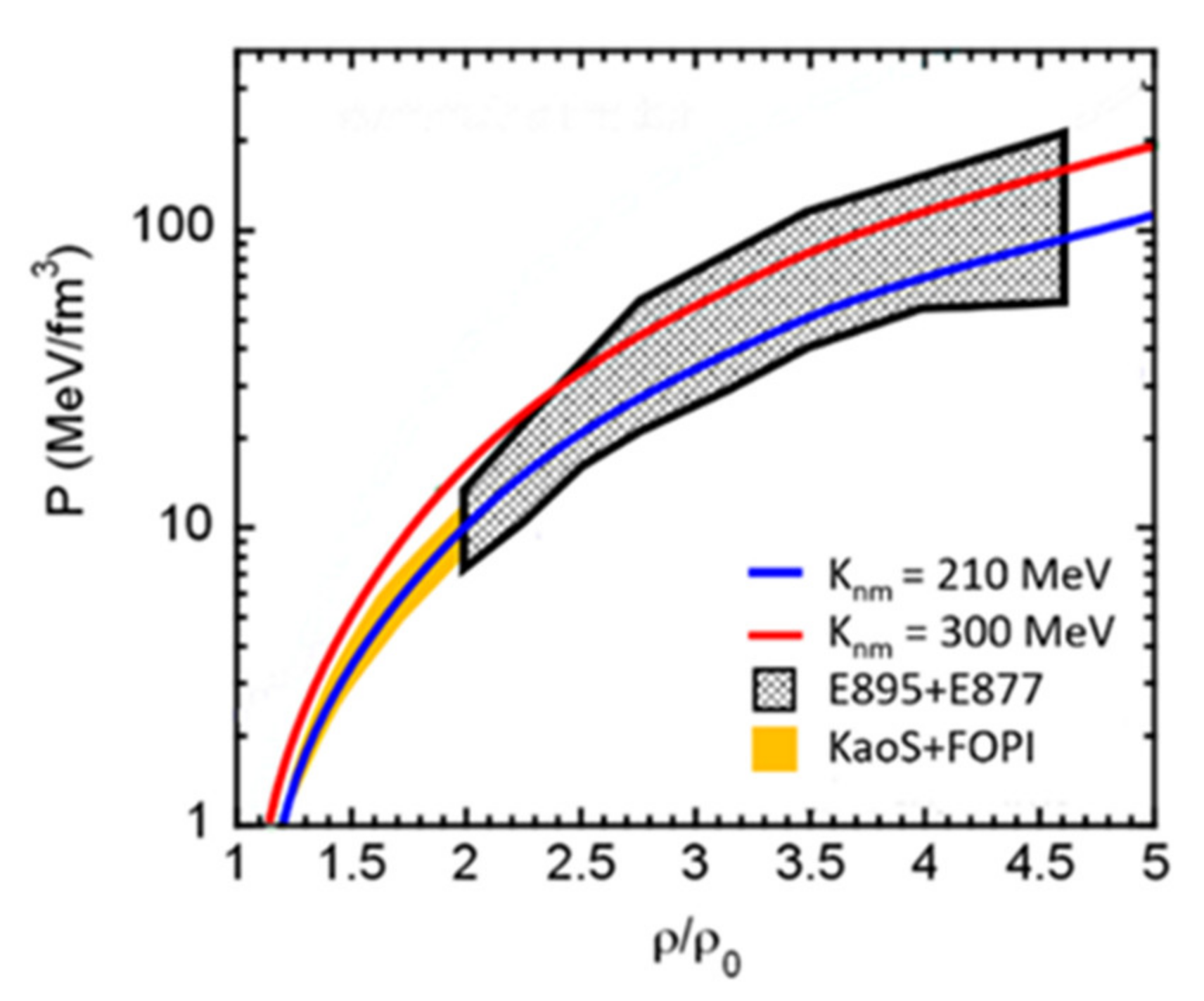}
    \includegraphics[width=0.55\textwidth]{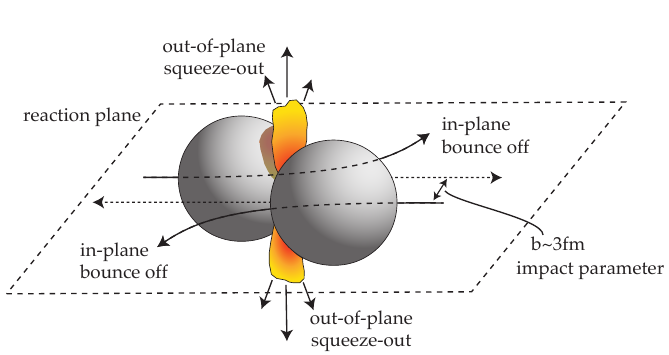}
    \caption{Left panel:  Pressure as function of baryon density for symmetric nuclear matter. Selected constraints on
      the symmetric EOS obtained from comparisons of experimental flow data to hadronic transport simulations, see text for the details.
      The figure is taken from \cite{sengereos}. Right panel:  Illustration of a semi-central collision of two nuclei with an impact in the $\snn$ = 2.0-3.5 GeV
      energy regime, with the direction of the flow phenomena indicated with arrows into
      ($v_1>0$ or ``bounce-off'' of spectators) or perpendicular ($v_2<0$ or ``squeeze-out'') to the reaction plane.}
    \label{fig:eos}
\end{figure}

Relativistic heavy-ion collisions   can directly generate the
high density and/or temperature strong interacting matter, and thus provide the opportunity to explore the
strong interaction properties at extreme conditions. One
of the interests is the exploration of
nuclear Equation of State (EoS) as well as the symmetry energy for asymmetric nuclear
matter at high densities \cite{eos}.  The  anisotropic collective flow  of final
state particles is a direct reflection of the pressure and its gradients created
in relativistic heavy-ion collisions and thus is closely related to the EoS of dense
matter. The anisotropic flow can be quantified by Fourier coefficients $v_n$ \cite{flow0,flow1,flow2,flow3} in
the expansion of the particle azimuthal distribution relative to the reaction plane given by the angle $\Psi_{R}$:
\begin{eqnarray}
dN/d\phi \propto 1 + \sum_{n=1} 2 v_{n} \cos(n(\varphi-\Psi_{R})),
\end{eqnarray}
where $n$ is the order of the harmonic and  $\varphi$ is the azimuthal angle of a particle of the given type. 
The  flow coefficients $v_n$ can be calculated as $v_{n} = \langle{\cos[n(\varphi - \Psi_R)]}\rangle$, where the brackets denote the average over
the  particles and events.  
The directed ($v_1$) and elliptic  ($v_2$) flows are dominant and most studied  signals in
the energy range of 2 < $\snn$  < 5~GeV~\cite{e895v1,e895v2,e895v22,fopi,hadesvn,star45,star3gev}.
The comparison of existing measurements of $v_1$ and  $v_2$  of protons and light fragments in Au+Au collisions at  $\snn$ = 2.07-4.72 GeV (corresponding to
beam energies $E_{beam}$=0.4-10 AGeV) with results from hadronic transport simulations provides the most stringent
currently available constraints on the high-density EOS of symmetric nuclear matter \cite{eos,daneos,sengereos},
see the right panel of Figure~\ref{fig:eos}. 
At densities between 1 and 2 times saturation density $\rho_{0}$,  the $v_2$ data for protons, deuterons and  tritons
in Au+Au collisions measured at $E_{beam}$ = 0.4–1.49 AGeV ($\snn$ = 2.07-2.51 GeV) by the FOPI experiment at GSI \cite{fopi}
have been used together with IQMD model transport calculations to constrain the nuclear incompressibility $K_{nm}$ \cite{iqmdeos}. The model that
takes into account momentum-dependent interactions,  can explain the data with a fairly
soft EOS ($K_{nm}$=190 $\pm$ 30 MeV) \cite{sengereos}, see the solid yellow region in the right panel of  Figure~\ref{fig:eos}.

At densities $\sim$2-5 $\rho/\rho_{0}$, the comparison of the existing $v_1$ and $v_2$ measurements of protons in Au+Au at $E_{beam}$ = 2–8 AGeV
(2.5 < $\snn$  < 4.5 GeV) by the E895 experiment at AGS \cite{e895v1,e895v2,e895v22} with results of microscopic transport models leads to
the values of nuclear incompressibility $K_{nm}$ = 200-380 MeV  \cite{daneos}, depicted by  the  grey hatched region in Figure~\ref{fig:eos}.
The description of $v_1$ results from E895 experiment
requires  a soft EOS with the incompressibility $K_{nm}$ = 200 MeV, while
reproducing the $v_2$ data required larger values of $K_{nm}$ = 380 MeV (and therefore
a harder EOS). A Bayesian analysis study \cite{neweos} suggests a difference  between the E895~\cite{e895v1,e895v2,e895v22} and recently obtained STAR
\cite{star45,star3gev}
data from RHIC Beam Energy Scan program. Using only
the STAR measurements, the study \cite{neweos} further found that the slope of the directed
flow and the elliptic flow of protons can be described by thу transport model with the same EOS.
The E895 flow measurements~\cite{e895v1,e895v2,e895v22} have been performed 15-20 years ago by the standart event plane method,  which do not take into account 
the influence of non-flow effects on $v_n$ measurements ~\cite{mamaev}. 
Therefore, high-precision measurements of anisotropic flow at  2 < $\snn$  < 5 GeV with a modern methods of analysis
are required, in order to further constrain the EOS of symmetric matter from
model comparisons~\cite{sengereos,mamaev}.\\
The  important characteristic of this energy range is that
the compressed overlap zone expands at the time $t_{exp}$ comparable to the passage time $t_{pass}$, at which the accelerated nuclei interpenetrate each other.
The expansion time $t_{exp} \sim R/c_{s}$ is governed by a fundamental property, the speed of sound $c_s$ which connects to
the EOS \cite{e895v22,daneos}. The passage time $t_{pass}$  can be estimated as
$t_{pass}=2R/sinh(y_{beam})$, where $R$ is the radius of the nucleus and $y_{beam}$ is the beam rapidity
\cite{e895v2,e895v22,daneos}. 
For Au+Au collisions at 2.4 < $\snn$  < 5 GeV, the $t_{pass}$  decreases from 18 fm/c to 6 fm/c. 
If the passage time is long compared to the expansion time, spectator nucleons serve to block the path of produced hadrons emitted towards the reaction plane. 
Such rather complex collision geometries result in strong change in the resulting flow patterns. 
For example, for Au+Au collisions at  $\snn$  < 3.3-3.5 GeV, the  nuclear matter is ``squeezed-out'' perpendicular to the reaction plane
giving rise to negative elliptic flow ($v_2$ < 0) and  squeeze-out contribution should then reflect the ratio  $c_s/sinh(y_{beam})$
\cite{e895v22,daneos}, see the right panel of Figure~\ref{fig:eos}. The $t_{pass}$ depends on the size of colliding system and beam energy.
Therefore, the study of the system size dependence of anisotropic flow  may help to estimate the participant-spectator contribution
and  improve our knowledge of EOS of symmetric nuclear matter.\\
The Baryonic Matter at the Nuclotron (BM@N)\cite{bmn}  is a fixed target experiment at JINR (Dubna),
In February 2023, the first physics run of the BMN experiment was completed with recorded   Xe + Cs(I) collision events  at $E_{beam}$ = 3 AGeV ($\snn$= 3.02 GeV) and
3.8 AGeV ($\snn$ = 3.26 GeV).  In this analysis note, we present first results on directed flow ($v_1$) of protons in 10-30\% central Xe + Cs(I) collisions
at $E_{beam}$ = 3.8 AGeV. The note is organized as follows. Section 2 brieﬂy  discusses   the existing data on flow of protons and transport
model predictions.  Section 3 introduces the BM@N experimental set-up, QA study, the centrality and  the particle identification methods,
while section 4 discusses the procedures used to determine the flow coefficients and systematic uncertainty study. Section 5
presents the main results on directed flow $v_1$  of protons.

\section{Directed and elliptic  flow of protons}
\begin{figure}[hbt]
        \begin{center}
            \includegraphics[width=0.5\linewidth]{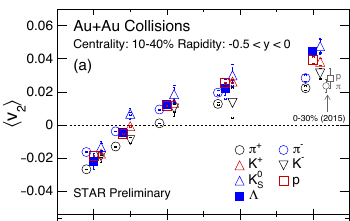}
            \includegraphics[width=0.51\linewidth]{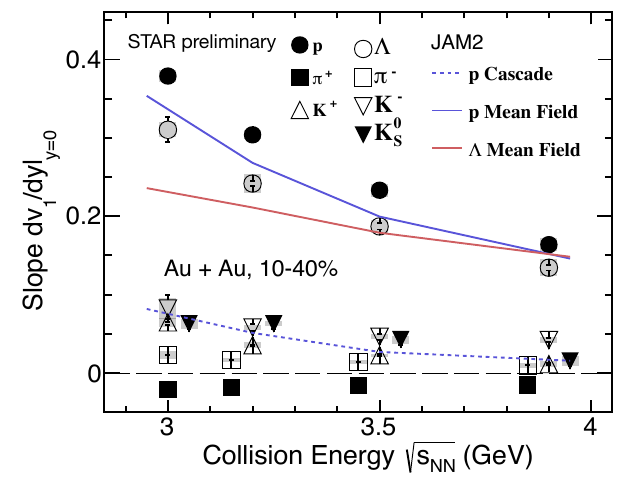}
            \vspace{-3mm}
            \caption{Elliptic flow $v_2$ (upper panel) and slope of the directed flow at mid-rapidity $dv_1/dy|_{y=0}$ (low panel) for different paricle
              species from 10-40\% central Au+Au collisions at $\sqrt{s_{NN}}$=3.0, 3.2, 3.5 and 3.9 GeV from the STAR Beam Energy Scan II program \cite{starv1,starbes2}}
        \end{center}
        \label{v12bes2}
        \vspace{-5mm}
    \end{figure}

\begin{figure}[htb]
	\centering
	\includegraphics[width=0.75\textwidth] {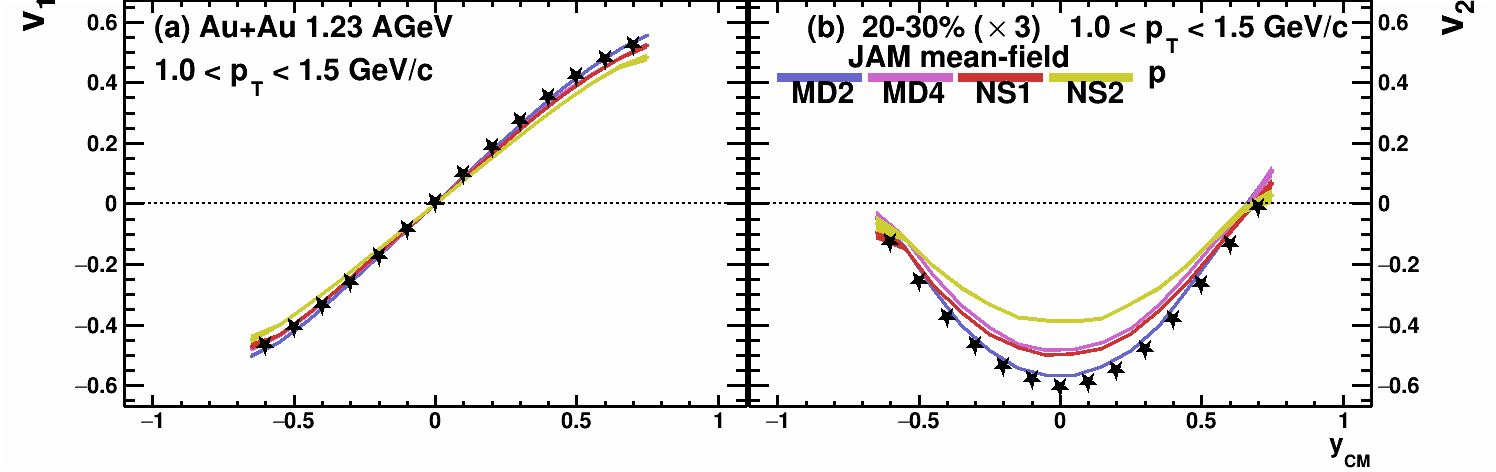}
	\caption{Rapidity ($y_{cm}$) dependence of   $v_1$ (left)  and  $v_2$ (right) of protons with $1.0 < p_{T} < 1.5$ Gev/c
          in the 20-30\% central  Au+Au collisions  at $\sqrt{s_{NN}}=2.4$~GeV.  The  closed star symbols represent the published
           HADES data \cite{hadesvn}.
 The  blue (MD2), purple (MD4), red (NS1)  and  yellow (NS2) bands represent the results from the mean-field  mode of the JAM model with different EOS, as indicated.
 The figure is taken from  \cite{parfenov1}. }
	\label{v12hades}
\end{figure}

\begin{figure}[hbt]
        \begin{center}
            \includegraphics[width=0.49\linewidth]{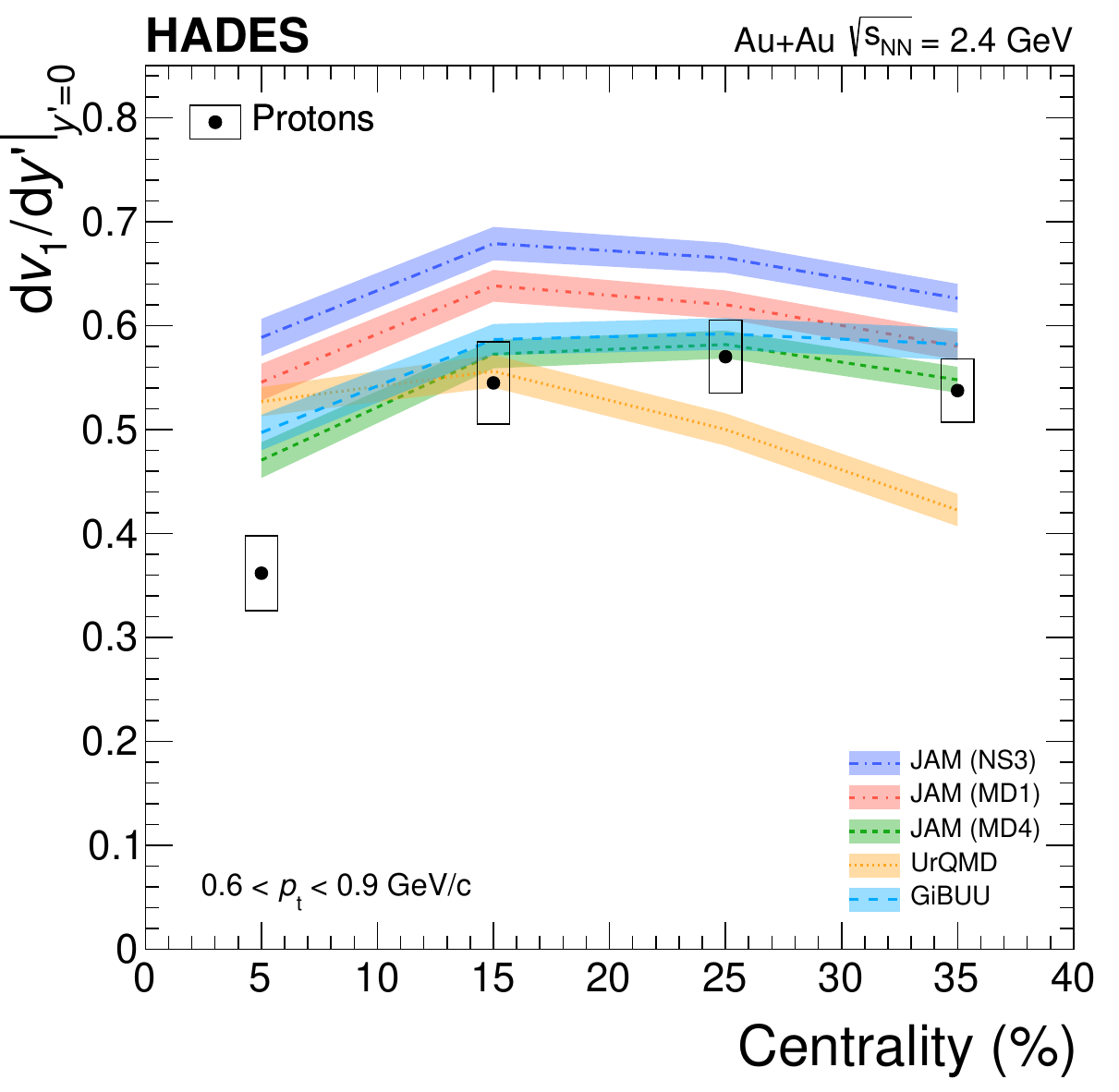}
            \includegraphics[width=0.49\linewidth]{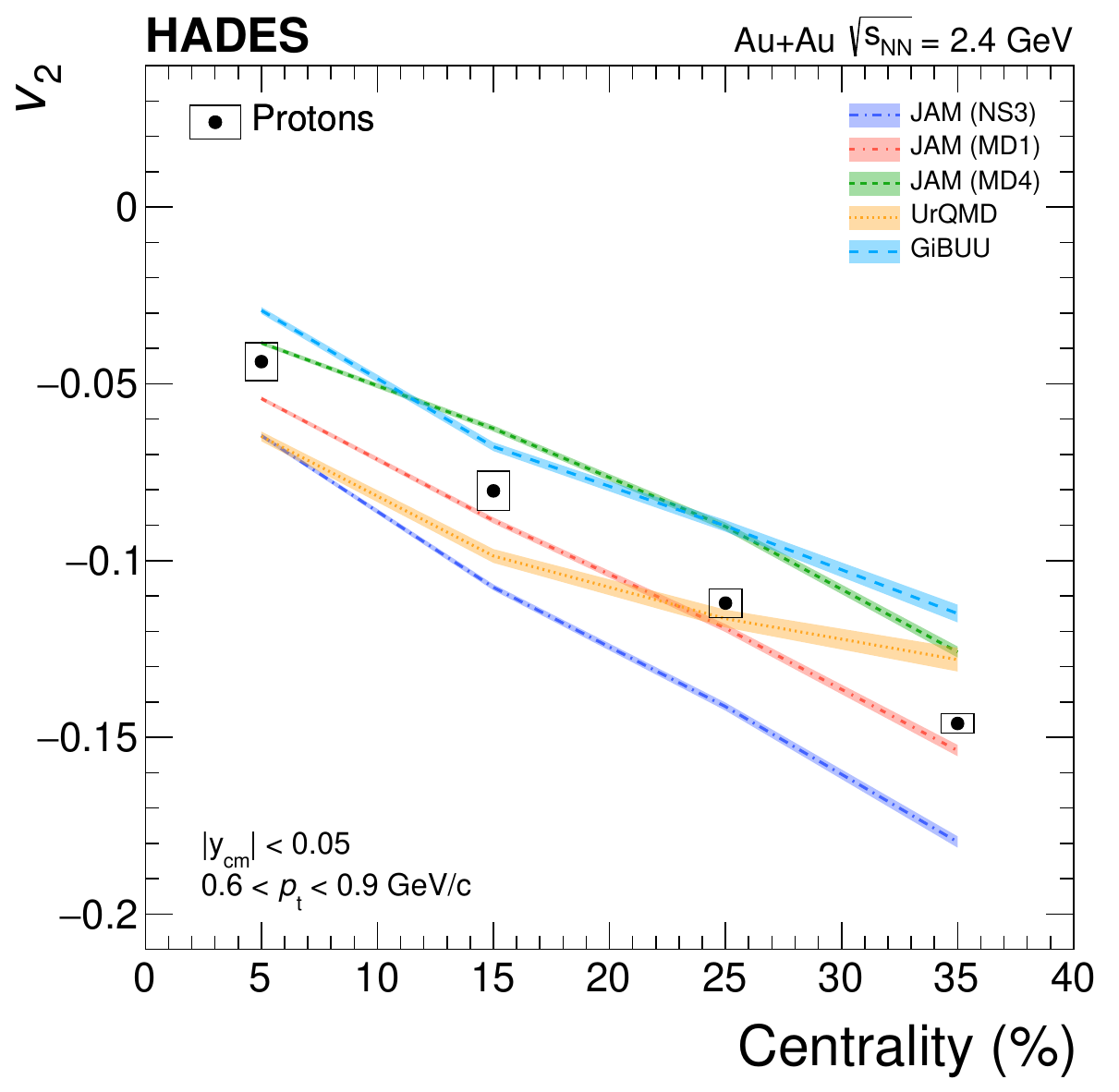}
            \vspace{-3mm}
            \caption{The slope  $dv_1/dy'|_{y'=0}$  (left panel) and
              elliptic $v_2$ (right panel) flow of protons in the interval
             0.6 $< p_T <$ 0.9 GeV/c at mid-rapidity in Au+Au collisions at $\sqrt{s_{NN}}$= 2.4 GeV  for four centrality classes. The HADES
            data are compared to several model predictions. The figure is taken from \cite{hadesvn}} \label{fig:hades2}
        \end{center}
       
        \vspace{-5mm}
 \end{figure}

\begin{figure}[hbt]
        \begin{center}
            \includegraphics[width=0.69\linewidth]{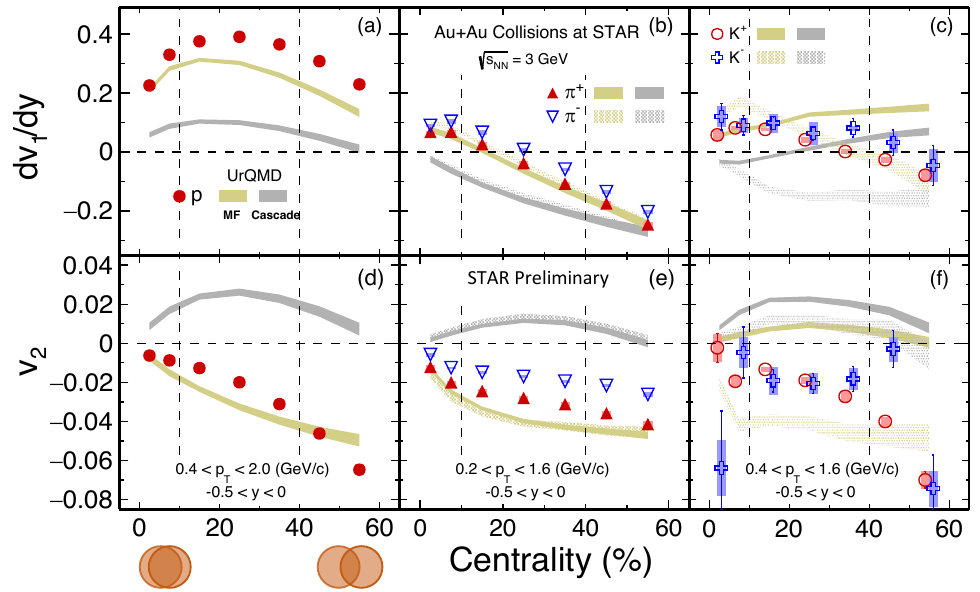}
            \vspace{-3mm}
            \caption{The centrality dependence of the slope  $dv_1/|_{y=0}$
              (upper panel) and  elliptic $v_2$ (lower panel) flow of protons, pions and kaons
              at mid-rapidity in Au+Au collisions at $\sqrt{s_{NN}}$= 3.0 GeV. The STAR
            data are compared to UrQMD model prediction. The figure is taken from \cite{star3gev,starv1}} \label{fig:star3gev}
        \end{center}
       
        \vspace{-5mm}
    \end{figure}

A large amount of data on measurements of directed $v_1$  and elliptic $v_2$  flow of protons
in nucleus-nucleus collisions in the energy region of  $\sqrt{s_{NN}}$= 2.4-5.0 GeV has been accumulated
over the past 20 years \cite{e895v1,e895v2,e895v22,fopi,hadesvn,star45,star3gev,starv1}.
At the moment, the main source of new experimental $v_n$ data is the analysis of Au + Au collision events,
which were collected by the STAR experiment as part of the Beam Energy Scan II program at RHIC \cite{star3gev,starv1,starbes2}, see Figure~\ref{v12bes2} as an example.
The main results of measurements of $v_1$  and $v_2$  of  protons can be summarized as follows:\\
 {\bf 1):} The relatively long passing time $t_{pass}$ leads to the interaction of particles with 
spectator nucleons, which flow predomently in the reaction plane. 
For Au+Au collisions at 2.4  $< \snn <$ 5 GeV, the $t_{pass}$  decreases from 18 fm/c to 6 fm/c. As result the  $v_1$, the slope of
the $v_1$ at mid-rapidity $dv_1/dy|_{y=0}$ and  $v_2$ of protons  decrease with increasing collision energy, see Figure~\ref{v12bes2}.
The $v_2$ signal is undergo the transition from $v_2<0$ (``out-of-plane'') to   $v_2>0$ (``in-plane'') at $\sqrt{s_{NN}}\sim$ 3.3 GeV \cite{e895v2,starbes2}.
All existing measurements of $v_1$ and $v_2$ of protons were performed with respect to the first-order event plane,
which is determined by the directed flow $v_1$ of the spectator nucleons. It is the dominant flow
signal in amplitude and does not change sign in this collision energy range.\\
{\bf 2):} While $v_1$ of protons is consistent with zero at mid-rapidity ($y_{cm}$=0), it rises towards forward
and decreases towards backward rapidities (see left panel of Figure~\ref{v12hades}). The rapidity dependence $v_2$ is opposite to $v_1$ ,
i.e. the absolute value of $v_2$ is largest at mid-rapidity and decreases towards forward and backward rapidities, see right panel of Figure~\ref{v12hades}.
The $v_1(p_T)$ of protons exhibits an almost linear rapid rise in the region $p_T < $ 0.6 GeV/c and then increases only moderately or even saturates for
$p_T > $ 1 GeV/c, see \cite{hadesvn} for plots. The $v_2$ values around mid-rapidity decrease (increase) continuously with $p_T$ for collision energies
below (above) $\sqrt{s_{NN}}\simeq$ 3.3 GeV.\\
{\bf 3):} The slope of $v_1$ of protons at mid-rapidity $dv_1/dy|_{y=0}$ exhibits no significant centrality dependence for all  $p_T$ intervals,
except for the very central class where $dv_1/dy|_{y=0}$ is smaller than for the other centralities, see left panel of Figures~\ref{fig:hades2} and upper panel of
Figure~\ref{fig:star3gev}. In contrast, the $v_2$ signal of protons has a strong
(almost linear) dependence on centrality, see right panel of  Figure~\ref{fig:hades2} and lower panel of Figure~\ref{fig:star3gev}. The fluctuations of $v_1$ and $v_2$ may lead to non-zero values in
the most central collisions. 
The strong $p_T$ and centrality dependence of $v_2$ can be explained in a simple way. 
A specific  particle  moving with transverse velocity $v_t$  will be shadowed by the spectator matter during the passage time $t_{pass}$.
 The simple  geometrical estimate then leads to the condition \cite{larionov}:   $v_t > (2R-b)/t_{pass}$, where $R$ is the radius of the nucleus
 and $b$ is the impact parameter.
  it is easier to fulfill this condition for the particle with high $p_T$ and for peripheral collisions.  \\

  {\bf 4):} The  detailed multi-differential study of flow coefficients $v_n$ of protons in relativistic heavy-ion collisions at
  $\snn$ = 2.4-5.0 GeV using several hadronic transport models: UrQMD \cite{urqmd}, PHQMD \cite{phqmd}, DCM-QGSM-SMM \cite{dcm}, SMASH \cite{eos} and
  JAM \cite{jam1,jam2,jam3}  and comparison with published   HADES/STAR proton flow data
  can found in \cite{eos,star3gev,hadesvn,mamaev,parfenov1}. 
  The cascade mode of all models (UrQMD, DCM-QGSM-SMM, SMASH, JAM)  failed to describe the existing experimental flow data~\cite{parfenov1}.
  The absence of a repulsive potential significantly reduces the $v_1$ and $v_2$ signals and results in essentially zero signals for the higher
  order ($v_3$ and $v_4$ ) flow coefficients for protons. However, by including the meanfield potential, the
  JAM and UrQMD models can qualitatively reproduce the HADES and STAR data for $p_T$ and rapidity ($y_{cm}$) dependence of
  anisotropic flow coefficients $v_1$ and  $v_2$  of protons in Au+Au collisions at $\snn$ = 2.4 and 3.0 GeV\cite{eos,star3gev,hadesvn,mamaev,parfenov1}.
  In the present work, we use the Jet AA Microscopic transport model (JAM) \cite{jam1,jam2,jam3} as the main event generator
  to simulate Xe+Cs(I) collisions for anticipated performance of the BM@N spectrometer for flow measurements of protons and for the comparison with first $v_1$ data.
   The nuclear mean field is simulated based on the relativistic version of the QMD model (RQMD.RMF)\cite{jam3}. We have used the version
JAM 1.9092  which includes five different EOS implementations, see \cite{parfenov1,mamaev} for details.

\section{Analysis details}

This section briefs about the related details for analysis of the experimental data for Xe+Cs(I) collisions at 3.8 AGeV (BM@N run8), such as the selection of good events
and tracks for analysis, particle identification used for selecting protons,
and the definition of collision centrality for geometry of collisions. 
Prior to conducting physical analysis, the data underwent a thorough evaluation to ensure that only good runs were included,
called run-by-run Quality Assurance (QA).

\subsection{The layout of the BM@N experiment}

\begin{figure}[hbt]
        \begin{center}
            \includegraphics[width=0.76\linewidth]{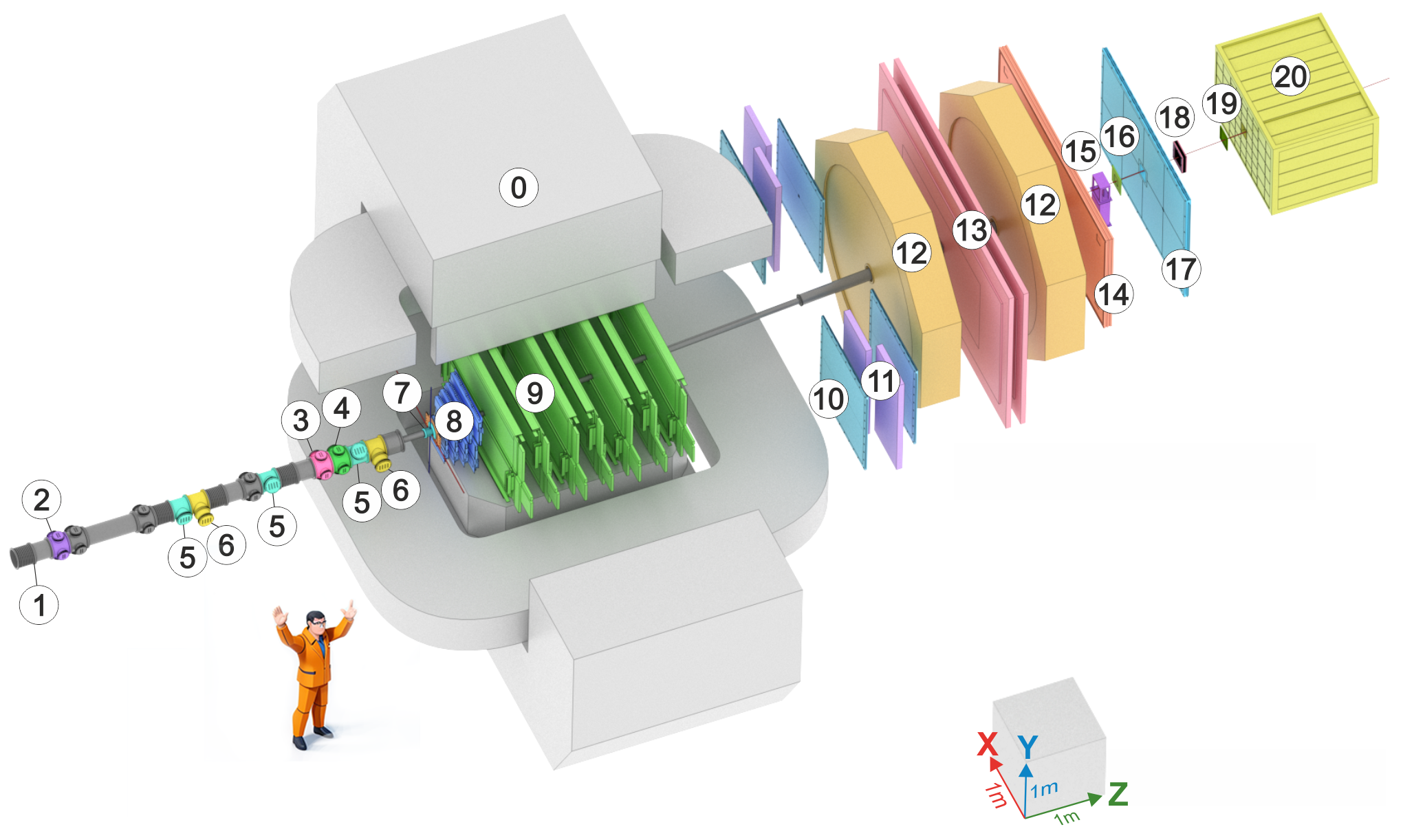}
            \vspace{-3mm}
            \caption{The layout of the BM@N experiment for the Xe+Cs(I) run8 2022-2023. Main components: (0) SP-41 analyzing magnet, (1) vacuum beam pipe, (2) BC1 beam counter, (3) Veto counter (VC), (4) BC2 beam counter, (5) Silicon Beam Tracker (SiBT), (6) Silicon beam profilometers, (7) Barrel Detector (BD) and Target station, (8) Forward Silicon Detector (FSD), (9) Gaseous Electron Multiplier (GEM) detectors, (10) Small cathode strip chambers (Small CSC), (11) TOF400 system, (12) drift chambers (DCH), (13) TOF700 system, (14) Scintillation Wall (ScWall), (15) Fragment Detector (FD), (16) Small GEM detector, (17) Large cathode strip chamber (Large CSC), (18) gas ionization chamber as beam profilometer, (19) Forward Quartz Hodoscope (FQH), (20) Forward Hadron Calorimeter (FHCal). The figure is taken from \cite{bmn}}  \label{fig:bmn2023}
        \end{center}
      
        \vspace{-5mm}
    \end{figure}

The BM@N detector is a forward spectrometer that covers
the pseudorapidity range 1.6$\le \eta \le$ 4.4 \cite{bmn}. The layout of the BM@N experiment for the Xe+Cs(I) run8
 is shown in the Figure~\ref{fig:bmn2023}. The main subsystems of the BM@N \cite{bmn} are the  tracking system for charged hadron tracking, the Time Of Flight (TOF)
system for charged particle identification and the set of forward detectors for
centrality and reaction plane estimations.
The tracking system is comprised of  4 stations of  the Forward Silicon Detector (FSD)
and 7 stations of Gaseous Electron Multipliers (GEM) chambers mounted downstream of the silicon sensors, see left part of  Figure~\ref{fig:bmn2023}.
Both the silicon tracking system (FSD)
and the GEM stations will be operated in the magnetic field (at maximum value of 1.2 T) of a large aperture
dipole magnet and allow the reconstruction of the momentum $p$  of charged particles.
The $z$ axis of the BM@N coordinate system
is directed along the beam line, while the magnetic field is directed along the $y$ axis.
The FSD+GEM system provides also
the measurements of the multiplicity of the produced charged particles $N_{ch}$, which can be used as an  estimator of the collision centrality.

The TOF-system consists of 3 planes of multi-gap Resistive Plate Chambers (mRPC) placed
at $z=400$ and $z=700$ cm (TOF-400 and TOF-700, respectively) from the target, see the central
part of Figure~\ref{fig:bmn2023}. The detectors BC1 and BC2 define the start time for the time-of-flight system.
Three forward detectors: Forward Hadronic Calorimeter (FHCal), quartz hodoscope (Hodo) and Scintillator Wall (ScWall)   provide the information
about the spectator fragments, see the  right part of the Figure~\ref{fig:bmn2023}.
FHCal  provides the information about the energy of spectators fragments and consists of 54 modules.
The  modules have sampling structure and consist of a set of lead and scintillator plates
compressed together by a steel band. FHCal has a 15 x 15 cm$^2$ square beam hole in the center. The beam hole leads to the leakage of
the fragments with small transverse momenta. As a result,
the deposited energy in the FHCal is comparable for the central and peripheral events. This creats  an ambiguity in the dependence of energy deposition on the collision centrality.
New forward quartz hodoscope  (Hodo) has  been developed to be placed in the beam hole to measure the energy of spectator fragments.
It helps to compensate the effect due to the 
leakage of the heavy fragments mostly in the peripheral collisions.
ScWall has a wider acceptance than FHCal and provides information about the charge of spectator fragments.

\subsection{Quality Assurance (QA) study}

The collection of events for a collision energy is done over several discrete time spans.
Each of these time spans where the detector was continuously recording events is called
a "run" and it can be selected by RunId. Each
run consists of event and track information of the heavy-ion collisions recorded by the
BM@N detector. We perform quality assurance (QA) checks for the selection of good runs.  
Averaged QA observables like: $N_{ch}$  (charged particle multiplicity in FSD+GEM system), $E_{tot}$ (total energy
of spectator fragments in the FHCal), $N_{vtx}$ (number of tracks in the  vertex reconstruction), etc.,
are calculated for each run.  Then, the  mean ($\mu$) and standard deviation ($\sigma$) are calculated for the
distribution of selected observables $Y$ as a function of RunId:
    \begin{equation}\label{QaRunByRun_mu}
        \mu = \frac{1}{N} \sum_{i=1}^{N} Y_i  
    \end{equation}
    \begin{equation}\label{QaRunByRun_sigma}
        \sigma = \sqrt{ \frac{1}{N} \sum_{i=1}^{N} (Y_i - \mu)^{2} },
    \end{equation}
    where i - RunId number and N  - total numbers of runs.
 The runs for which the
averaged QA observables lie beyond $\pm 3\sigma$ away from their global means are identified as
bad runs, and all the events from that run are removed from the analysis.
 \begin{verbatim}
    Converter (DST to QA tree): https://github.com/DemanovAE/convertBmn.git
    QA code: https://github.com/DemanovAE/QA_bmn.git
    DST run8 data: /eos/nica/bmn/exp/dst/run8/24.04.0 (May 2024)
    QA Data .tree.root at Clusters
    NICA: /nica/mpd1/demanov/data_bmn/run8_vf_24.04.0
    HybriLIT: /lustre/home/user/a/ademanov/bmn/data/run8_vf_24.04.0
\end{verbatim}

\begin{figure}[H]
        \begin{center}
        \subfigure[]{
            \includegraphics[width=0.45\linewidth]{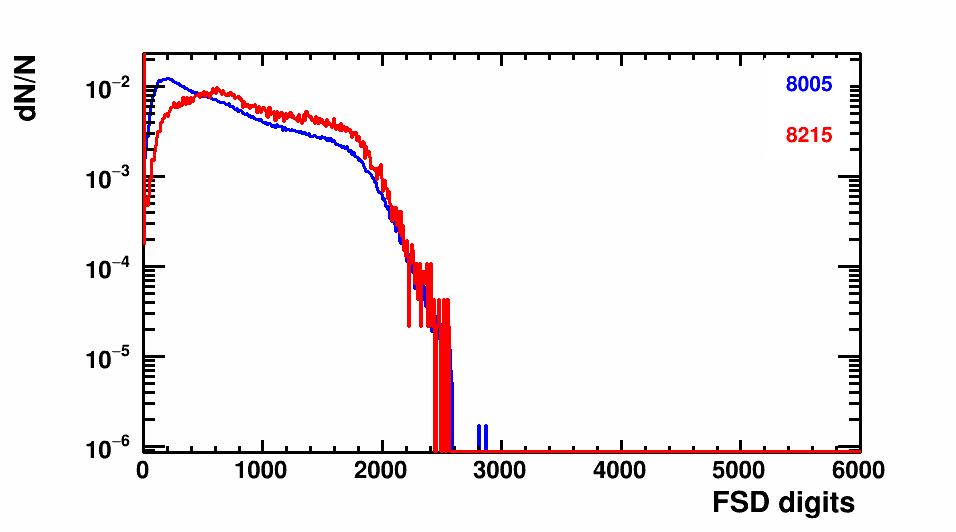} 
            \label{fig:FSDdigits} }  
        \subfigure[]{
            \includegraphics[width=0.45\linewidth]{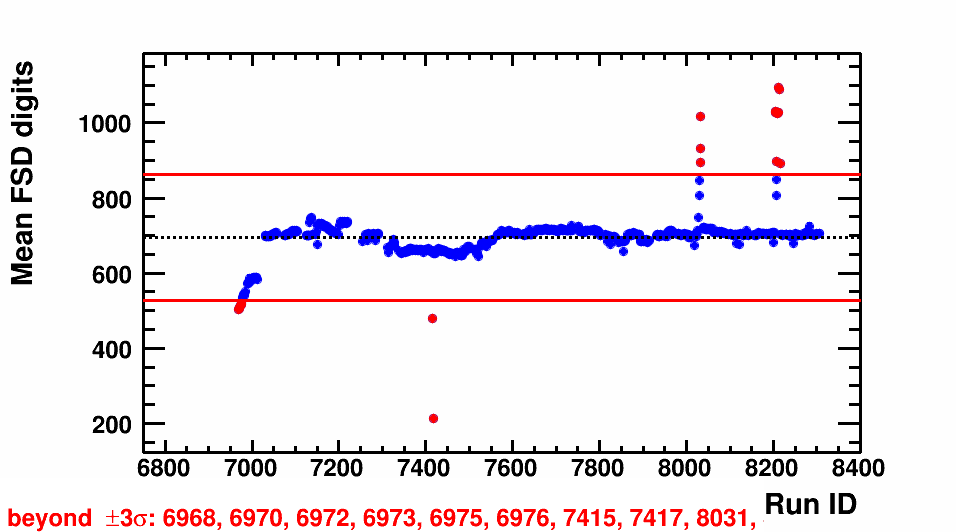} 
            \label{fig:FSDdigitsRun} }
        \subfigure[]{
            \includegraphics[width=0.45\linewidth]{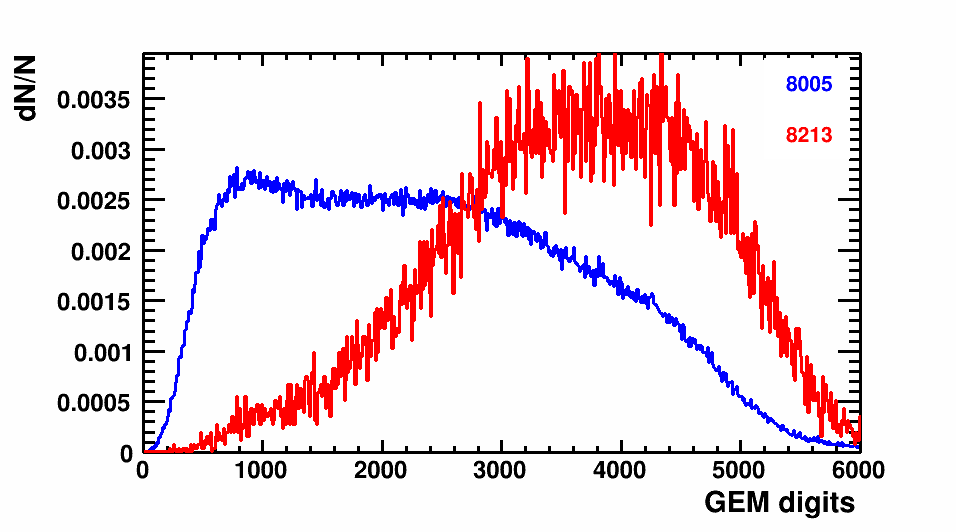} 
            \label{fig:GEMdigits} }  
        \subfigure[]{
            \includegraphics[width=0.45\linewidth]{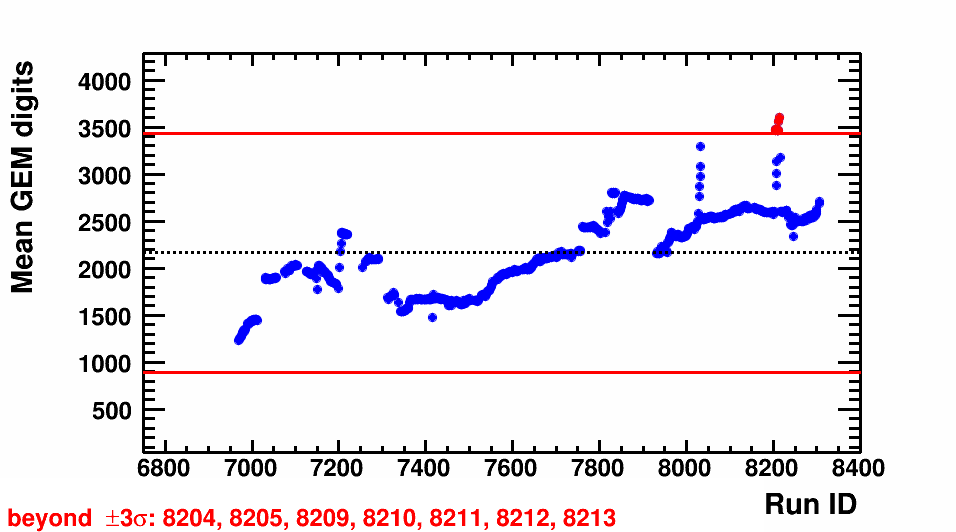}
            \label{fig:GEMdigitsRun} }
        \vspace{-3mm}
        \caption{Distribution of the number of digits in the FSD \subref{fig:FSDdigits} and GEM \subref{fig:GEMdigits} detectors.
          The red marker corresponds to the distribution from the "outlier" RunId. Mean number of FSD digits \subref{fig:FSDdigitsRun} and
          GEM digits \subref{fig:GEMdigitsRun} as a function of RunID (right panel). Black dotted horizontal
          line and red horizontal lines represent $\mu$ and $\pm3\sigma$, respectively.} \label{ris:DigitsQA}
        \end{center}
       
        \vspace{-5mm}
    \end{figure}

\begin{figure}[H]
        \begin{center}
        \subfigure[]{
            \includegraphics[width=0.45\linewidth]{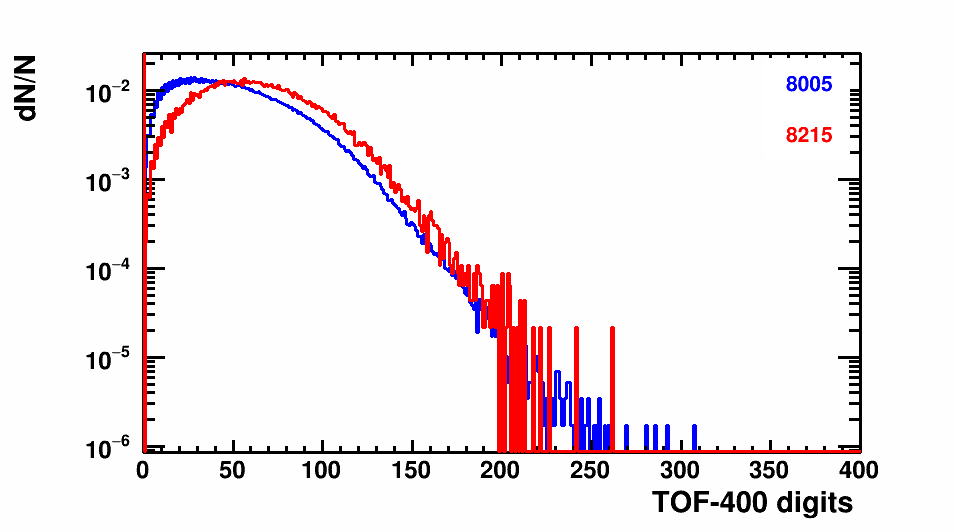} 
            \label{fig:TOF400digits} }  
        \subfigure[]{
            \includegraphics[width=0.45\linewidth]{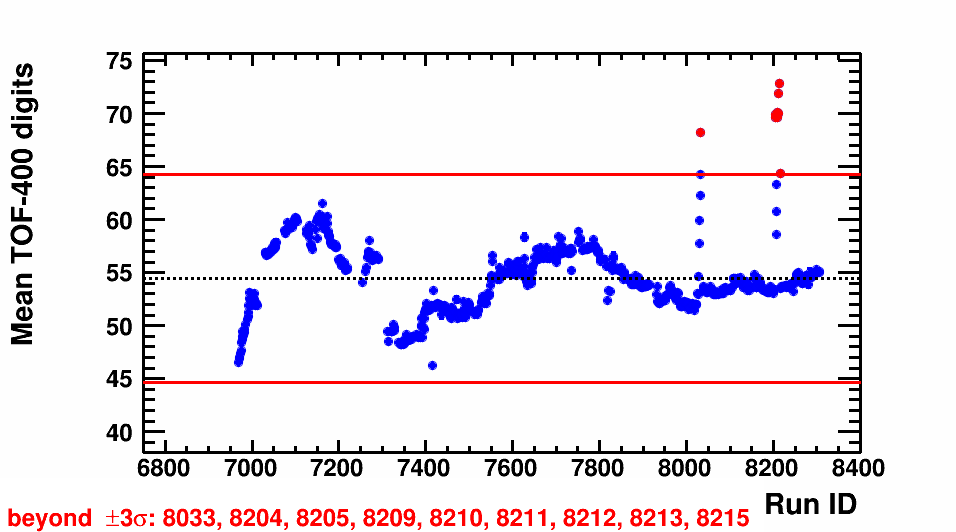} 
            \label{fig:TOF400digitsRun} }
        \subfigure[]{
            \includegraphics[width=0.45\linewidth]{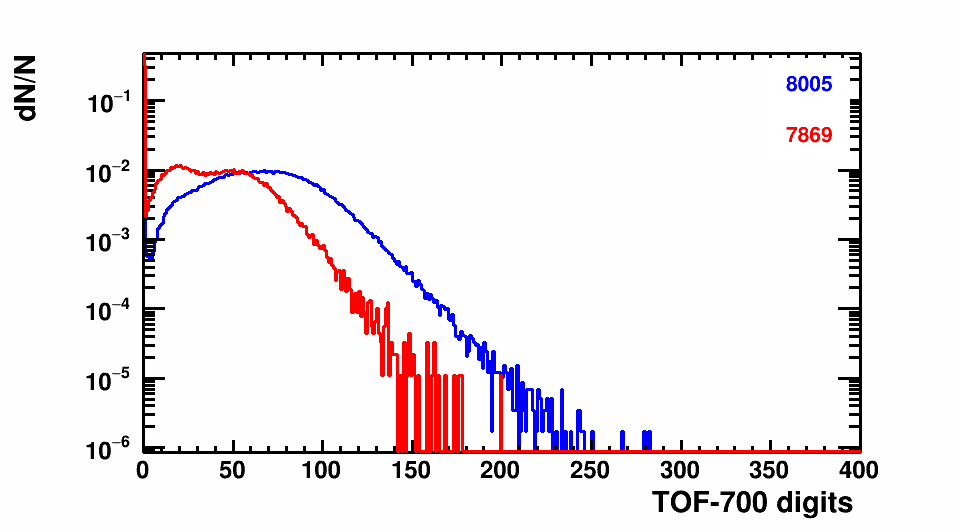} 
            \label{fig:TOF700digits} }  
        \subfigure[]{
            \includegraphics[width=0.45\linewidth]{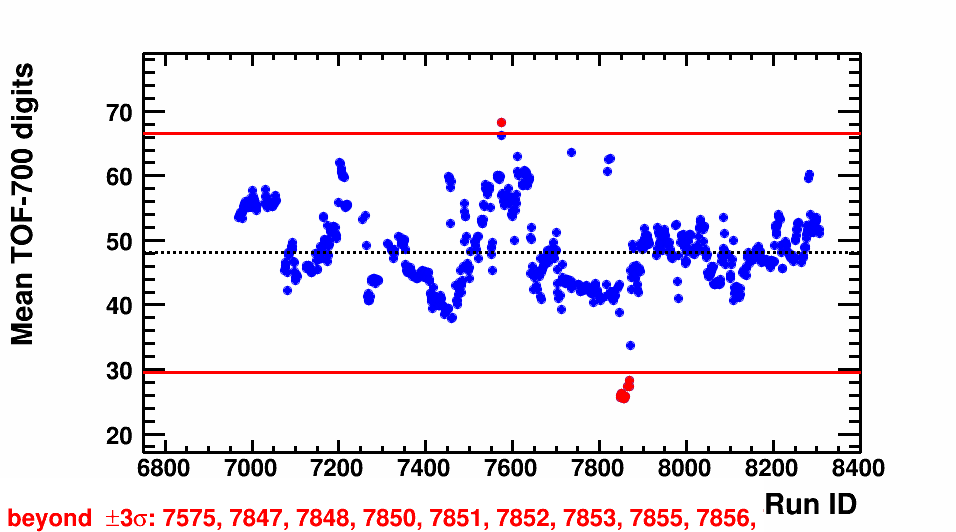}
            \label{fig:TOF700digitsRun} }
        \vspace{-3mm}
        \caption{Distribution of the number of digits in the TOF400 \subref{fig:TOF400digits} and TOF700 \subref{fig:TOF700digits} detectors.
          The red marker corresponds to the distribution from the "outlier" RunId. Mean number of TOF400 digits \subref{fig:TOF400digitsRun}
          and TOF700 digits \subref{fig:TOF700digitsRun} as a function of RunID (right panel).
          Black dotted horizontal line and red horizontal lines represent $\mu$ and $\pm3\sigma$, respectively.} \label{ris:DigitsQA2}
        \end{center}
       
        \vspace{-5mm}
    \end{figure}

\begin{figure}[H]
        \begin{center}
            \includegraphics[width=0.49\linewidth]{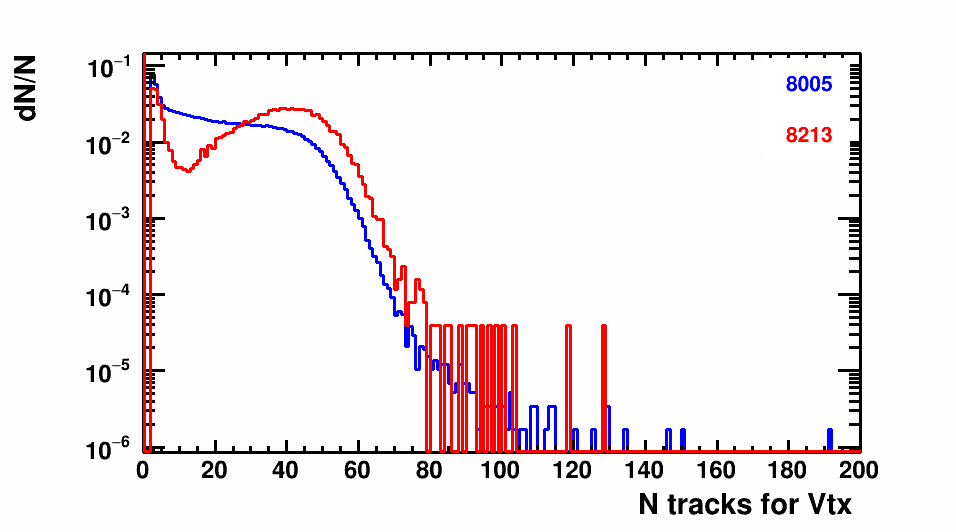}
            \includegraphics[width=0.49\linewidth]{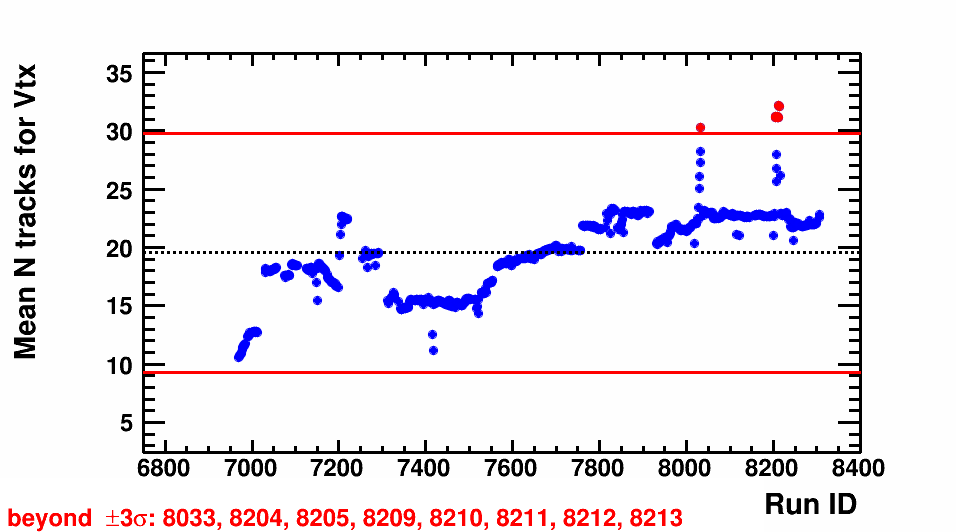}
            \vspace{-3mm}
            \caption{Left panel: distribution of the number of tracks in the vertex reconstruction.
              The red marker corresponds to the distribution from the "outlier" RunId. Right panel:
              Mean the number of tracks in vertex reconstruction as a function of RunID.
              Black dotted horizontal line and red horizontal lines represent $\mu$ and $\pm3\sigma$, respectively.} \label{VtxNTracks}
        \end{center}
       
        \vspace{-5mm}
    \end{figure}

\begin{figure}[H]
        \begin{center}
            \includegraphics[width=0.32\linewidth]{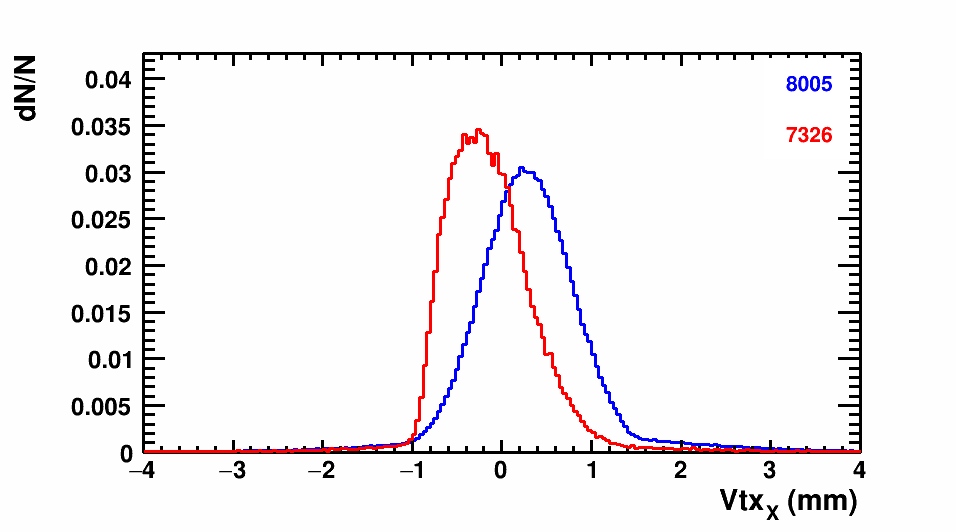}
            \includegraphics[width=0.32\linewidth]{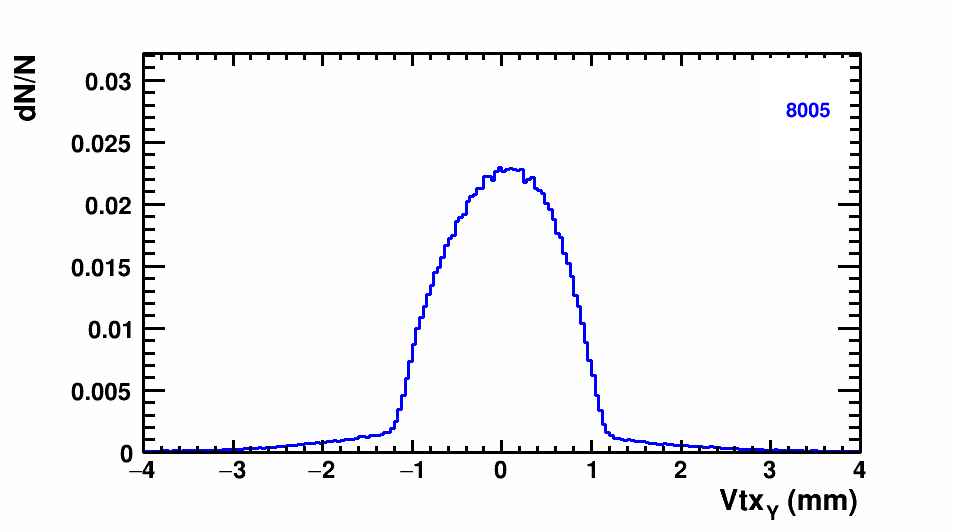}
            \includegraphics[width=0.32\linewidth]{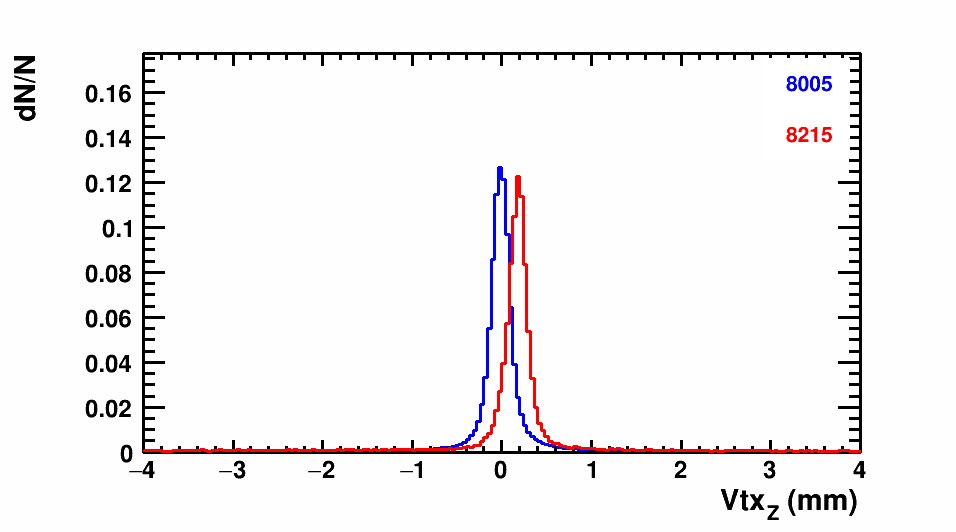}
            \includegraphics[width=0.32\linewidth]{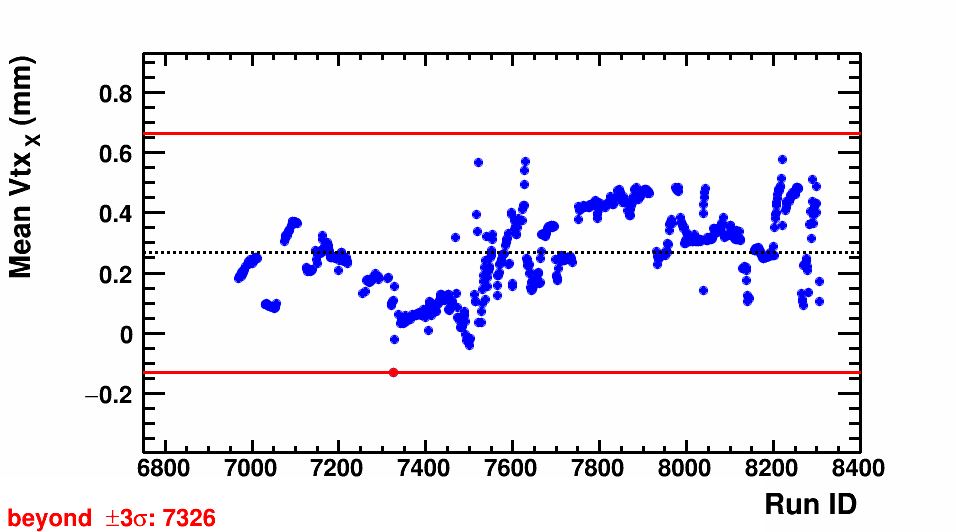}
            \includegraphics[width=0.32\linewidth]{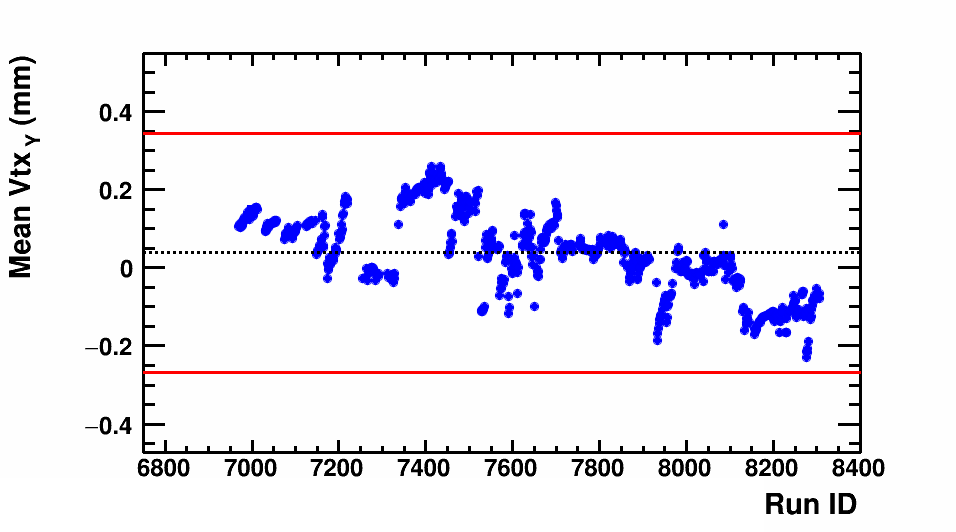}
            \includegraphics[width=0.32\linewidth]{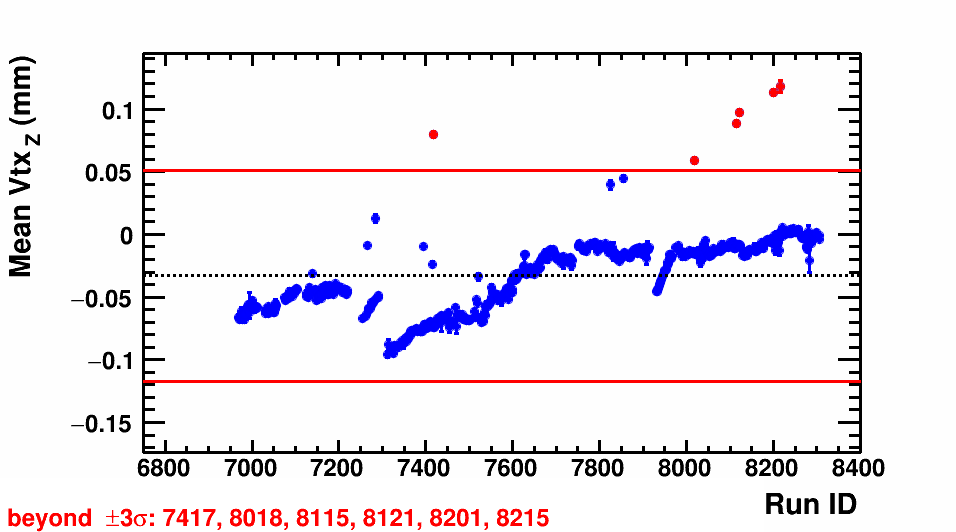}
            \vspace{-3mm}
            \caption{Upper panels: distribution of the x, y and z positions of vertex. The red marker corresponds to the
              distribution from the "outlier" RunId. Bottom panels: Mean of the x, y and z positions of the
              vertex as a function of  RunID. Black dotted horizontal line and red horizontal lines
              represent $\mu$ and $\pm3\sigma$, respectively.} \label{Vertex}
        \end{center}
       
        \vspace{-5mm}
    \end{figure}

\begin{figure}[H]
        \begin{center}
            \includegraphics[width=0.49\linewidth]{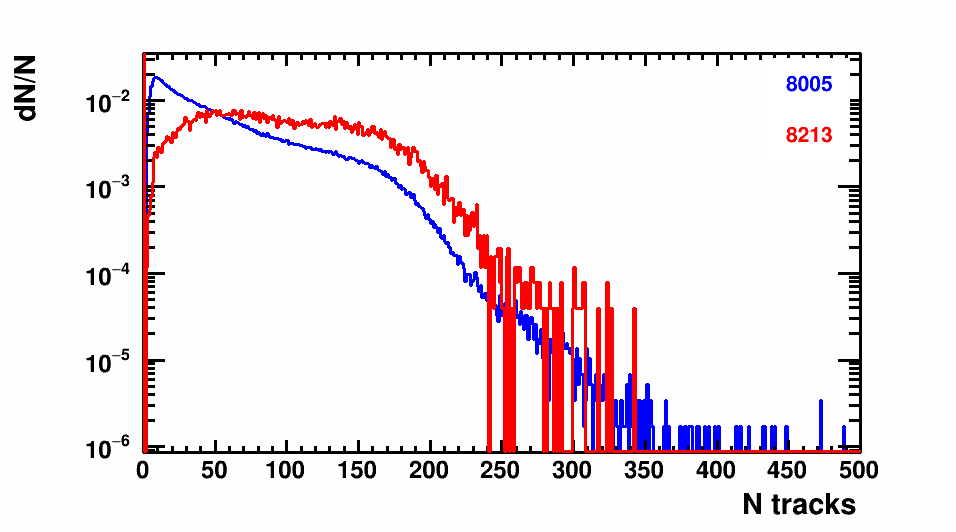}
            \includegraphics[width=0.49\linewidth]{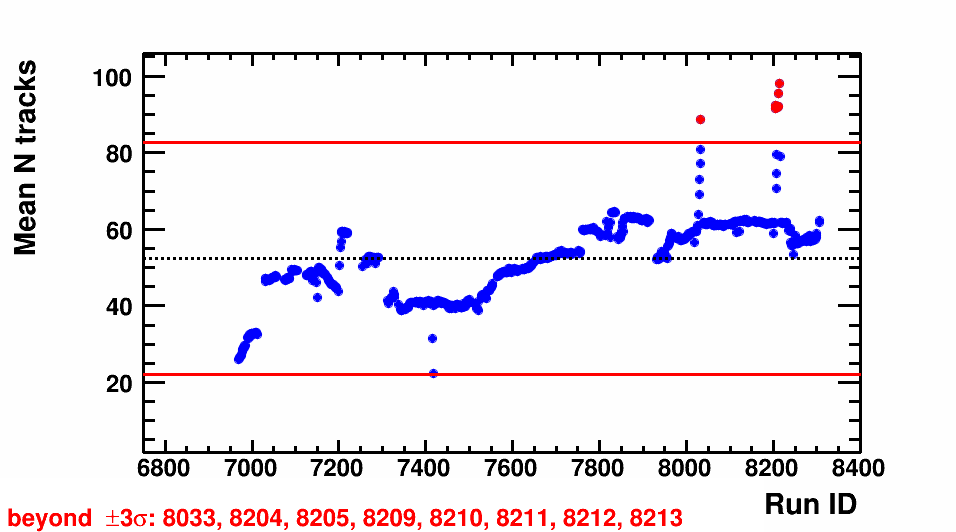}
            \vspace{-3mm}
            \caption{Upper panels: Distribution of the number of charged particles $N_{ch}$ in the tracking
              system (FSD + GEM). The red marker corresponds to the distribution from the "outlier" RunId.
              Bottom panels: Mean multiplicity as a function of  RunID. Black dotted horizontal line and
              red horizontal lines represent $\mu$ and $\pm3\sigma$, respectively.}  \label{mult}
        \end{center}
      
        \vspace{-5mm}
    \end{figure}

 Several examples of the application of the QA checks for different BM@N  observables which provide the
 event and track information can be found below. \\
{\bf 1)} Figures~\ref{ris:DigitsQA}--\ref{ris:DigitsQA2} show the RunId dependence of the mean number of   FSD, GEM, TOF400 and TOF700 digits.
Black dotted horizontal line and red horizontal lines represent $\mu$ and $\pm3\sigma$, respectively.\\
{\bf 2)} Figure~\ref{VtxNTracks} shows the RunId dependence of the mean number of tracks used  in the vertex reconstruction. 
Figure~\ref{Vertex} shows the  RunId dependence of the mean  x, y and z positions of the  reconstructed vertex.  \\
{\bf 3)} Figure~\ref{mult} shows the RunId dependence of the mean multiplicity of charged particles in the tracking
    system (FSD + GEM)\\

\begin{figure}[H]
        \begin{center}
            \includegraphics[width=0.49\linewidth]{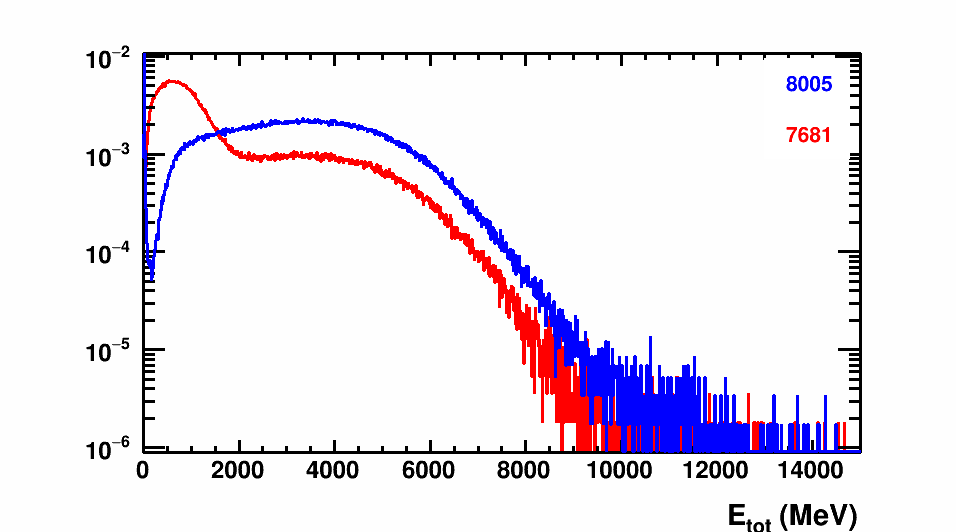}
            \includegraphics[width=0.49\linewidth]{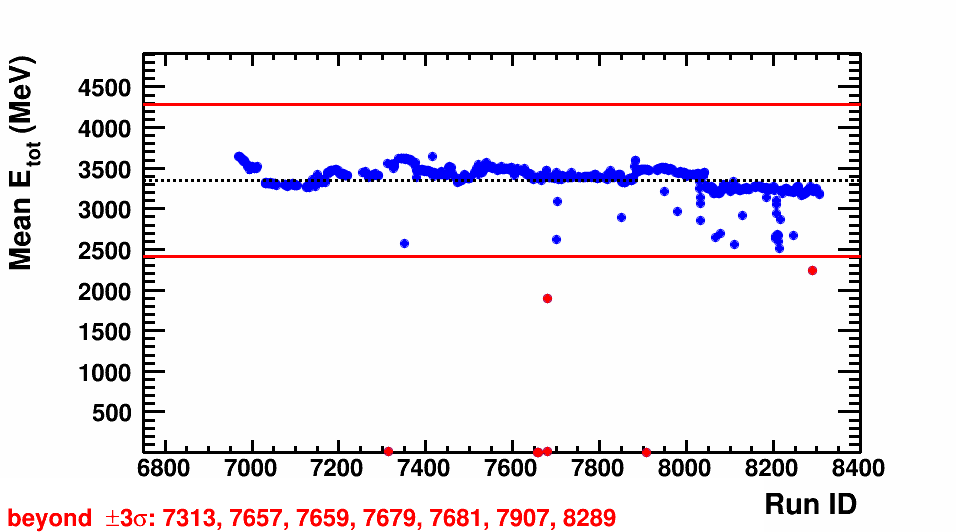}
            \vspace{-3mm}
            \caption{Left panel: distribution of the total energy $E_{tot}$ of spectator fragments in the FHCal.
              The red marker corresponds to the distribution from the "outlier" RunId. Right panel:
              Mean $E_{tot}$ as a function of  RunID. Black dotted horizontal line and red horizontal lines
              represent $\mu$ and $\pm3\sigma$, respectively.} \label{FHCal}
        \end{center}
       
        \vspace{-5mm}
    \end{figure}

\begin{figure}[H]
        \begin{center}
            \includegraphics[width=0.49\linewidth]{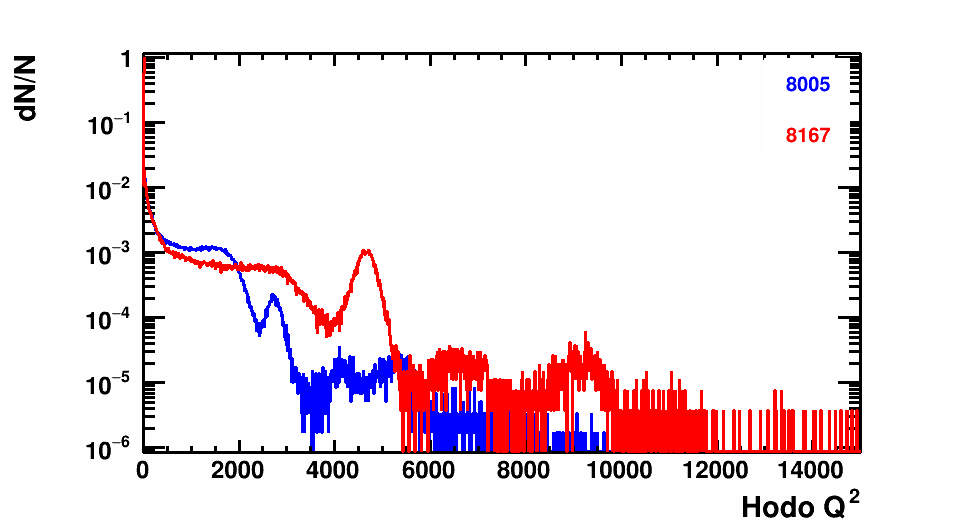}
            \includegraphics[width=0.49\linewidth]{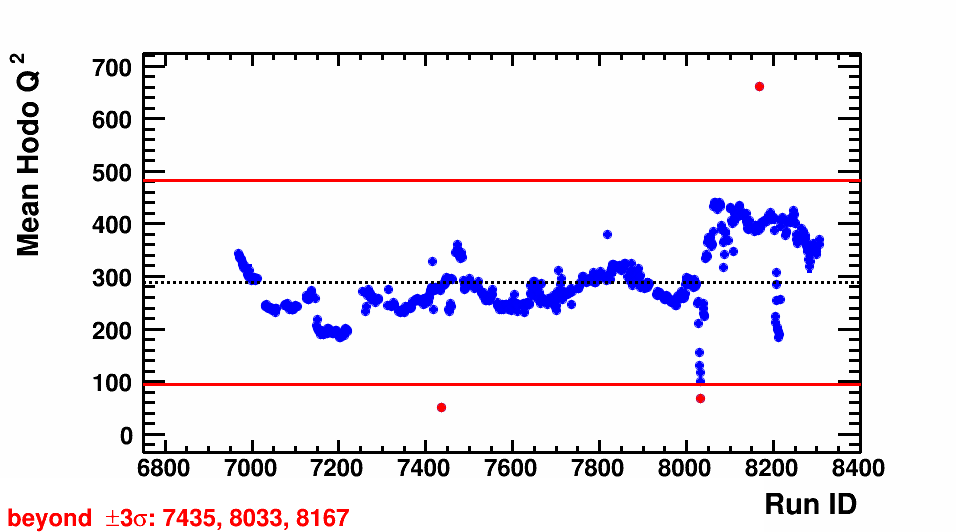}
            \vspace{-3mm}
            \caption{Left panel: distribution of the charge ($Q^2$) of spectator fragments in
              the forward quartz hodoscope (FQH). The red marker corresponds to the distribution from the "outlier"
              RunId. Right panel: Mean $Q^2$ as a function of RunID. Black dotted horizontal line
              and red horizontal lines represent $\mu$ and $\pm3\sigma$, respectively.} \label{FQH}
        \end{center}
       
        \vspace{-5mm}
    \end{figure}

\begin{figure}[H]
        \begin{center}
            \includegraphics[width=0.32\linewidth]{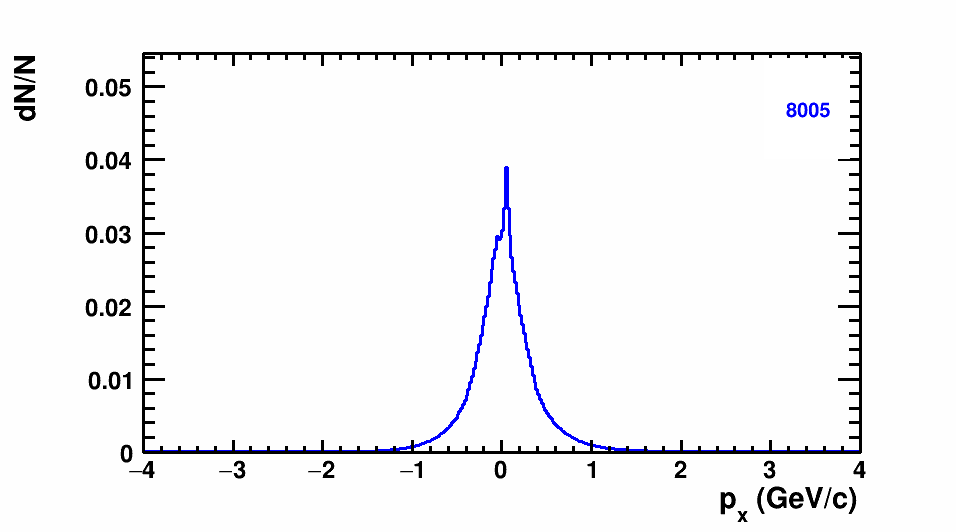}
            \includegraphics[width=0.32\linewidth]{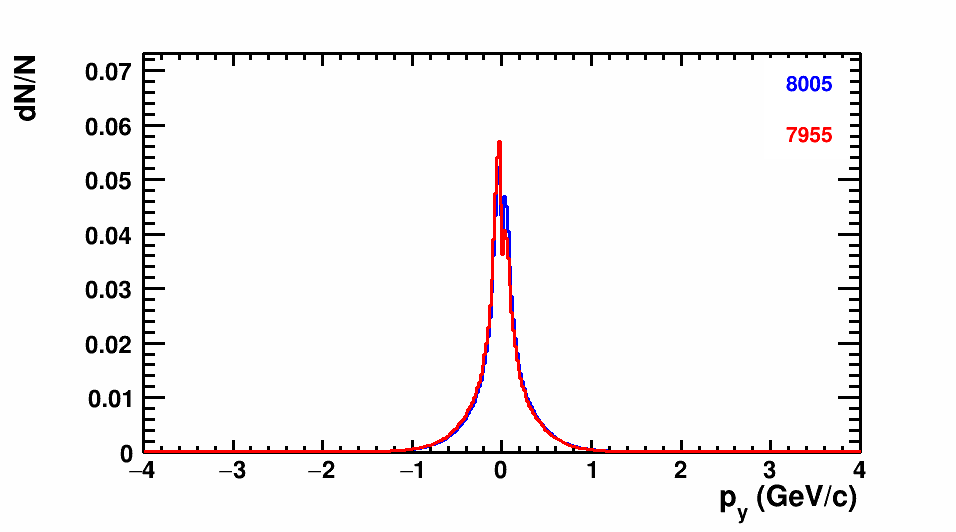}
            \includegraphics[width=0.32\linewidth]{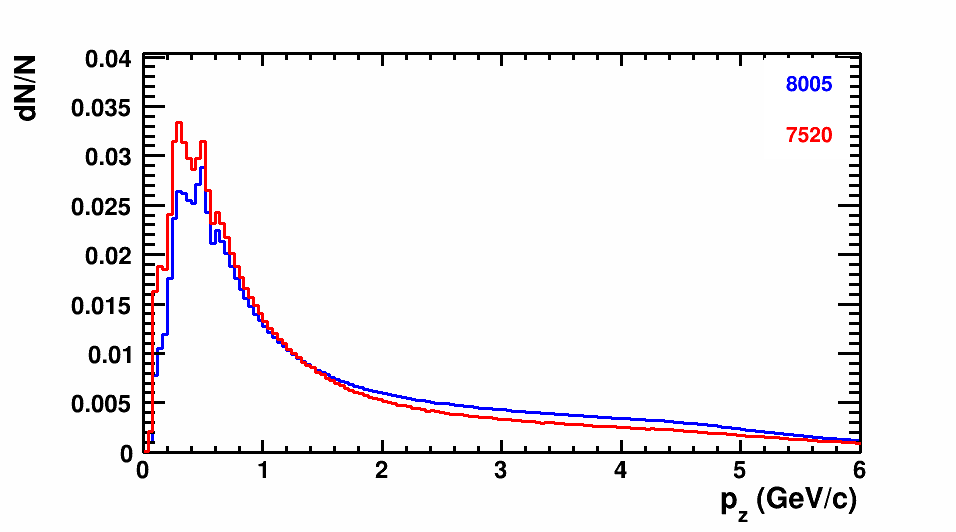}
            \includegraphics[width=0.32\linewidth]{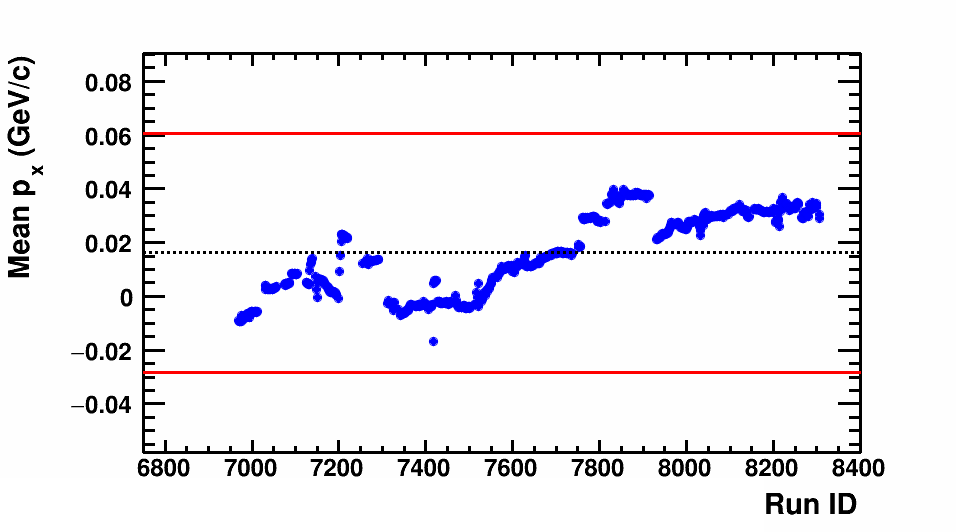}
            \includegraphics[width=0.32\linewidth]{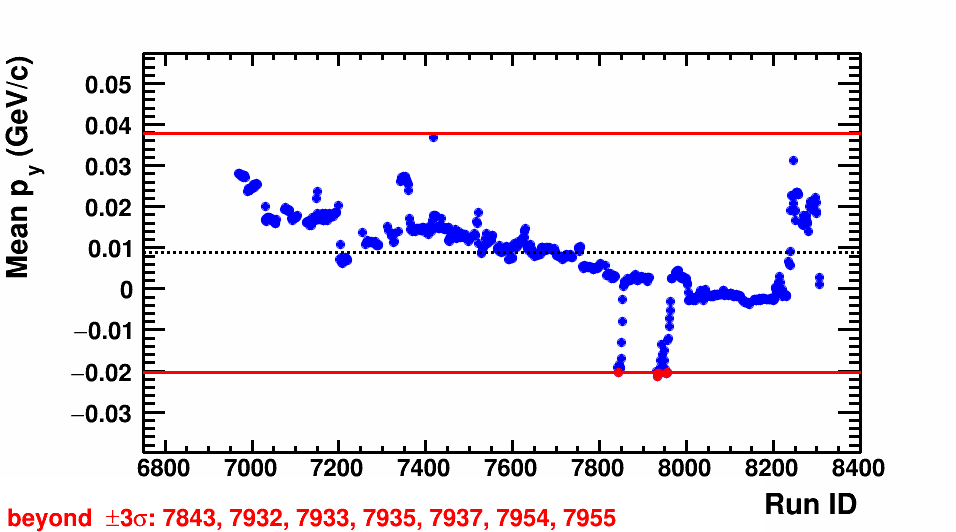}
            \includegraphics[width=0.32\linewidth]{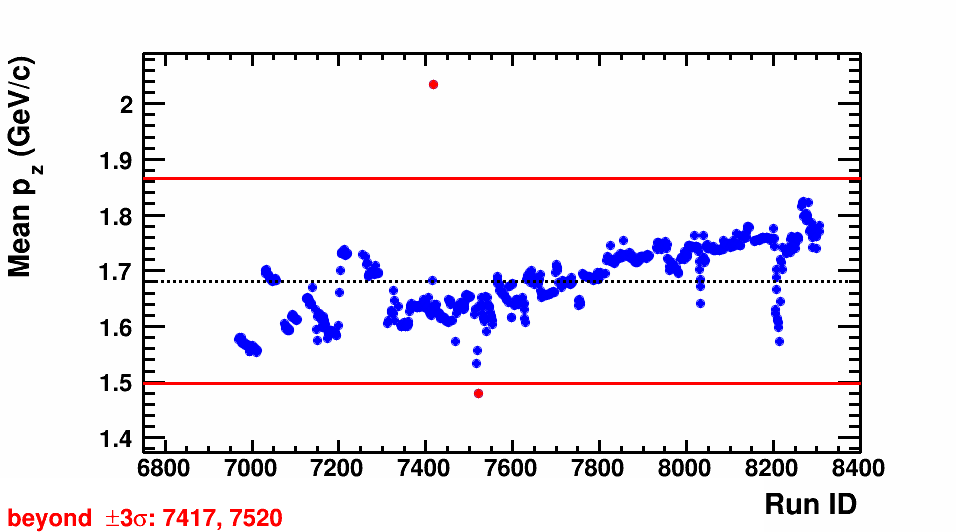}
            \vspace{-3mm}
            \caption{Upper panels: Distribution of the x, y and z components of momentum of charged particles.
              The red marker corresponds to the distribution from the "outlier" RunId. Bottom panels:
              Mean of  x, y and z components of momentum as a function of RunID.
              Black dotted horizontal line and red horizontal lines represent $\mu$ and $\pm3\sigma$, respectively.} \label{Momentum}
        \end{center}
       
        \vspace{-5mm}
    \end{figure}

\begin{figure}[H]
        \begin{center}
            \includegraphics[width=0.32\linewidth]{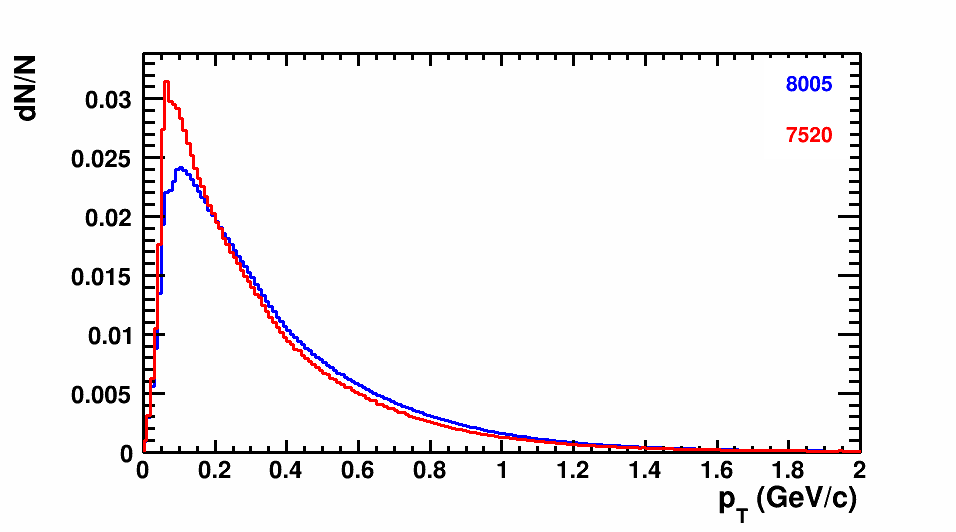}
            \includegraphics[width=0.32\linewidth]{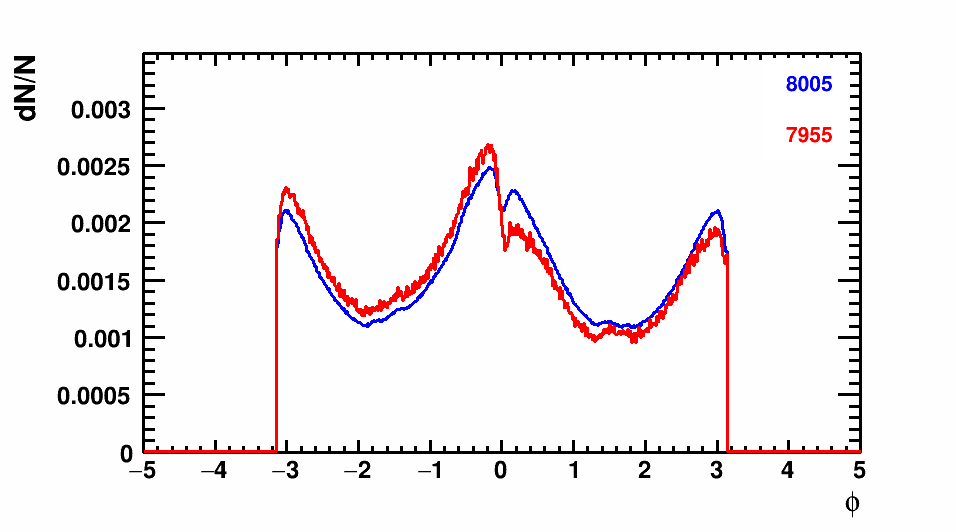}
            \includegraphics[width=0.32\linewidth]{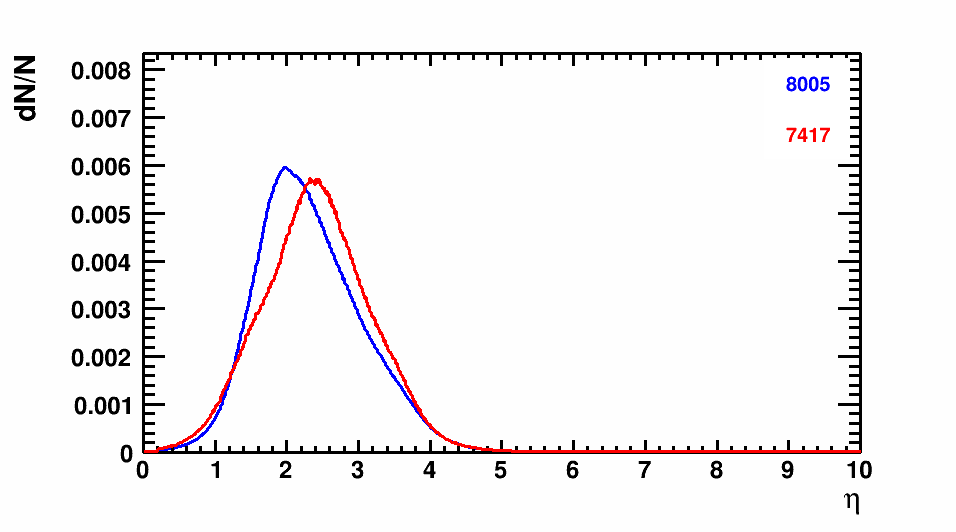}
            \includegraphics[width=0.32\linewidth]{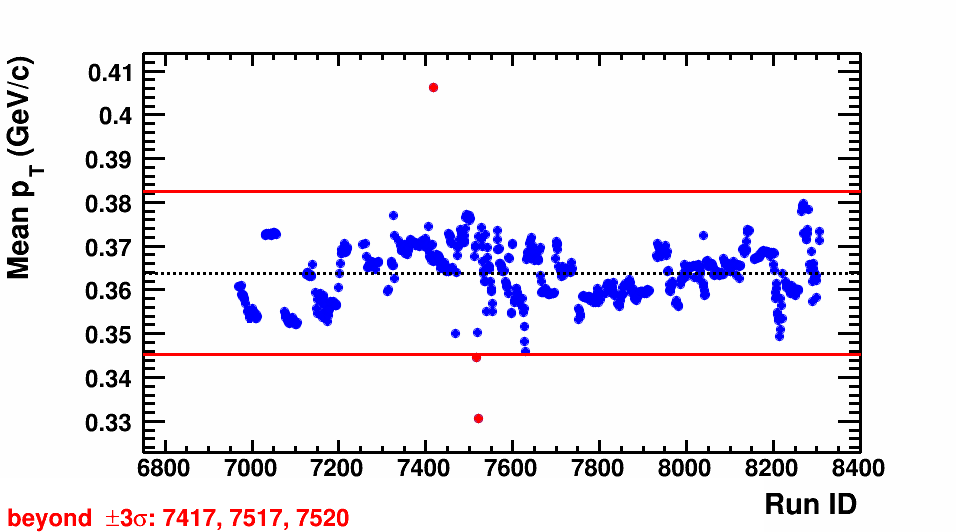}
            \includegraphics[width=0.32\linewidth]{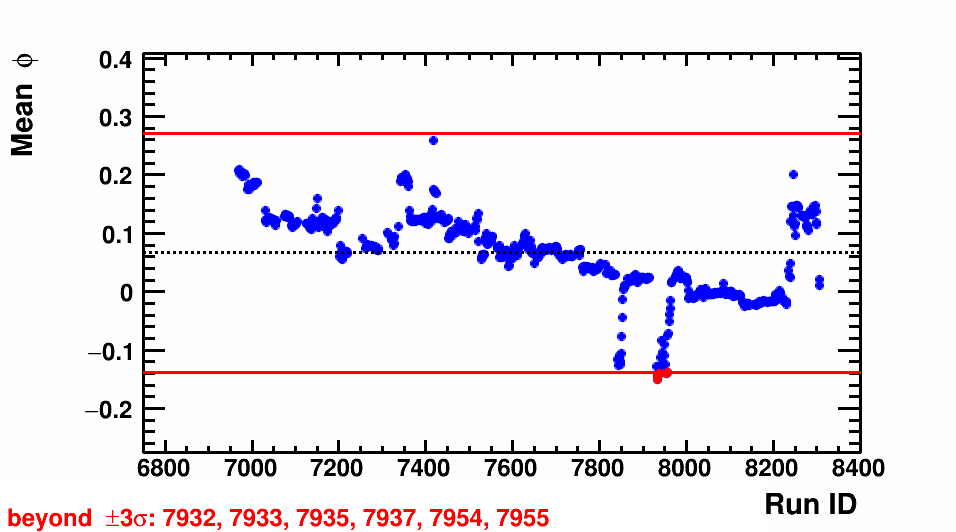}
            \includegraphics[width=0.32\linewidth]{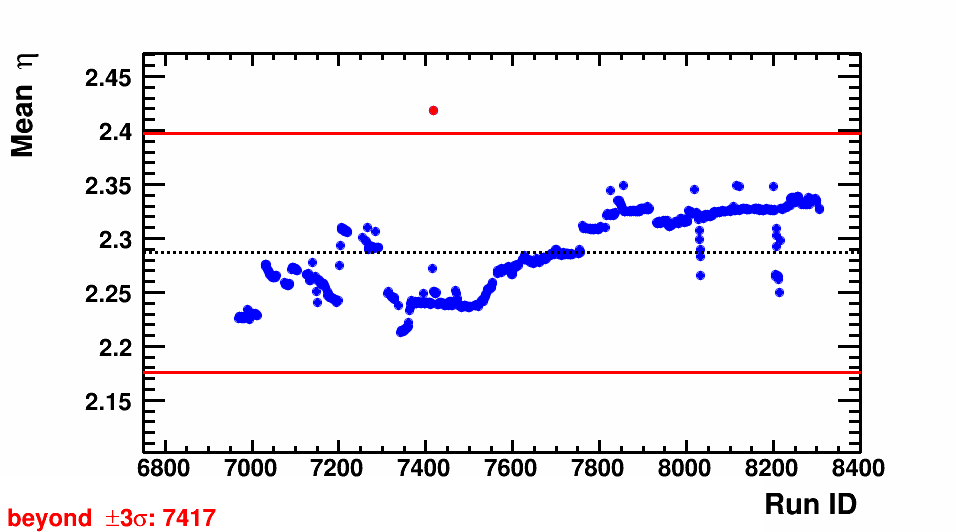}
            \vspace{-3mm}
            \caption{Upper panels: Distributions of the $p_T$ (left), azimuthal angle $\phi$ (center) and $\eta$ (right) of charged particles.
              The red marker corresponds to the distribution from the "outlier" RunId. Bottom panels: Mean
              $p_T$, $\phi$ and $\eta$ as a function of  RunID. Black dotted horizontal line and red horizontal
              lines represent $\mu$ and $\pm3\sigma$, respectively.} \label{track_par}
        \end{center}
       
        \vspace{-5mm}
    \end{figure}

\begin{figure}[H]
        \begin{center}
            \includegraphics[width=0.32\linewidth]{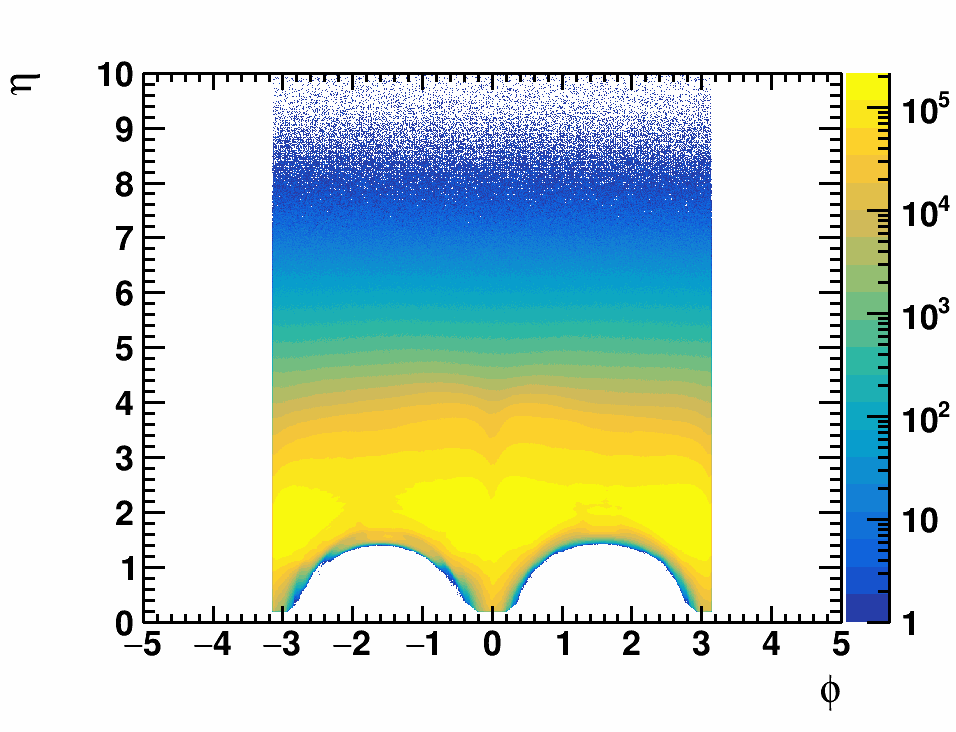}
            \includegraphics[width=0.32\linewidth]{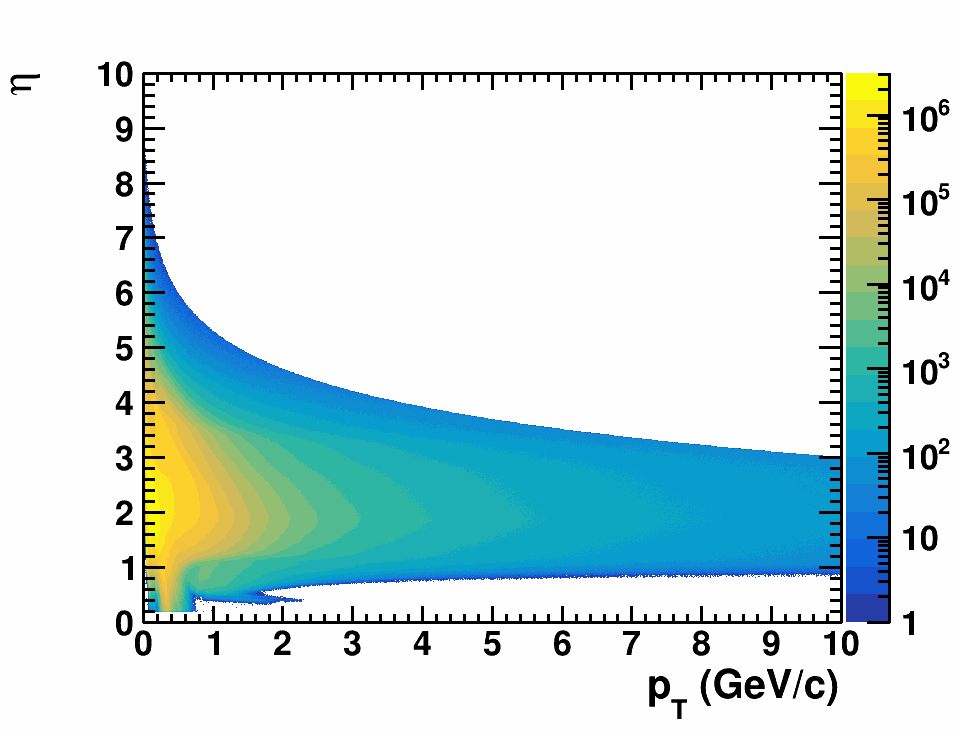}
            \includegraphics[width=0.32\linewidth]{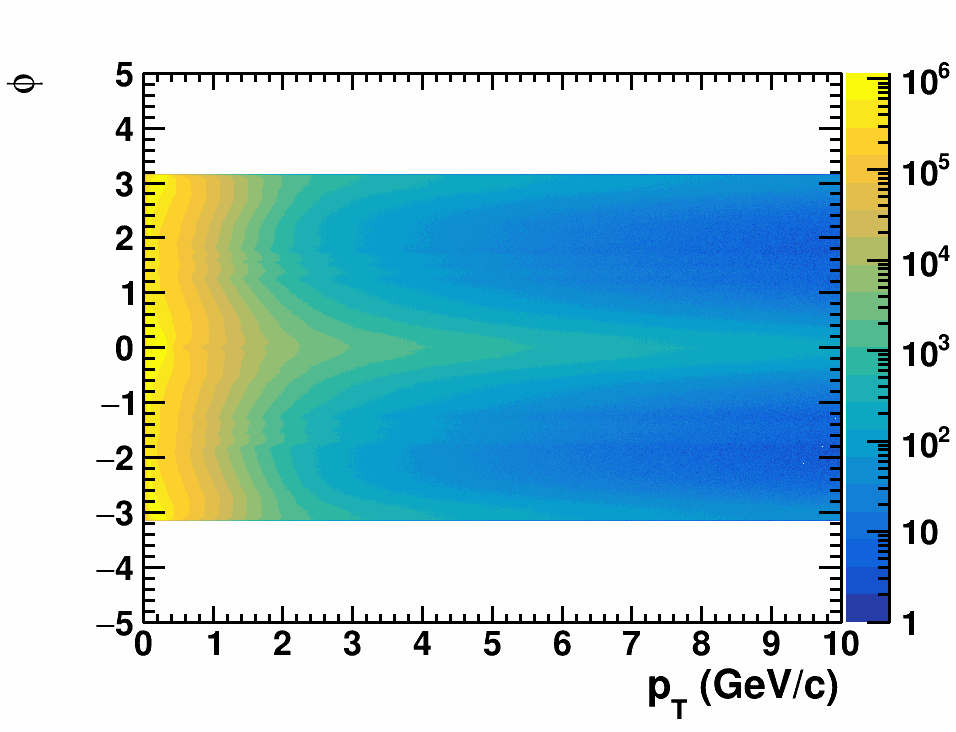}
            \vspace{-3mm}
            \caption{Correlation between the $\eta$ and the $\phi$ (left), $\eta$ and $p_T$ (center), $\phi$ and $p_T$ (right)
              for charged particles} \label{2dplots}
        \end{center}
       
        \vspace{-5mm}
    \end{figure}

\begin{figure}[H]
        \begin{center}
            \includegraphics[width=0.49\linewidth]{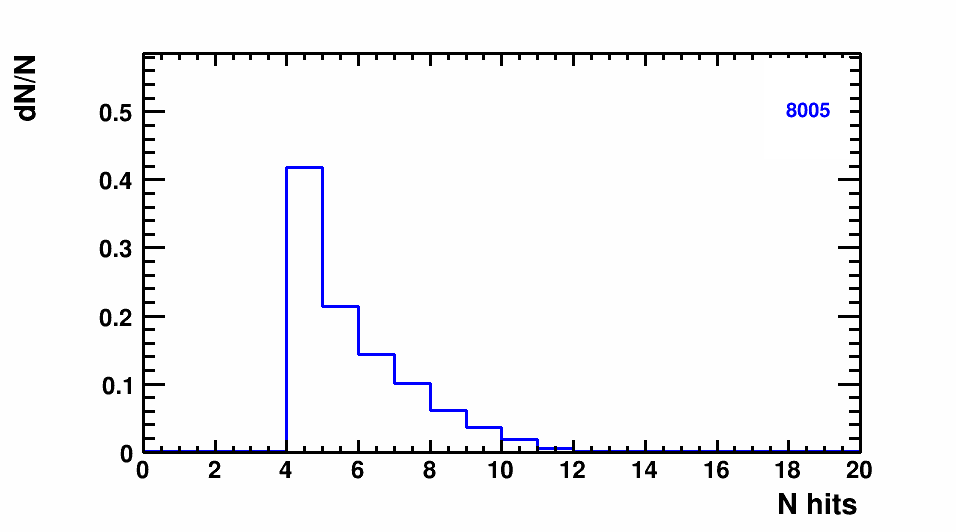}
            \includegraphics[width=0.49\linewidth]{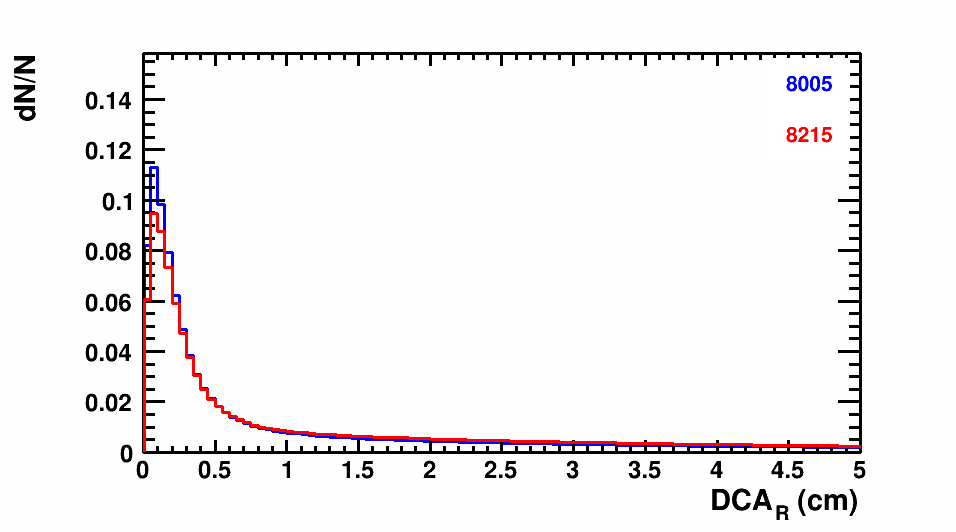}
            \includegraphics[width=0.49\linewidth]{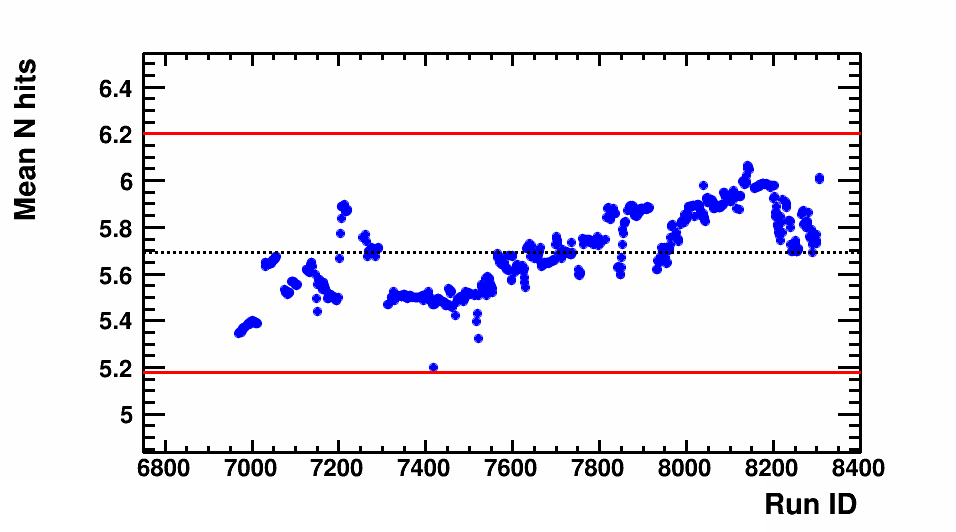}
            \includegraphics[width=0.49\linewidth]{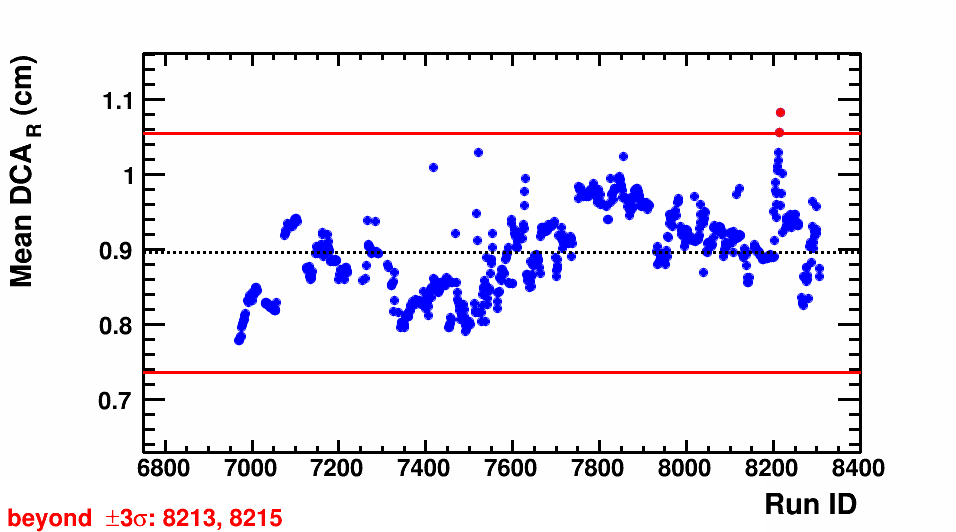}
            \vspace{-3mm}
            \caption{Upper panels: Distribution of the number of  nHits to  accurate the track momentum
              reconstruction (left) and the distance of closest approach $DCA_R$ (right). The red marker corresponds
              to the distribution from the "outlier" RunId. Bottom panels: Mean nHits and $DCA_R$
              as a function of RunID. Black dotted horizontal line and red horizontal lines
              represent $\mu$ and $\pm3\sigma$, respectively.}  \label{nHits_dca}
        \end{center}
        \vspace{-5mm}
    \end{figure}

    {\bf 4)} Figures~\ref{FHCal}--\ref{FQH} shows the RunId dependence of the mean of the total energy $E_{tot}$ of
    spectator fragments in the FHCal and  mean of the
    charge ($Q^2$) of spectator fragments in the forward quartz hodoscope (FQH).  Black dotted horizontal
    line and red horizontal lines represent $\mu$ and $\pm3\sigma$, respectively.\\
    {\bf 5)} Figure~\ref{Momentum} shows the  RunId dependence of the mean
    of  x, y and z components of the momentum of the charged particles. The upper panels of  Figure~\ref{track_par} show
    the typical distributions of the transverse momentum $p_T$ (left), azimuthal angle $\phi$ (center) and
    pseudorapidity $\eta$ (right) of charged particles. Bottom panels of Figure~\ref{track_par} show the  RunId dependence of the mean
    $p_T$, $\phi$ and $\eta$ distributions.  Figure~\ref{2dplots} shows the
    correlations between the $\eta$ and  $\phi$ (left), $\eta$ and $p_T$ (center), $\phi$ and $p_T$ (right)  for charged particles.
The upper panels of  Figure~\ref{nHits_dca} show the typical distributions for the number of  nHits to  accurate the track momentum
reconstruction (left) and the distance of closest approach $DCA_R$ (right).  The bottom panels show the
RunId dependence of the mean nHits and $DCA_R$.

\begin{figure}[H]
        \begin{center}
            \includegraphics[width=0.49\linewidth]{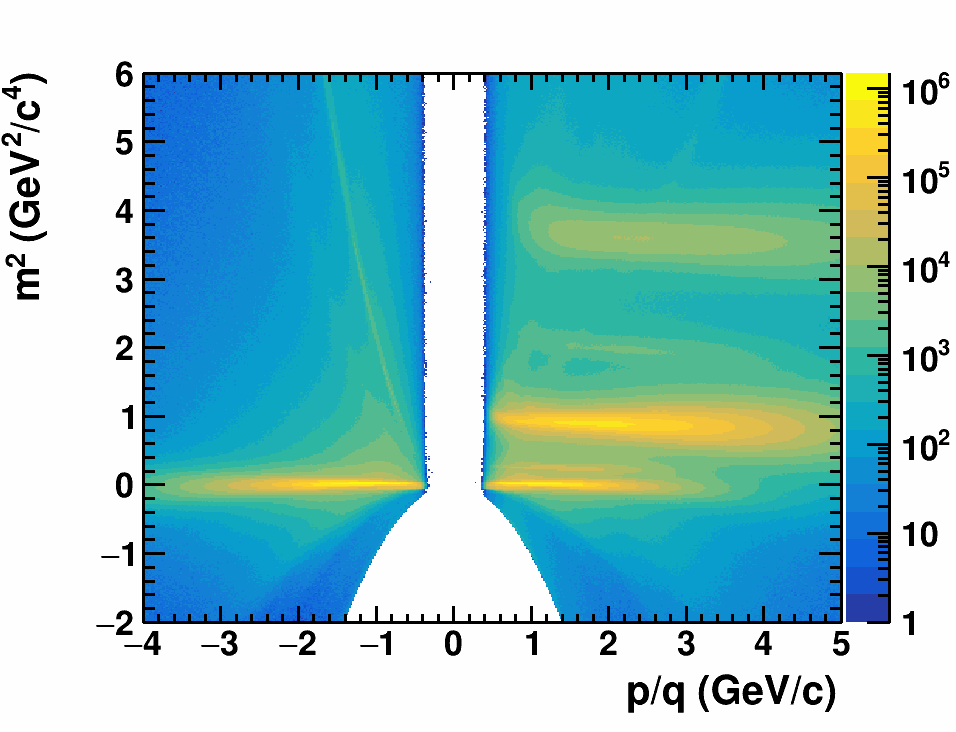}
            \includegraphics[width=0.49\linewidth]{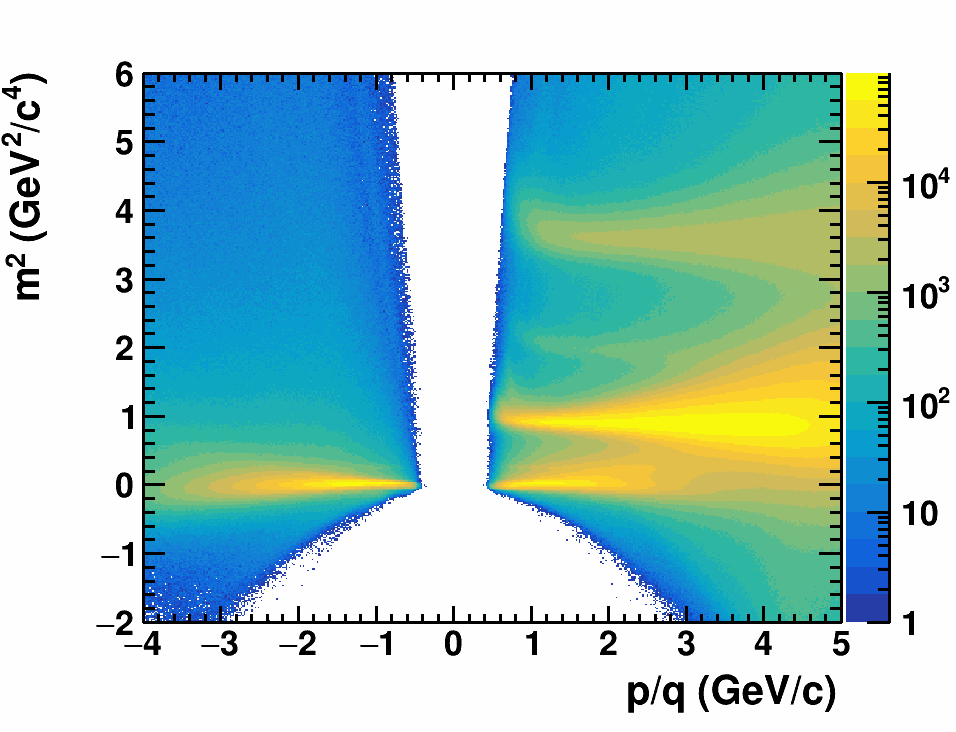}
            \vspace{-3mm}
            \caption{Population of charged particles in the mass squared ($m^2$) vs. laboratory momentum over charge
(p/q) plane for the TOF-400  (left panel) and TOF-700 (right panel) detectors.}  \label{m2_TOF}
        \end{center}
        \vspace{-5mm}
    \end{figure}

\begin{figure}[H]
        \begin{center}
            \includegraphics[width=0.9\linewidth]{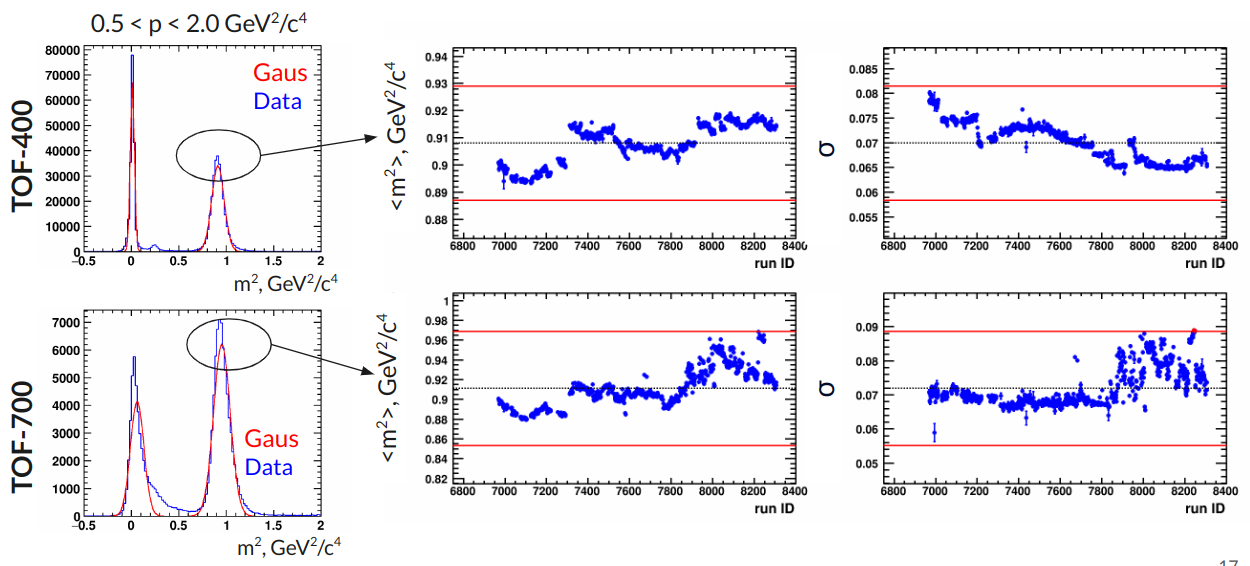}
            \vspace{-3mm}
            \caption{Distribution of the mass squared ($m^2$) and Gaussian fit of the proton peak in the TOF-400 (left upper panel)
              and TOF-700 (left bottom panel) detectors. Center and right panels:
              mean of the mass squared of proton and $\sigma_{m^2}$ as a function of RunID.
              Black dotted horizontal line and red horizontal lines represent $\mu$ and $\pm3\sigma$, respectively. }  \label{mass2_proton}
        \end{center}
      
        \vspace{-5mm}
    \end{figure}

     {\bf 6)} As an example, the  Figure~\ref{m2_TOF}   shows the population of all charged particles in the plane
    spanned by their  mass squared ($m^2$) vs. laboratory momentum divided by charge (rigidity) for
    the TOF-400  (left panel) and TOF-700 (right panel) detectors. The left panels of  Figure~\ref{mass2_proton}
    show the distributions of the mass squared ($m^2$) and Gaussian fit of the proton peak for the TOF-400 (left upper panel)
    and TOF-700 (left bottom panel) detectors. Center and right panels of  Figure~\ref{mass2_proton} show the  RunID
    dependence of mean of the mass squared ($m^2$) of proton  and the width of the peak $\sigma_{m^2}$.

    The preliminary  list of bad runs based on QA study [18M events] RunId:
    6968, 6970, 6972, 6973, 6975, 6976, 6977, 6978, 6979, 6980, 6981, 6982,
    6983, 6984, 7313, 7326, 7415, 7417, 7435, 7517, 7520, 7537, 7538, 7542,
    7543, 7545, 7546, 7547, 7573, 7575, 7657, 7659, 7679, 7681, 7843, 7847,
    7848, 7850, 7851, 7852, 7853, 7855, 7856, 7857, 7858, 7859, 7865, 7868,
    7869, 7907, 7932, 7933, 7935, 7937, 7954, 7955, 8018, 8031, 8032, 8033,
    8115, 8121, 8167, 8201, 8204, 8205, 8208, 8209, 8210, 8211, 8212, 8213,
    8215, 8289.

\subsection{Data, Event and Track Selection}

\begin{figure}[H]
        \begin{center}
            \includegraphics[width=0.32\linewidth]{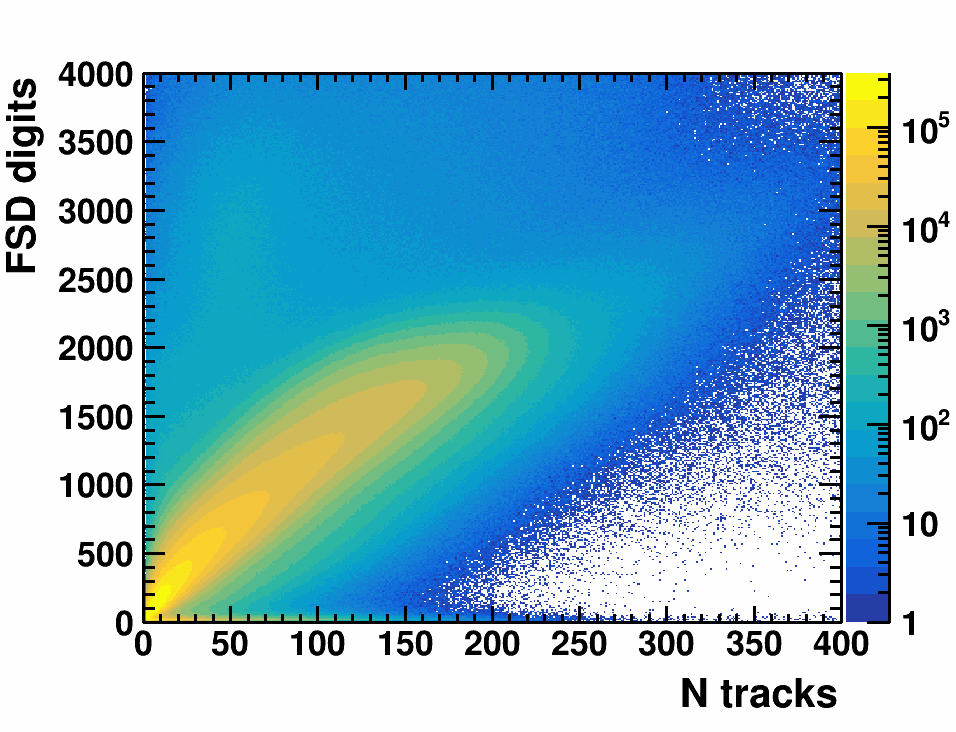}
            \includegraphics[width=0.32\linewidth]{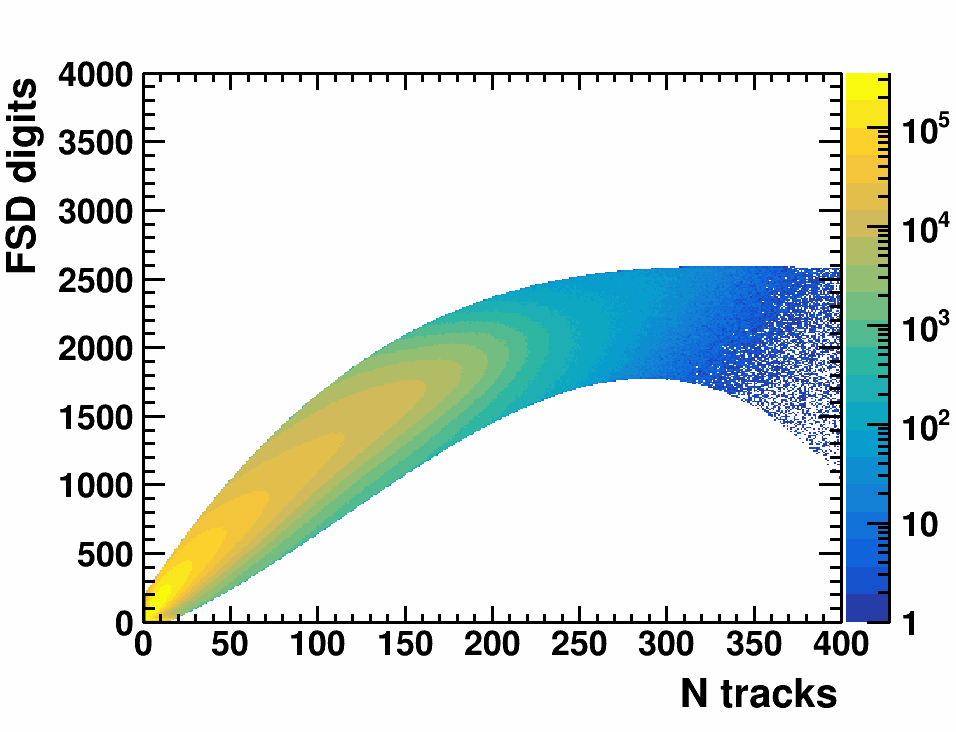}
            \includegraphics[width=0.32\linewidth]{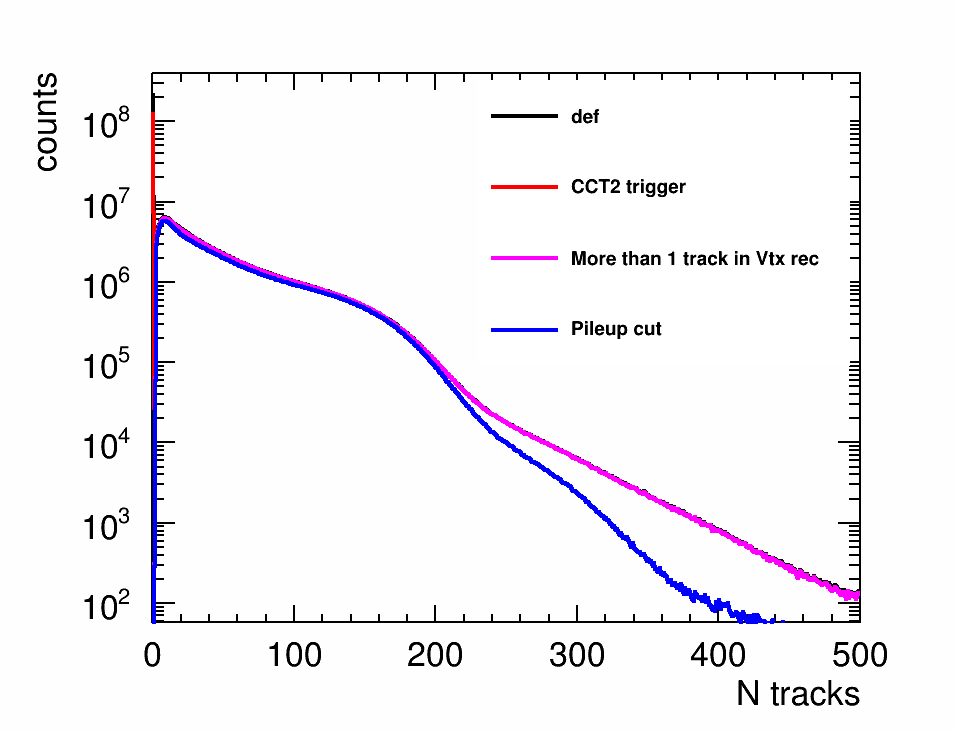}
            \vspace{-3mm}
            \caption{Left and center panels: Dependence of the number of FSD digits 
              and the number of charged particles in the tracking system (FSD + GEM) before and after application
              of the pileup rejection cut.  Right panel: tracks multiplicity distribution
              before and after applying the pileup rejection cut.}  \label{fig:pileup}
        \end{center}
      
        \vspace{-6mm}
    \end{figure}

In total approximately 500 million events of Xe+Cs(I) collisions at the beam
energy of 3.8A GeV were collected by the BM@N experiment in the January of 2023.\\
{\bf 1)} We don’t consider runs below RunId=6924 due to unstable
operation of the GEM and FSD detectors (BM@N Electronic Logbook).\\
{\bf 2)} We removed 74 runs [18M events]
based on QA study, see  section 3.2\\
{\bf 3)} We used events from Physical runs and  CCT2 trigger \cite{bmn}. \\
{\bf 4)} at least 2 tracks  in vertex reconstruction \\
{\bf 5)} The pileup events were rejected based on the  $\pm 3\sigma$ cut on the correlation
between the number of FSD digits and the
number of charged particles in the tracking system (FSD + GEM),
see the left and center panels of the Figure~\ref{fig:pileup}.

  \begin{table}[h]
    \caption{ Statistics after applying the selection criteria }
    \begin{center}
        \begin{tabular}{|c|c|c|}
            \hline
            Cuts & no. of events & $\%$ \\
            \hline
            \hline
            def. &  530 M & 100$\%$ \\
            \hline
            CCT2 trigger & 437 M & 82$\%$ \\
            \hline
            at least 2 tracks  in vertex reconstruction & 315 M & 59$\%$ \\
            \hline
            Pileup rejection cuts & 285 M &  53$\%$ \\
            \hline
            QA study & 267 M &  50$\%$ \\
             \hline
        \end{tabular}
    \end{center}
    \label{EventCutTable}
\end{table}

\begin{figure}[hbt]
        \begin{center}
            \includegraphics[width=0.7\linewidth]{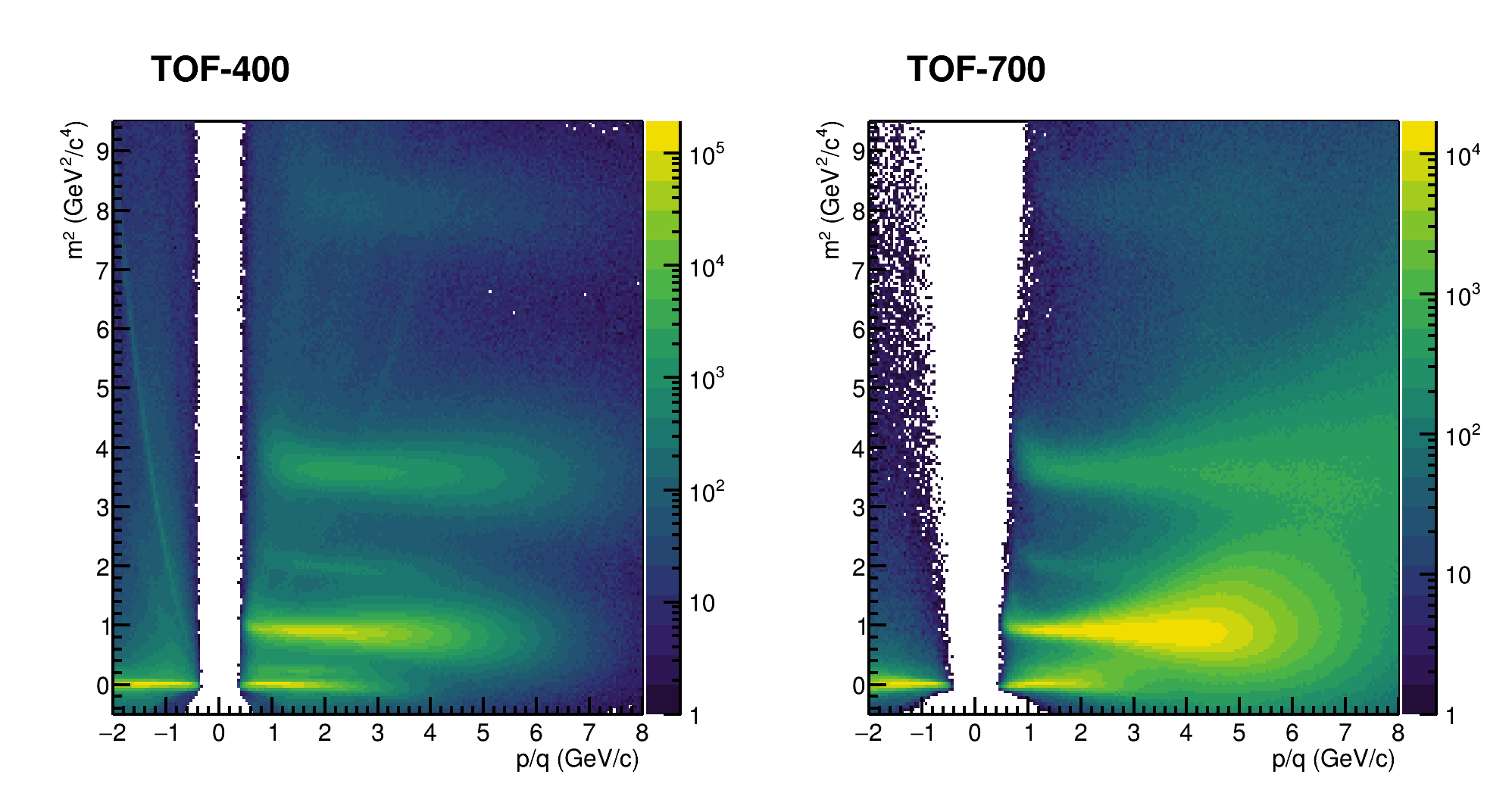}
            \vspace{-3mm}
            \caption{Population of charged particles in the $m^2$ vs.  rigidity
            (p/q) plane for the TOF-400  (left panel) and TOF-700 (right panel) detectors.}  \label{m2TOF}
        \end{center}
        \vspace{-5mm}
    \end{figure}

\begin{figure}[hbt]
        \begin{center}
            \includegraphics[width=0.7\linewidth]{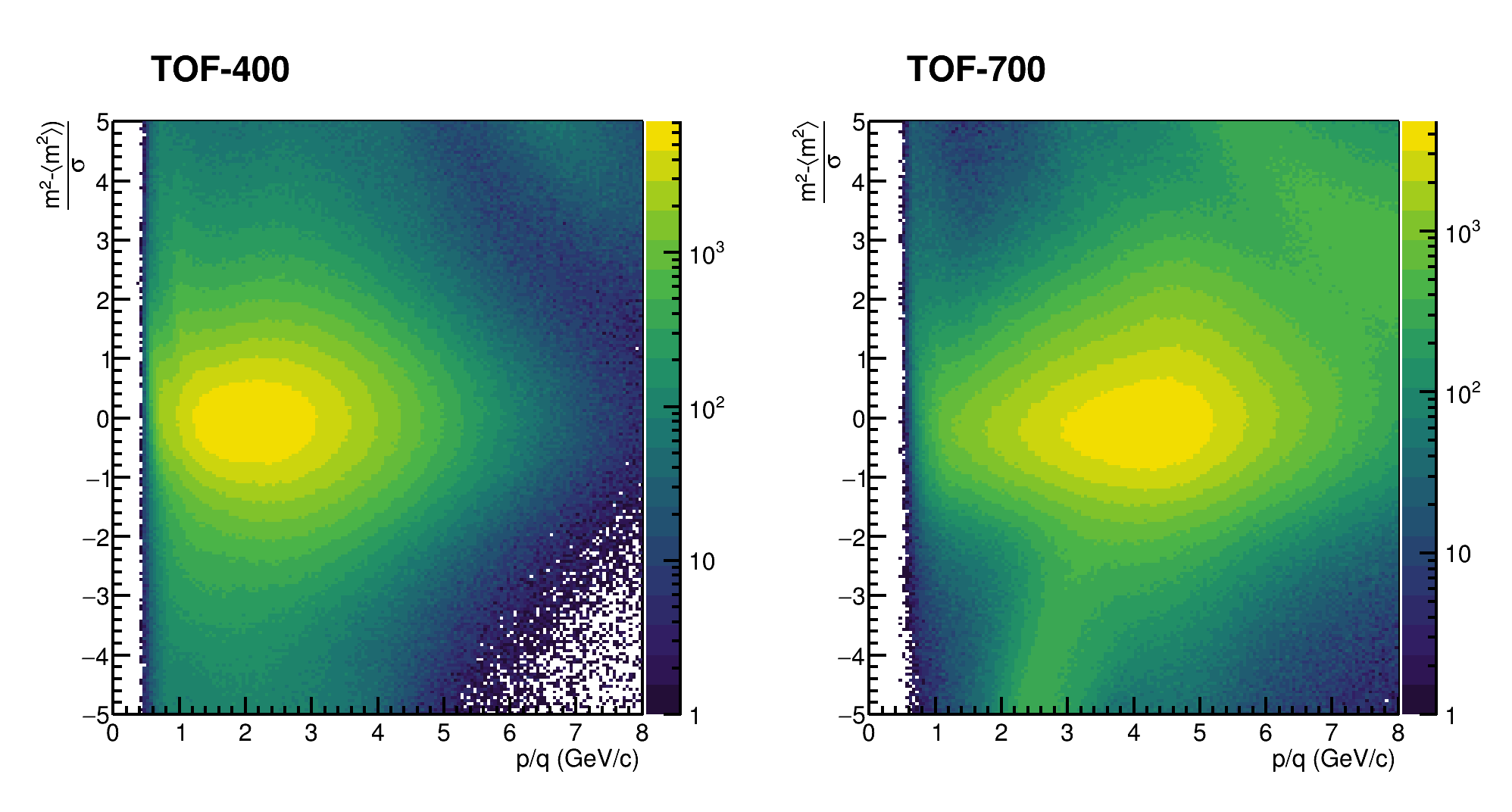}
            \vspace{-3mm}
            \caption{Population of charged particles in the n-sigma($m_{p}^2$) =
              ($m^2$-$\left\langle m_{p}^{2} \right\rangle$)/$\sigma_{m_{p}^2}$  vs.  rigidity
            (p/q) plane for the TOF-400  (left) and TOF-700 (right) detectors.}  \label{nsigma}
        \end{center}
        \vspace{-5mm}
    \end{figure}

\begin{figure}[hbt]
        \begin{center}
            \includegraphics[width=0.7\linewidth]{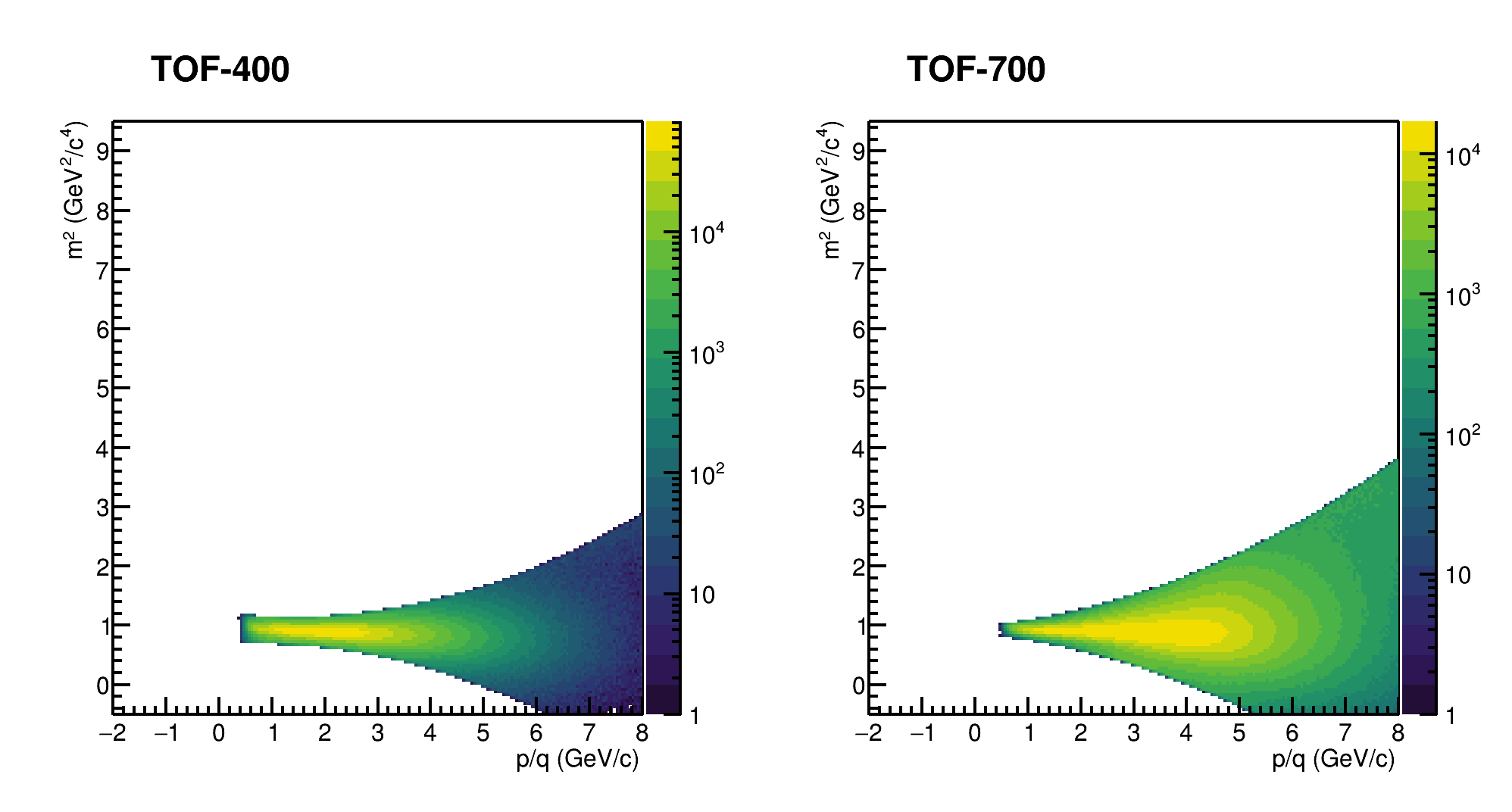}
            \vspace{-3mm}
            \caption{Population of selected protons  in the $m^2$ vs.  rigidity
            (p/q) plane for the TOF-400  (left) and TOF-700 (right) detectors. The protons were selected by 
             ($m^2$-$\left\langle m_{p}^{2} \right\rangle$)  $< 3 \sigma_{m_{p}^2}$ cut.}  \label{3sigmacut}
        \end{center}
        \vspace{-5mm}
    \end{figure}

Selection criteria are also imposed on tracks to ensure good tracks for analysis. The
selection cuts applied are as follows:\\
{\bf 1)} Tracks of charged particles were selected based on the number of stations $N_{hits}$ in
the BM@N inner tracking system used for track reconstruction. At least 6 were
required  to satisfy the criteria of a good track: $N_{hits}>6$.\\
{\bf 2)} Only tracks with fit quality $\chi^2/NDF <$ 5 were analyzed.\\
{\bf 3)} Distance of the closest approach (DCA)  of tracks from the primary vertex in the direction perpendicular
to the beam: DCA $<$ 5 cm\\
Protons  are identified using the time of flight $\Delta t$ measured
between T0 and the ToF detectors, the length of the trajectory $\Delta L$  and the momentum $p$
reconstructed in the BM@N central tracker. Then the squared mass $m^2$ of a particle is calculated.
For each bin in momentum   the position  $\left\langle m_{p}^{2} \right\rangle$
and the width $\sigma_{m_{p}^2}$ of the proton  $m^2$  peak was extracted from the  Gaussian fit. The procedure
was done separately for TOF-400 and TOF-700 as they have different timing resolution.
The proton  samples selected by the requirements of ($m^2$-$\left\langle m_{p}^{2} \right\rangle$)  $< 3 \sigma_{m_{p}^2}$, see
Figures~\ref{m2TOF}--\ref{3sigmacut}.

\begin{figure}[hbt]
        \begin{center}
            \includegraphics[width=0.7\linewidth]{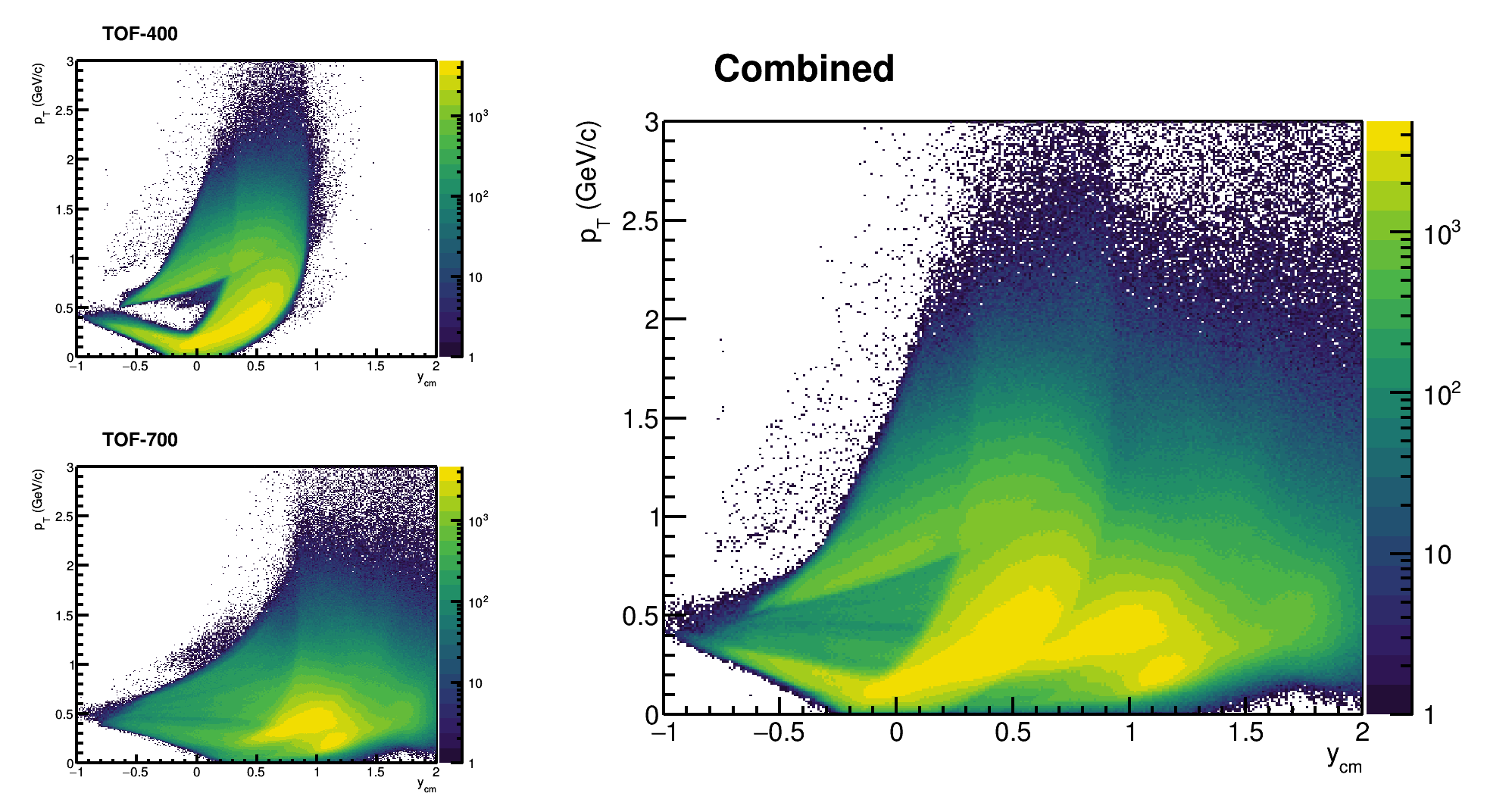}
            \vspace{-3mm}
            \caption{The phase space coverage of identified protons  as a function of the centre-of-mass rapidity $y_{cm}$
and transverse momentum $p_T$.}  \label{tofacc}
        \end{center}
        \vspace{-5mm}
    \end{figure}

\begin{figure}[hbt]
    \centering
    \includegraphics[width=0.6\linewidth]{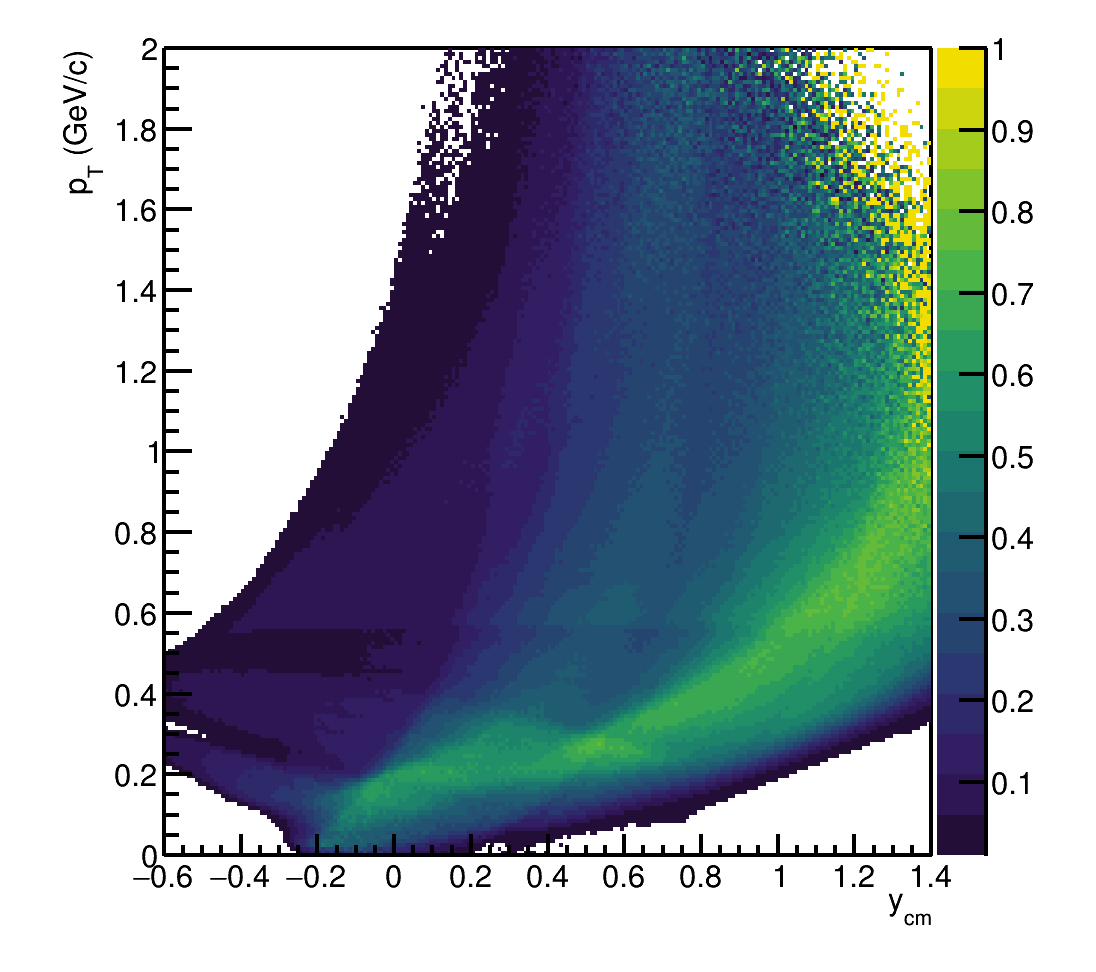}
    \caption{ Efficiency of the proton reconstruction in the phase space of rapidity $y_{cm}$ and transverse momentum $p_T$ }
    \label{fig:pT_y_efficiency}
\end{figure}

Figure~\ref{tofacc} shows the phase space coverage of identified protons  as a function of rapidity $y_{cm}$
and transverse momentum $p_T$ for TOF-400, TOF-700 and for the combined system.
Efficiency of the proton reconstruction was calculated using the realistic Monte-Carlo modelling
of the BM@N experiment using GEANT4 transport code and JAM in the mean field mode events as an input.
Efficiency of the proton reconstruction with the TOF-detectors acceptance applied is shown in the Figure.~\ref{fig:pT_y_efficiency}.

\subsection{Centrality determination}
\begin{figure}[ht]
\center
\includegraphics[width=16 cm]{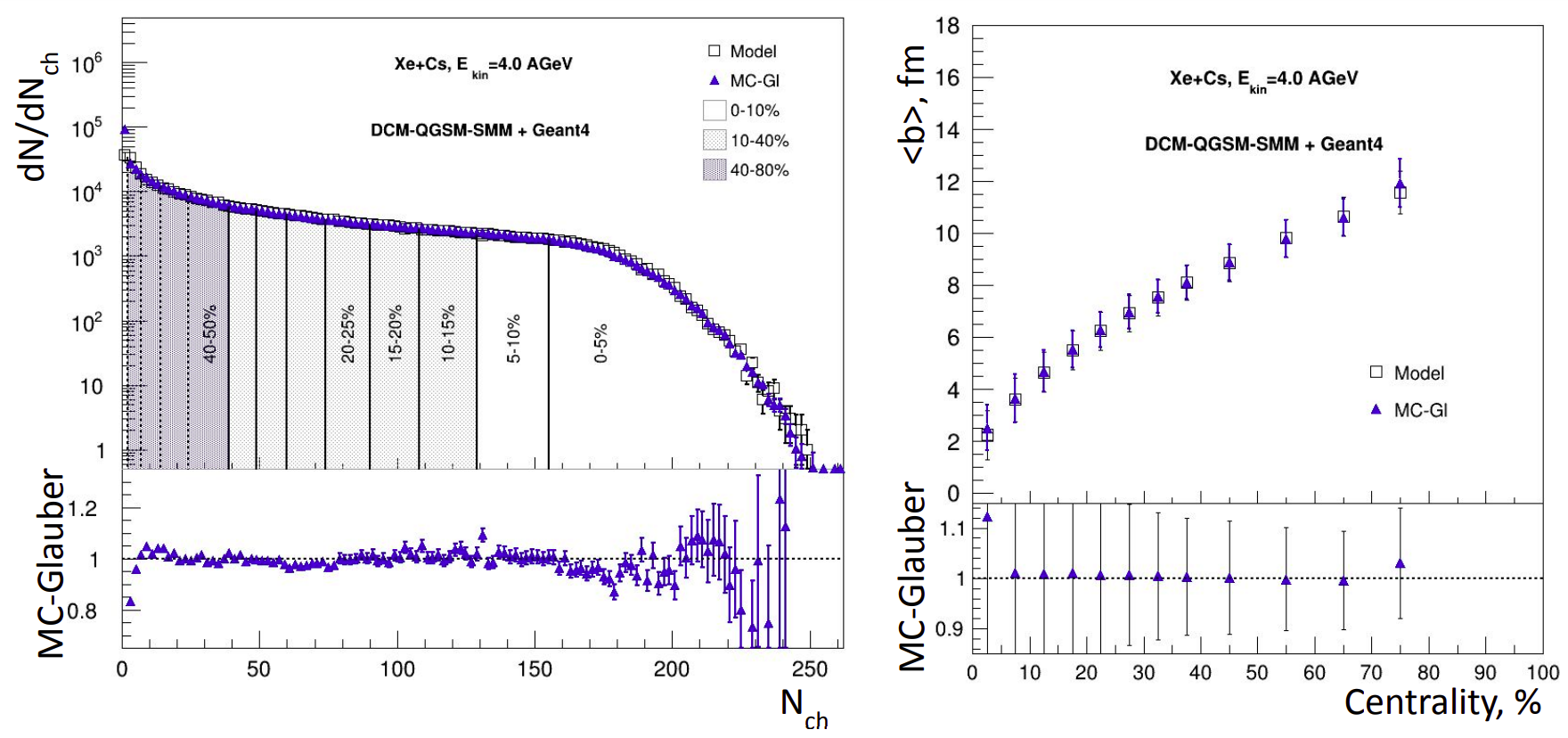}
\caption{ Left panel: FSD+GEM multiplicity distribution $N_{ch}$ from the fully reconstructed DCM-QGSM-SMM  model events  (open squares) for
  Xe+Cs collisions  compared to the fitted distribution using MC-Glauber  approach (solid triangles). The centrality classes defined with MC-Glauber
  normalization are indicated with
black vertical lines. Right panel: centrality dependence of the $\left\langle b \right\rangle$ from MC-Glauber approach (closed symbols) and
directly from the model (open symbols) \label{fig:Glauber}.}
\end{figure}

The size and evolution of the matter created in relativistic heavy-ion collisions strongly
depend on collision geometry defined by the impact parameter. Since the impact parameter
$b$ of collisions (defined as the distance between the geometrical centers of the colliding nuclei in the transverse plane) cannot
be accessed directly, the centrality classification can be based on the number of produced
charged particle multiplicity $N_{ch}$ in an event.  Usually the correlation between the impact parameter $b$ and
the multiplicity $N_{ch}$ is determined using the Monte-Carlo Glauber (MC-Glauber) method
combined with a simple particle production model \cite{glauber}. The modeled multiplicity is assumed to be a function of the number of participating
nucleons ($N_{part}$) and the number of binary interactions
between nucleons ($N_{bin}$), which one obtains from the output of the  MC-Glauber model. The particle multiplicity distribution $N_{ch}^{fit}$ can then be
fitted to the experimentally measured one \cite{alicecent,cbmcent}. Centrality classes are defined by sharp cuts on $N_{ch}$ and corresponding mean  values of $\left\langle b \right\rangle$  for each class determined from  MC-Glauber events.
While this approach offers a convenient parametrization of the measured  $N_{ch}$ distributions and the main   classifier for centrality determination in the
STAR \cite{star45,star3gev} and HADES \cite{hadescent} experiments,
it may suffer from large systematic uncertainties at low multiplicities
and assumptions about the particle production mechanism \cite{milov}. In contrast to the MC-Glauber method, the recently proposed
$\Gamma$-fit method  does not require any modeling of the
collision dynamics and can be used over a broad range of collision energies: from $\sqrt{s_{NN}}$=5.44 TeV~\cite{gfit1}  to the
bombarding energy of 25 AMeV \cite{gfit2}. The $\Gamma$-fit method is  based on the assumption that the
relation between the measured $N_{ch}$ and $b$ is purely probabilistic and can be inferred from data
without relying on any specific model of collisions. This typical inverse
problem  can be solved by a deconvolution method. A gamma distribution is used for
the fluctuation kernel $P(N_{ch}|b)$  to model fluctuations of $N_{ch}$ at a
fixed impact parameter. The parameters of the gamma distribution  were then
extracted  by fitting the measured distribution of $N_{ch}$ \cite{gfit1,gfit2}. The application of both methods for centrality determination at NICA
energies can be found  in \cite{mamaev,cbmcent,gfitnica1,gfitnica2}.\\
In the first step, the validity of the procedures for  centrality determination by the MC-Glauber and  $\Gamma$-fit methods was assessed using the
simulated  data for Xe+Cs collisions at beam kinetic energy of 4 AGeV. The DCM-QGSM-SMM model \cite{dcm} has been used  to simulate around 2 M
minimum bias Xe+Cs collision events. At the next step, the sample of events was made as
an input for the full chain of realistic simulations of the BM@N detector subsystems based on the GEANT4 platform 
and reconstruction algorithms built in the BMNROOT framework for run8. The fully reconstructed events were used to generate the
distributions of the multiplicity $N_{ch}$ of the produced charged particles detected by FSD+GEM system, see left panel of Figure~\ref{fig:Glauber}.

\begin{figure}[ht]
\center
\includegraphics[width=16 cm]{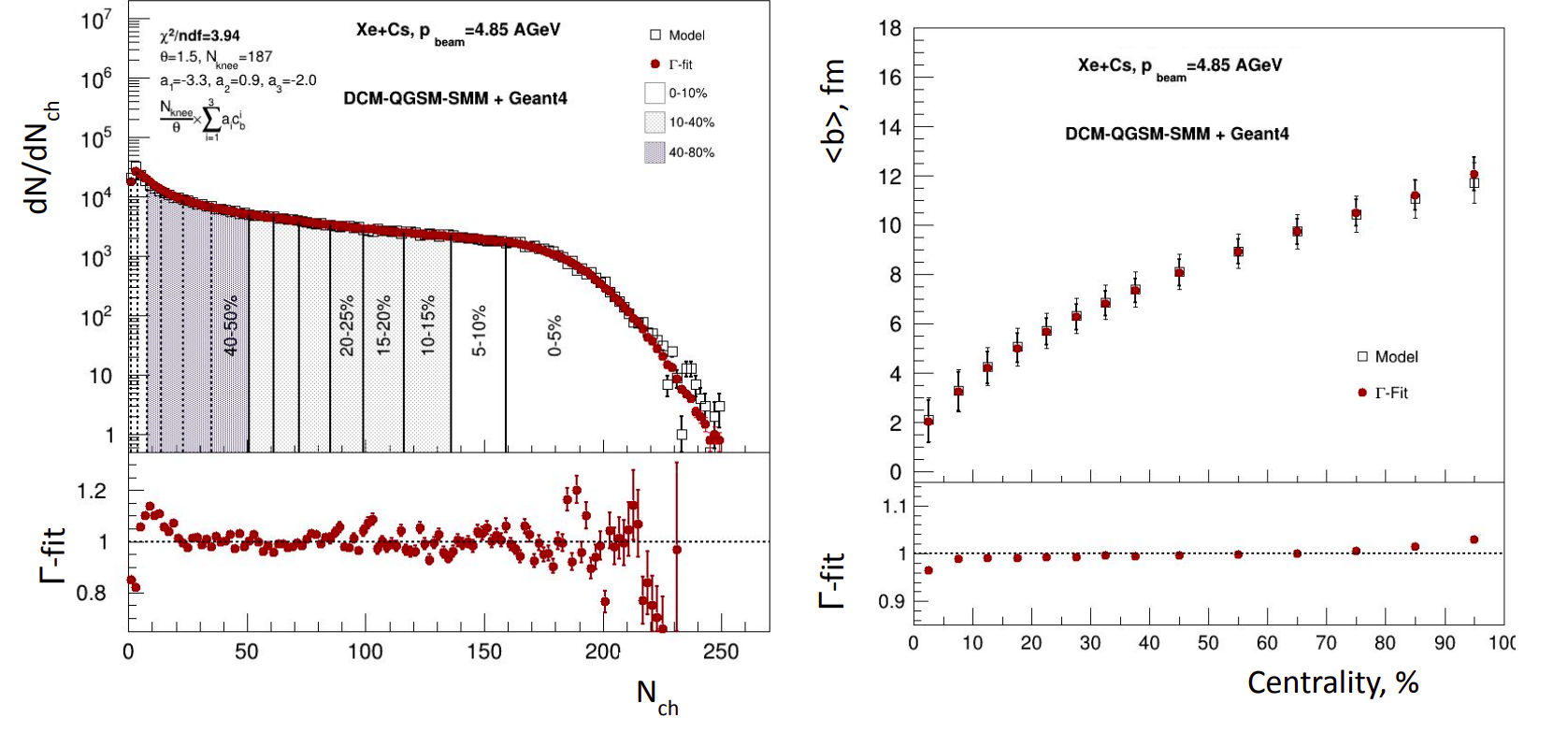}
\caption{ Left panel: FSD+GEM multiplicity distribution $N_{ch}$ from the fully reconstructed DCM-QGSM-SMM  model events  (open squares) for
  Xe+Cs collisions  compared to the fitted distribution using  $\Gamma$-fit method   (solid circles).
  The centrality classes defined with MC-Glauber  normalization are indicated with
black vertical lines. Right panel: centrality dependence of the $\left\langle b \right\rangle$ from  $\Gamma$-fit method (closed symbols) and
directly from the model (open symbols) \label{fig:Gamma-Fit}.}
\end{figure}

The 3.2 version of the PHOBOS MC-Glauber model ~\cite{glauber} has been used  to compose
two nuclei out of nucleons and simulate their collision process event-by-event.
An input of the MC-Glauber
model is the nucleon density $\rho(r)$ inside the nucleus. It is usually parametrized by
Fermi distribution:
 \begin{eqnarray}
	\label{eq:McGlauberFermi}
	\rho(r) = \rho_0 \frac{1+w\left( \frac{r}{R} \right)^2}{1+\exp{\frac{r-R}{a}}},
  \end{eqnarray}
  where $R$ is the radius of the nucleus, the constant
  $\rho_0$ corresponds to the density  in the center of the nucleus. The skin thickness of the nucleus $a$ 
  defines how abruptly the density falls at the edge of the nucleus. The following parameters have been used:
    Xe (A=129, Z=54, R=5.46 fm, a=0.57 fm) and Cs (A=133, Z=55, R=6.125 fm, a=0.5). The nucleus-nucleus collision is treated as a sequence of independent
binary nucleon-nucleon collisions, where the nucleons travel on straight-line trajectories and the
inelastic nucleon-nucleon cross section $\sigma_{\text{NN}}^{\text{inel}}$ assumed to depends only on the collision energy:  $\sigma_{\text{NN}}^{\text{inel}}$=27.7 mb.
Two nucleons from different nuclei are assumed to collide if the relative transverse distance $d$
between centers is less than the distance corresponding to the inelastic nucleon-nucleon cross section:
$d < \sqrt{\sigma_{\text{NN}}^{\text{inel}}/\pi}$.
Geometrical properties of the collision, such as the impact
parameter $b$, number of participating nucleons ($N_{part}$ ), and number of binary nucleon-nucleon collisions ($N_{coll}$ ), are calculated 
by simulating around 2 M minimum bias Xe+Cs collision events.
The procedure for centrality determination includes fitting experimentally measured
particle multiplicity $N_{ch}$  with a MC-Glauber model based function $N_{ch}^{fit} (f, \mu, k)$ \cite{alicecent,cbmcent,gfitnica1,gfitnica2} : 
\begin{equation}
N_{ch}^{fit} (f, \mu, k) = N_a(f) \times P_{\mu,k},\ N_a(f) = fN_{part} + (1-f)N_{coll},
\end{equation}
where  $P_{\mu,k}$ is the negative binomial distribution  (NBD) with  mean $\mu$  and width $k$. 
$N_{a}(f)$ is a number of ancestors (number of independent sources), $f$ characterizes the fraction of hard processes, $N_{part}$ and $N_{coll}$ are the number of
participants and the number of binary collisions from  MC-Glauber model output. The optimal 
set of parameters $f$, $\mu$ and $k$ can be found from the minimization procedure is applied to
find the minimal value of the $\chi^2$,wich defined as follows:

\begin{eqnarray}
	\label{eq:McGlauberChi2}
	\chi^2 = \sum \limits_{i=n_{low}}^{n_{high}} \frac{\left( F_{fit}^i - F_{data}^i \right)^2}{\left( \Delta F_{fit}^i \right)^2 + \left( \Delta F_{data}^i \right)^2},
\end{eqnarray}
where $F_{fit}^i$ and $F_{data}^i$ are values of the fit function and fitted histogram at a given bin $i$, $\Delta F_{fit}^i$ and $\Delta F_{data}^i$ are
corresponding uncertainties, $n_{low}$ and $n_{high}$ are the lowest and highest fitting ranges correspondingly. A grid of $k$ and $f$ parameters was
formed with corresponding $\chi^2$ values for each ($k$,$f$) combination:
$k \in [1,50]$ with step of $1$ and $f \in [0,1]$ with step of $0.01$. The framework and documentation
for centrality determination by the  MC-Glauber approach can be found in:
{\small \it https://github.com/FlowNICA/CentralityFramework/tree/master/Framework/McGlauber}.
As an example, left panel of Figure~\ref{fig:Glauber} shows by blue solid triangles the resulting $N_{ch}^{fit}$ distribution from MC-Glauber fit.
The ratio $(N_{ch}^{fit}/N_{ch})$ of the fit to the data shows the quality of the procedure, see the bottom part of Figure~\ref{fig:Glauber}.
After finding the optimal  set of the fit parameters one can easily estimate the
total cross-section and  all events can be divided into
groups with a given range of total cross-section (0-5\%, 5-10\% etc), see the black solid vertical lines in Figure~\ref{fig:Glauber}.
High multiplicity events have a
low average $b$ (central collisions) and low multiplicity events have a large average $b$ (peripheral collisions). 
For each centrality class the mean value of the impact parameter $\left\langle b \right\rangle$ and its corresponding standard deviation was found using
 the information from the simulated MC-Glauber model events. 
Figure~\ref{fig:Glauber} (righ panel) shows the centrality dependence of  $\langle b \rangle$ for the  model events
denoted by open symbols. The $\langle b \rangle$ from  MC-Glauber approach (closed symbols) are presented  for comparison.\\
In the $\Gamma$-fit method \cite{gfit1,gfit2,gfitnica1,gfitnica2} the main ingredient is the fluctuation kernel which is used to model multiplicity
fluctuations $P(N_{ch}|b)$ at a fixed impact parameter $b$.  
The fluctuations of the multiplicity can be described by the gamma
distribution ~\cite{gfit1,gfit2}:

\begin{eqnarray}
	\label{eq:GammaP}
	P(N_{ch}|b) = \frac{1}{\Gamma(k)\theta^k} N_{ch}^{k-1} e^{-N_{ch}/\theta}
\end{eqnarray}
where $\Gamma(k)$ is gamma function and two parameters $k(b)$ and $\theta(b)$ corresponding to the mean, $\left\langle N_{ch} \right\rangle$,
and to the variance, $\sigma_{N_{ch}}$:  $\left\langle N_{ch} \right\rangle = k \theta,\ \sigma_{N_{ch}} = \sqrt{k} \theta$.
Similar to  the multiplicity $N_{ch}$, which is always positive,  the gamma distribution is only defined for $N_{ch}\geq$ 0. It can be 
considered as a continuous version of the negative binomial distribution (NBD), which 
has long been used to fit multiplicity distributions in heavy-ion collisions ~\cite{gfitnica1,gfitnica2}.
The normalized measured
multiplicity distribution, $P(N_{ch})$, can be obtained by summing the contributions to multiplicity
at all impact parameters:

\begin{equation}
  \label{convolution}
 P(N_{ch})=\int_0^{\infty} P(N_{ch}|b)P(b)db=\int_0^1 P(N_{ch}|c_b)dc_b, \ P(b) = \frac{2\pi b}{\sigma_{inel}} P_{inel}(b),
\end{equation}
where $P(b)$ is the probability distribution of the impact parameter, and $c_{b}$ denotes the centrality: $c_b\equiv\int_0^bP(b')db'$.
$P(b)$  depends on the probability  $P_{inel}(b)$ for an inelastic collision to occur at given $b$, and $\sigma_{inel}$ is
the inelastic nucleus-nucleus cross section. $P_{inel}(b)\simeq 1$ and $c_b\simeq \pi b^{2} /\sigma_{inel}$, except for peripheral collisions.
For  the variable $k$, one can  use the following parameterization:
\begin{eqnarray}
	\label{eq:k(cb)}
	k(c_{b}) = k_0 \cdot exp \left[ - \sum \limits_{i=1}^{3} a_i \left(c_{b} \right)^i \right],
\end{eqnarray}
We fit $P(N_{ch})$ to the experimental  distribution of $N_{ch}$ using Eqs. (5) and (6) ~\cite{gfit1,gfit2,gfitnica1,gfitnica2}.
The fit parameters $\theta$, $k_0$ and three coefficients $a_i$. The resulting parameters allow to reconstruct 
the probability of $N_{ch}$ at fixed $c_{b}$: $P(N_{ch}|c_b)$.
The fitting procedure has been tested for the same  charged particle multiplicity $N_{ch}$ distribution, see left panel of  Figure~\ref{fig:Gamma-Fit}.
The result of the $\Gamma$-fit is shown as red solid circles. The ratio plot shows that the $\Gamma$-fit
method can reproduce the charged particle multiplicity distribution with a good accuracy.

\begin{figure}[ht!]
\center
\includegraphics[width=0.8\linewidth]{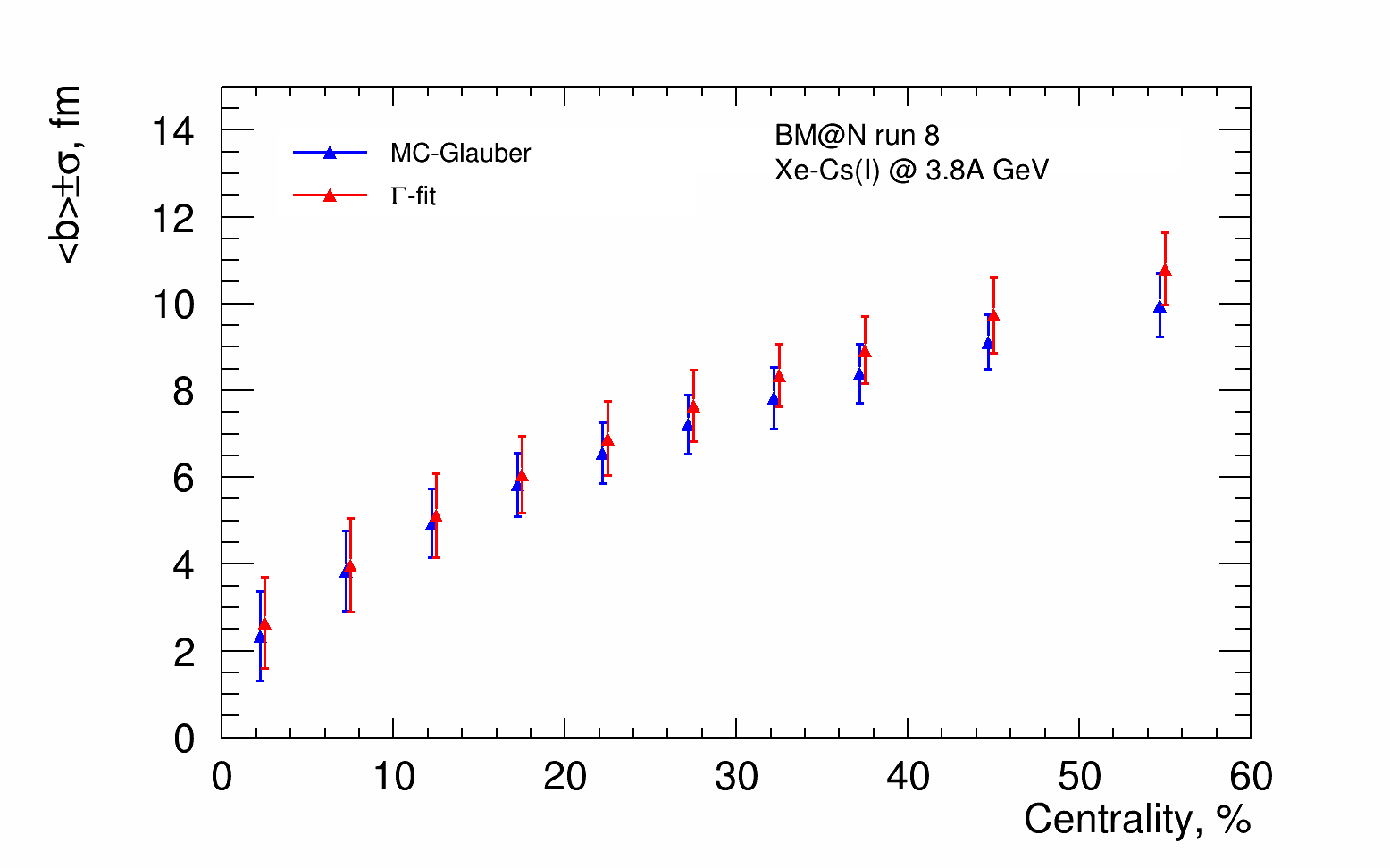}
\caption{Centrality dependence of the $\left\langle b \right\rangle$ from the MC-Glauber (blue solid triangles)  and  $\Gamma$-fit (red solid triangles)
  methods for BM@N run8 experimental data:  Xe+Cs(I) collisions at 3.8 AGeV.} 
\label{fig:ExpResultB}
\end{figure}

Once the probability of $N_{ch}$ at fixed $c_{b}$ is reconstructed, the probability distribution of $b$, at
fixed  $N_{ch}$ can be extracted  by Bayes’ theorem:
$P(b|N_{ch}) = P(N_{ch}|b)P(b)/P(N_{ch})$, 
where $P(N_{ch}|b) = P(N_{ch}|c_{b})$ and $c_b\simeq \pi b^{2} /\sigma_{inel}$ ~\cite{gfit1,gfit2}. Extending
this reconstruction to a finite centrality bin, corresponding to an interval  $N_{ch}^{low} < N_{ch} <  N_{ch}^{high}$ is very
straightforward upon integration over $N_{ch}$:

\begin{eqnarray}
	\label{eq:Bayes}
	P(b|N_{ch}^{low} < N_{ch} < N_{ch}^{high}) = P(b)\frac{\int \limits_{N_{ch}^{low}}^{N_{ch}^{high}} P(N'_{ch}|b) dN'_{ch}}{\int \limits_{N_{ch}^{low}}^{N_{ch}^{high}} P(N'_{ch}) dN'_{ch}}, 
\end{eqnarray}
where $\int \limits_{N_{ch}^{low}}^{N_{ch}^{high}} P(N'_{ch}) dN'_{ch}$ is the  width of the centrality bin $\Delta c_{b}$  (i.e., 0.1
for the 0-10\% centrality bin). 10\% centrality classes defined with $\Gamma$-fit normalization are indicated with black solid vertical lines
in Figure~\ref{fig:Gamma-Fit} (left).
The framework and documentation
for centrality determination by the  $\Gamma$-fit method can be found in:
{\small \it https://github.com/FlowNICA/CentralityFramework/tree/master/Framework/GammaFit}.\\
The centrality determination methods described above were applied to experimental BM@N data
for Xe+Cs(I) collisions at 3.8 AGeV. To construct the multiplicity of charged particles ($N_{ch}$), we selected all events that
satisfied the central collision trigger condition (CCT2), as well as events in which more than one track
was used to reconstruct the collision vertex. The pile-up events were removed as well. 
Figure ~\label{fig:ExpResultMult} shows the results for determining centrality based on
FSD+GEM multiplicity (open squares)  using the MC-Glauber method  (solid blue triangles) and the $\Gamma$-fit method (red solid  circles).
Both approaches describe the multiplicity distribution well up to 60\%.  The results of the $\Gamma$-fit
method describe the experimental data in the mid-central region somewhat better. Figure~\ref{fig:ExpResultB} shows the resulting 
centrality dependence of the $\left\langle b \right\rangle$ from the MC-Glauber (blue solid triangles)  and  $\Gamma$-fit (red solid triangles).
The results agree well for central collisions, but differ slightly for peripheral collisions. The results obtained provide a very preliminary estimate
of collision centrality. To obtain the final results, it is necessary to evaluate the efficiency  of the CCT2 trigger and take into account
changes in the average FSD+GEM multiplicity  during the run8, as well as to evaluate the systematics associated with the
use of the MC-Glauber and $\Gamma$-Fit methods.

\begin{figure}[h]
\center
\includegraphics[width=1.0\linewidth]{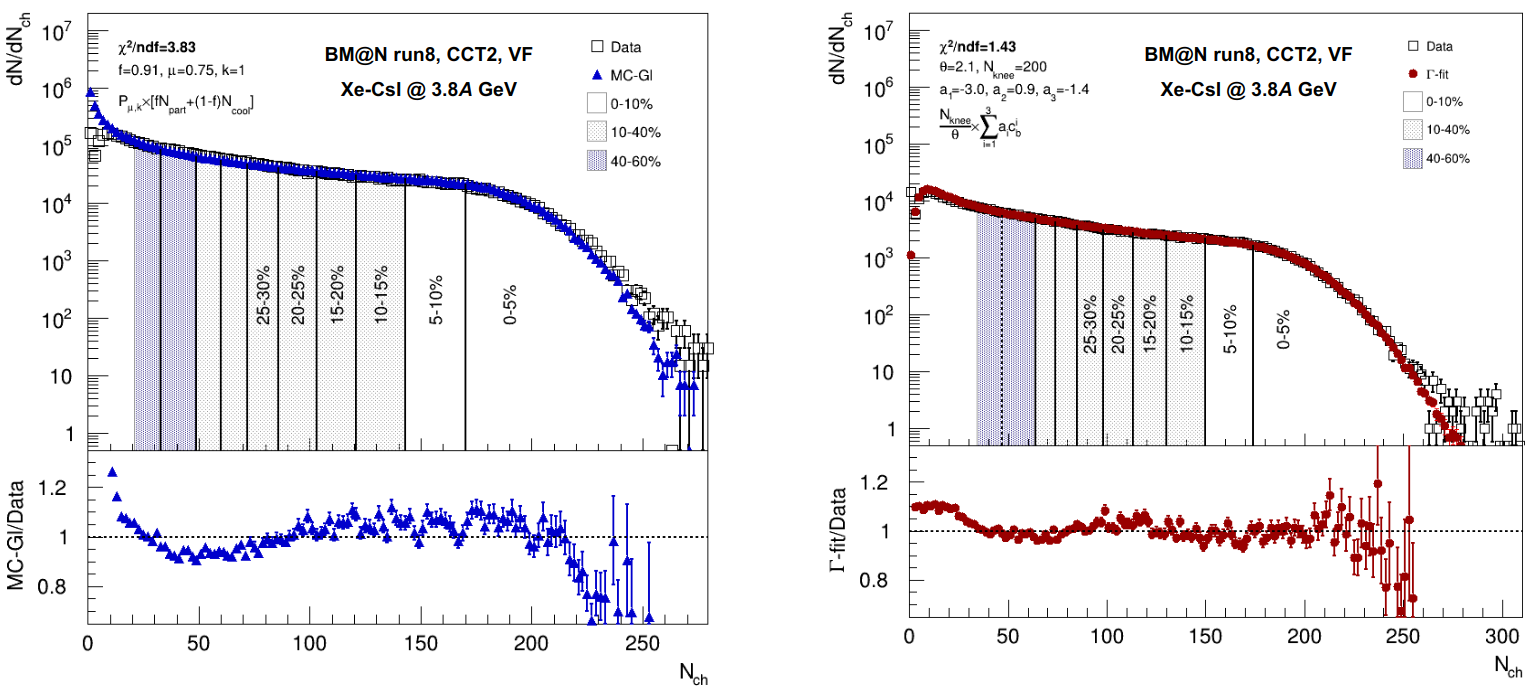}
\caption{FSD+GEM multiplicity distribution $N_{ch}$ from the BM@N run8 experimental data for  
  Xe+Cs(I) collisions at 3.8 AGeV (open squares) compared to the fitted distribution using the MC-Glauber method (solid blue triangles) and
   $\Gamma$-fit method (red solid  circles). The centrality classes  are indicated with
black vertical lines.} 
\label{fig:ExpResultMult}
\end{figure}

\section{Methods for analyzing anisotropic flow in BM@N}

\subsection{General framework for the flow measurements}
We start from the brief description of the general framework for the  measurements of flow coefficients $v_n$ in
the fixed target experiment \cite{flow0,flow1,flow2,flow3,mamaev}.
The observables for $v_n$ coefficients can be written in terms of flow $Q_n$ and unit $u_n$ vectors \cite{flow2,flow3,mamaev}.
For each particle $k$ in the event the unit  $u_{n,k}$  vector in the transverse (x,y) plane can be defined as:
\begin{equation}
    u_{n,k}  = e^{in\phi_k} = x_{n,k} + i y_{n,k} =  \cos n\phi_k + i \sin n\phi_k ,
    \label{eq:un_definition}
\end{equation}
where $\phi_k$ is the azimuthal angle of the particle's momentum.
A two dimensional symmetry-plane (flow)  $Q_n$-vector in the transverse
plane   is defined as a sum of unit $u_{n,k}$-vectors over
a group of particles in the event:
\begin{equation}
    Q_n  = \frac{ \sum_{k=1}^M w_k u_{n,k} }{ \sum_{k=1}^M w_k } = 
    X_n + iY_n =
    |Q_n| e^{in\Psi_n^E},
    \label{eq:Qn_definition}
\end{equation}
where $M$ is the multiplicity of particles in the selected group, $\Psi_n^E$ is the symmetry plane angle of $n$-th harmonic and  $w_k$ is the weight of 
particle, which is used either to correct the azimuthal anisotropy of the detector or to account for
the multiplicity of particles falling into a specific cell of a segmented detector \cite{flow2,flow3,mamaev}.\\
Detectors are not required to measure individual particles to be able to reconstruct the symmetry plane. As
long as the detector is sensitive to the shape of the particle distribution in the transverse plane,
the symmetry-plane (flow) $Q_n$ vector can be determined. For the case of a segmented detector, such as a calorimeter,
the mean position of the individual channels correspond to $u_{n,k}$. The channel amplitudes
correspond to the weights  $w_k$  assigned to the $u_{n,k}$ in Eq.~\ref{eq:Qn_definition}. A segmented detector needs a
segmentation which is larger than 2$n$ to be able to measure the  $Q_n$  vector of harmonic $n$.

At very large multiplicities in the selected group ($M \rightarrow \infty$) sum can be substituted by the integral and equation~\ref{eq:Qn_definition} can be transformed as follows:
\begin{multline}
    \lim_{M \rightarrow \infty}Q_n  = \frac{ 
        \int_{2\pi} d\phi w(\phi) e^{in\phi} \rho(\phi-\Psi_n^R)
    }{ 
        \int_{2\pi} d\phi w(\phi) e^{in\phi} \rho(\phi-\Psi_n^R)
    } = \\
    \frac{ 
        \int_{2\pi} d\phi w(\phi) e^{in\phi - \Psi_n^R} e^{in \Psi_n^R} \rho(\phi-\Psi_n^R)
    }{ 
        \int_{2\pi} d\phi w(\phi) e^{in\phi} \rho(\phi-\Psi_n^R)
    } =
    V_n e^{in\Psi_R}
    ,
    \label{eq:Qn_}
\end{multline}
where $V_n \propto v_n M$. 
From the equation above we can conclude that in limit of summation over very large group of particles in a event, $\Psi_n^E \rightarrow \Psi_n^R$ and $\Psi_n^E$ is the estimation of reaction plane orientation in the event. 
We will refer to this estimation as symmetry plane of the collision or event plane of the collision.\\
Measurements of the azimuthal flow   can be carried out projecting the $u_n$ vector of selected particles  onto symmetry plane of the collision (Scalar Product method):

\begin{multline}
    v_n^{obs} = \frac{1}{2\pi} \langle u_n Q_n^* \rangle = \int  d \Psi_n^R \int d\phi e^{in\phi} e^{ -in \Psi_n^E } \rho(\phi-\Psi_n^R) =  \\
     \langle \cos n ( \phi - \Psi_n^R ) V_n \cos n ( \Psi_n^E - \Psi_n^R )\rangle .
\end{multline}

Since the number of particles used for symmetry plane estimation is always limited, the cosine term with difference of symmetry plane angle and reaction plane angle will always be less than 1.
Therefore correction $R_n$ on the symmetry plane resolution is needed. 
This correction is provided using the resolution correction coefficient $R_n$ defined as follows:
\begin{equation}
    R_n = \langle V_n \cos n( \Psi_n^E - \Psi_n^R ) \rangle.
\end{equation}
Then the unbiased observable for the azimuthal flow of particles is defined by following equation:
\begin{equation}
    v_n = \frac{v_n^{obs}}{R_n} =  \frac{\langle u_n Q_n^* \rangle }{R_n}.
\end{equation}

Since the reaction plane of the collision is unknown, calculation of the resolution correction factor $R_n$ can be performed using the pairwise correlations of $Q_n$ vectors:
\begin{equation}
    \langle Q_n^a Q_n^b* \rangle =   \langle V_n^a \cos n(\Psi_n^a - \Psi_R) V_n^b \cos n(\Psi_n^a - \Psi_R) \rangle,
\end{equation}
where $a$ and $b$ indices indicate two groups of particles in each of which the symmetry plane $\Psi_n^{a,b}$ estimation was carried out separately.
In this work the resolution correction factor was calculated using the method of three sub-events.
Using three groups of particles, say $a$, $b$ and $c$, we can estimate resolution via this formula:
\begin{equation}
    R_n\{a(b,c)\} = \sqrt{ \frac{  \langle Q_n^a Q_n^b \rangle \langle Q_n^a Q_n^c \rangle }{  \langle Q_n^b Q_n^c \rangle } }
\end{equation}

To suppress the correlations not correspondent to the initial collective motion of the produced particles (non-flow) we suggest defining a group of particles with sufficient (pseudo-) rapidity separation between each of symmetry planes $a$, $b$ and $c$. 
In the case where this separation cannot be achieved (for example $a$ and $b$ or $a$ and $c$ are not separated) we can introduce additional symmetry plane vector $d$, and require (pseudo-) rapidity separation only between three of the event planes, say $a$ and $d$, $b$ and $d$, $c$ and $d$ and $c$ and $d$.
Slight modification of the three sub-event method produces the estimation of resolution correction factor produces which we going to call the method of four sub-events:
\begin{equation}
    R_n\{a(d)(b,c)\} = \langle Q_n^a Q_n^d \rangle \sqrt{ \frac{  \langle Q_n^d Q_n^b \rangle \langle Q_n^d Q_n^c \rangle }{  \langle Q_n^b Q_n^c \rangle } }
\end{equation}

In this  work we use the symmetry plane defined from the spectator energy deposition in a modular detector FHCal.
In this case the first-order symmetry plane $Q_1$ can be estimated using the modification of formula~\ref{eq:Qn_definition}:
\begin{equation}
Q_1 = \sum_{k=1}^{N} E_k e^{i\varphi_k} / \sum_{k=1}^{N} E_k,
\label{eq:module_vector}
\end{equation}
where $\varphi$ is the azimuthal angle of the $k$-th FHCal module, $E_k$ is the signal amplitude seen by the $k$-th FHCal module, which is
proportional to the energy of spectator.  $N$ denotes   the number of modules in the group. 
To suppress the auto-correlations between $u_n$ and $Q_n$ vectors we rejected protons with projected position in FHCal plane within the acceptance of FHCal.\\

Since the reaction plane orientation is random and uniform, in the case of the ideal detector acceptance, correlation of vectors can be substituted with the correlation of their components (for more details see~\cite{flow1,flow2,mamaev}):
\begin{equation}
    \langle Q_n^{a} Q_n^{b} \rangle = 
    2 \langle X_n^a X_n^b \rangle =
    2 \langle Y_n^a Y_n^b \rangle,
    \label{eq:components_2part}
\end{equation}
or similarly for the three-particles correlation:
\begin{equation}
    \langle Q_{2n}^{a} Q_n^{b}  Q_n^{c} \rangle = 
    4 \langle X_{2n}^a X_n^b  X_n^c \rangle =
    4 \langle X_{2n}^a Y_n^b  Y_n^c \rangle =
    4 \langle Y_{2n}^a X_n^b  Y_n^c \rangle =
    -4 \langle Y_{2n}^a Y_n^b  X_n^c \rangle.
    \label{eq:components_3part}
\end{equation}

Based on this, one can use only correlations of components of $Q_n$ and $u_n$ vectors to calculate flow coefficients.
\begin{equation}
    v_n = 2\frac{ \langle x_n X_n^* \rangle }{ R_n^x } = 2\frac{ \langle y_n Y_n^* \rangle }{ R_n^y }, \label{eq:v1_formula_2subevts}
\end{equation}
where $R_n^{x,y}$ notate values of resolution correction coefficient calculated using the $X$ and $Y$ components of $Q_n$-vectors.\\

For instance, equation~\ref{eq:v1_formula_2subevts} for $v_1$ can be rewritten as follows:
\begin{equation}
    v_1 = \frac{ 2 \langle y_1 Y_1^a \rangle }{ R_1^{y} \{ a \} },
\end{equation}
where $y_1$ and $Y_1^a$ are $y$-components of $u_1$ and $Q_1^a$ vectors respectively, and $R_1^{y} \{ a \}$ is the resolution correction
factor for $Y_1^{a}$:
\begin{equation}
    R_1^{y}\{a(b,c)\} = 
    \sqrt{
    \frac{ 2 \langle Y_1^b Y_1^c \rangle }
    { 2\langle Y_1^{a} Y_{1}^{b} \rangle 
    2 \langle Y_1^a Y_1^c \rangle }},
\end{equation}

\begin{figure}[h]
    \centering
    \includegraphics[width=0.98\textwidth]{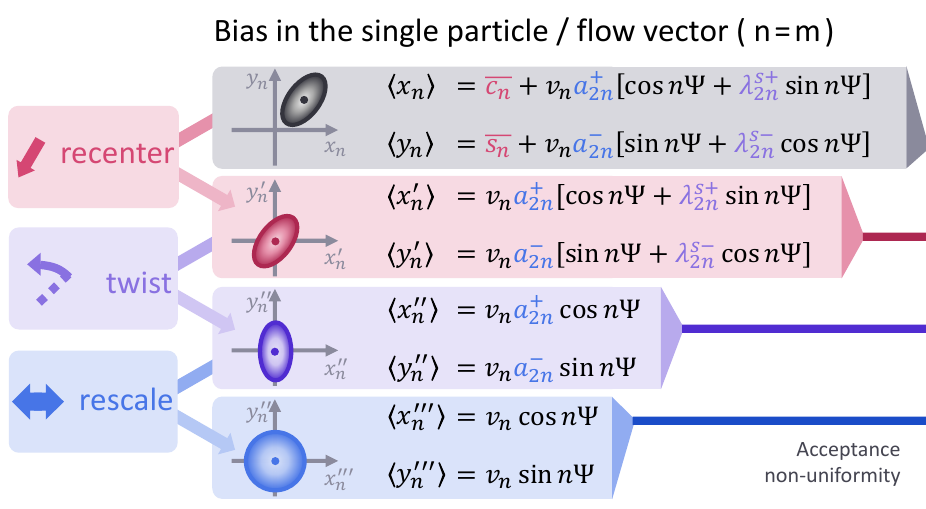}
    \caption{ Schematic illustration of recentering, twist and rescale correction steps for $q_n$-vector introduced in~\cite{flow2}.  }
    \label{fig:qn_corrections}
\end{figure}

In case of an ideal detector, the $Q_n$-vector relation to the symmetry plane is limited only by the multiplicity of the particles within the acceptance. 
In reality, the detector non-uniformity in $\phi$ and effects from the magnetic field, additional material etc., can bias the flow measurements.
This leads to equations~\ref{eq:components_2part} and~\ref{eq:components_3part} are being no longer valid.
Detector non-uniformities can  be treated on the level of the  flow $Q_n$ vectors. The following procedure was introduced in \cite{flow2}.
The advantages compared to re-weighting of the
azimuthal particle spectra is that the procedure also works with detectors that have holes in
the azimuthal acceptance. The necessary correction factors can be fully determined from the
data itself. Monte Carlo simulations are not needed.
The corrections (re-centering, twist and rescaling) can also be generalized to a generic normalized flow vector $q_n$ with the
components $x_n$  and $y_n$. Schematic representation of these corrections are shown in Figure~\ref{fig:qn_corrections}.\\

\begin{figure}[h]
    \centering
    \includegraphics[width=0.98\textwidth]{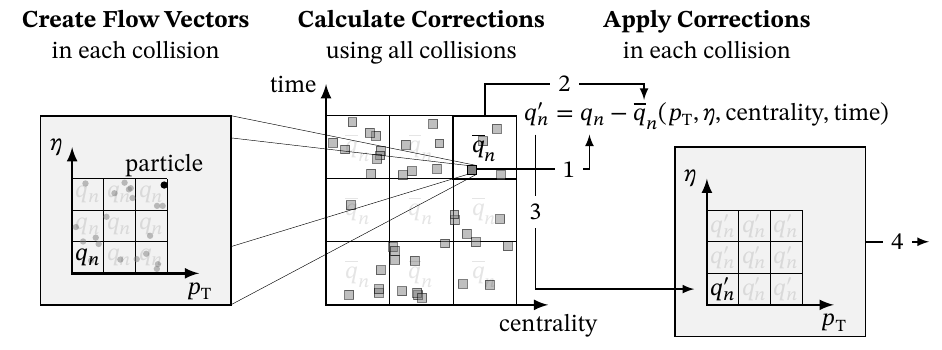}
    \caption{Sketch of the multi-dimensional correction procedure in the QnTools framework. As an example the recentering correction as a
function of $p_T$, $\eta$, centrality, and time is shown. }
    \label{fig:qn_corrections2}
\end{figure}

{\bf Re-centering:} A static shift of the detector signals can manifest in a shift of the average flow vector away
from the origin. This shift can be removed by subtracting the mean flow vector from the flow
vector in each collision.\\
{\bf Twist/Diagonalization:} The flow vector distribution can appear twisted, if $sin(n\Psi$), or $cos(n\Psi$) terms bias the $x_n$  and $y_n$
components of the flow vectors. The diagonalization corrections are calculated from the
averaged flow vector components and applied to the flow vector in each collision.\\
{\bf Rescaling:} A squashed flow vector distribution, which corresponds to different widths in the $x$ and $y$ direction, can be corrected with the rescaling
correction.\\
The formalism of these corrections has been implemented in a software framework known as QnTools\cite{qntool}, which
allows to perform the corrections of differential flow vectors, which may depend on a number of particle properties $q_n(p_T, \eta, PID,...)$, see
Figure~\ref{fig:qn_corrections2}.

\subsection{BM@N performance for flow measurements}
\begin{figure}[h]
    \centering
    \includegraphics[width=0.4\textwidth]{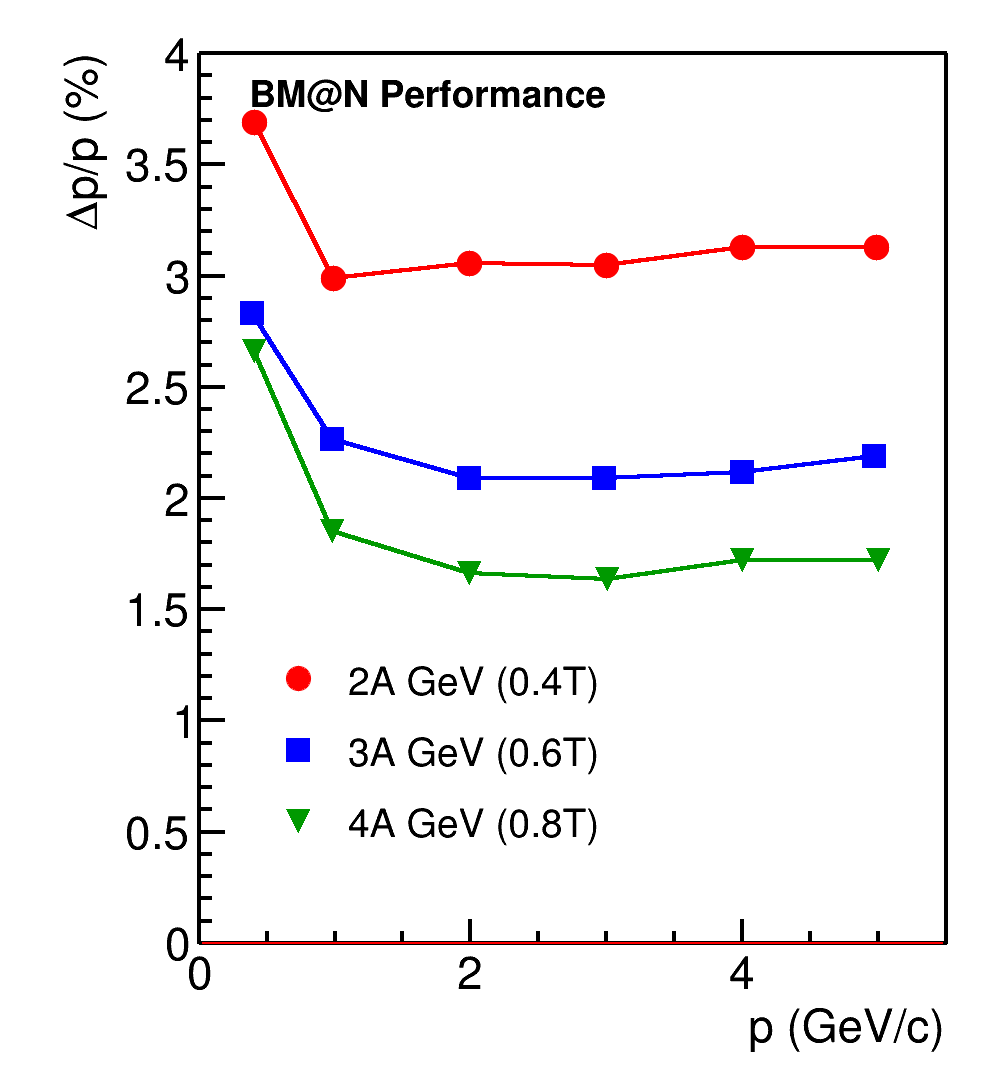}
    \includegraphics[width=0.4\textwidth]{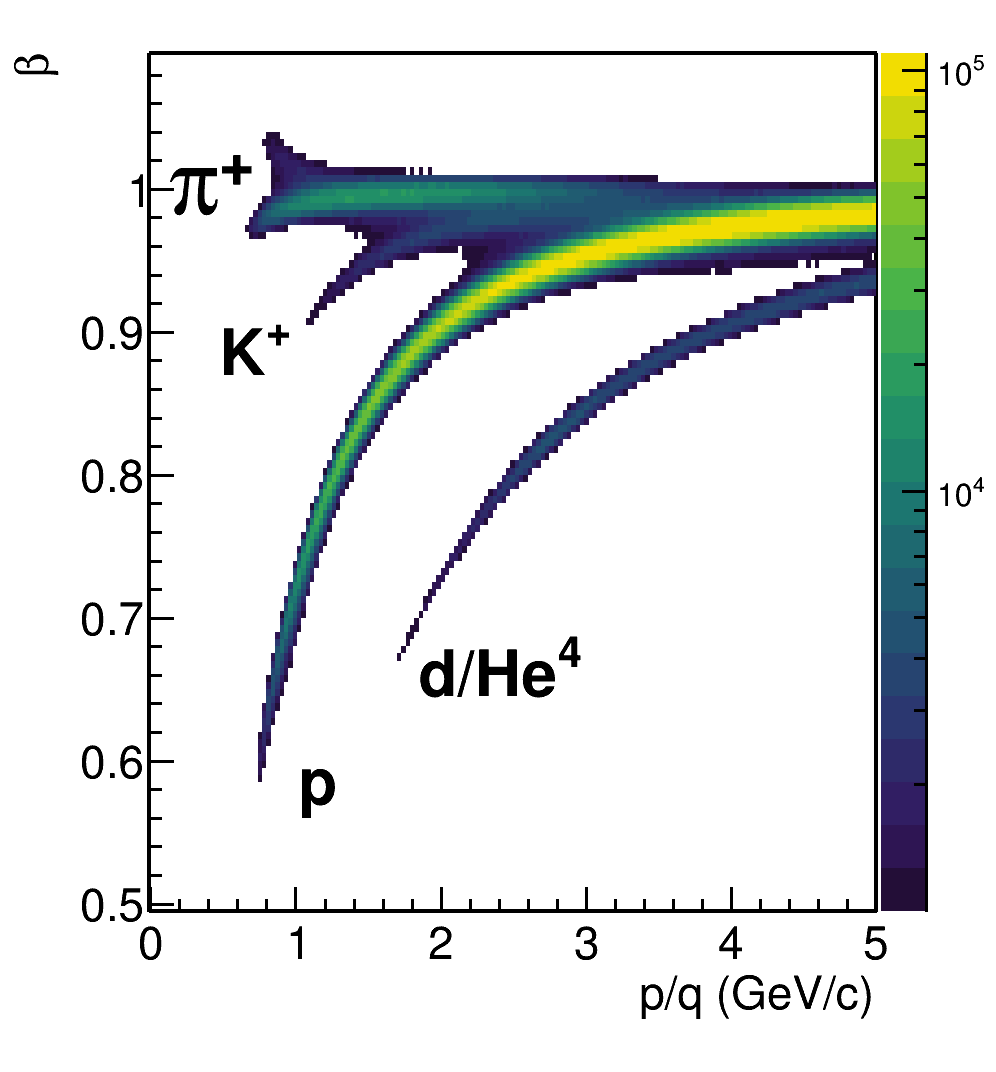}
    \caption{Left: Relative momentum resolution $\Delta p/p$ as a function of the momentum $p$ for fully reconstructed charged tracks from Xe+Cs(I)
      collisions generated using the JAM model at different kinetic energies:
      4 AGeV (triangles), 3 AGeV (boxes) and 2 AGeV (circles). Right: Population of the reconstructed charged particles in the
      velocity $\beta$ vs. laboratory momentum over charge ($p/q$) plane for the TOF-400.}
    \label{fig:momentum_resolution}
\end{figure}

In this subsection, we discuss  the
anticipated performance of BM@N experiment \cite{bmn} in the configuration for run8
for differential anisotropic flow measurements of identified hadrons at Nuclotron
energies $\snn$ = 2.3-3.5 GeV, see \cite{mamaev} for the details.  As the main event generator we have
used the JAM (RQMD.RMF) model \cite{jam1,jam2,jam3} with momentum dependent mean field \cite{jam3}, which
qualitatively describes the existing measurements of directed and elliptic flow of protons
at this energy range \cite{mamaev,parfenov1}. We generated  about 5 M minimum bias Xe+Cs(I) collision events for each beam energy: 2, 3 and 4 AGeV.
At the next step, the sample of JAM model events was made
as an input for the full chain of realistic simulations of the BM@N detector subsystems for run8
based on the GEANT4 platform and reconstruction algorithms built in the BMNROOT
framework. The fully reconstructed events were used to generate the distributions of
the multiplicity $N_{ch}$  of the produced charged particles detected by FSD+GEM system of the BM@N \cite{bmn} and estimate
the centrality, see the section 3.4 for the details.\\
The tracking system allows to reconstruct the momentum $p$ of the particle
with a momentum resolution of $\Delta p/p \sim$ 1.7-2.5$\%$ for the
kinetic energy 4A~GeV (magnetic field 0.8 T). For the experiment at lower kinetic energy 2~AGeV one needs to use the reduced magnetic field 0.4 T. This 
leads to a deterioration in the momentum resolution, see the left part of the Figure~\ref{fig:momentum_resolution}.
Charged-hadron identification is based on the time-of-flight measured with TOF-400 and TOF-700.
The time resolutions of the ToF-400 and ToF-700 systems are 80 ps and 115 ps, respectively.
Particle velocity is obtained from the measured flight time and flight path. Combining this information with the particle momenta $p$ allows
to identify charged hadrons with high significance. As an example, the right part of the Figure ~\ref{fig:momentum_resolution} shows the population of all
charged particles in the plane spanned by their $\beta$ and  momenta divided by charge (rigidity) for the TOF-400.

Symmetry plane estimation was carried out in assumption that spectator fragments are pushed
in reaction plane by the expanding overlap region of colliding
nuclei and they have positive directed flow signal $v_1>0$ in the forward rapidity resgion \cite{flow1,flow2}.
The Forward Hadron Calorimeter (FHCal) registers the energy deposition of spectator fragments in the BM@N experiment.
Modules of the FHCal
were divided into three groups according to the ranges of pseudorapidity in the laboratory frame $\eta$:
(F1) $4.4 < \eta < 5.5$; (F2) $3.9 < \eta < 4.4$; and (F3) $3.1 < \eta < 3.9$, see the left part of Figure~\ref{fig:fhcal_subevents}.
The $Q_1$ vectors for each sub-event (F1, F2, F3) in the FHCal  have been  obtained as follows:
\begin{equation}
Q_1 = \sum_{k=1}^{N} E_k e^{i\varphi_k} / \sum_{k=1}^{N} E_k,
\label{eq:module_vector}
\end{equation}
where $\varphi$ is the azimuthal angle of the $k$-th FHCal module, $E_k$ is the signal amplitude seen by the $k$-th FHCal module, which is
proportional to the energy of spectator.  $N$ denotes  the total number of modules  in the given sub-event. 
\begin{figure}
    \centering
    \includegraphics[width=0.37\textwidth]{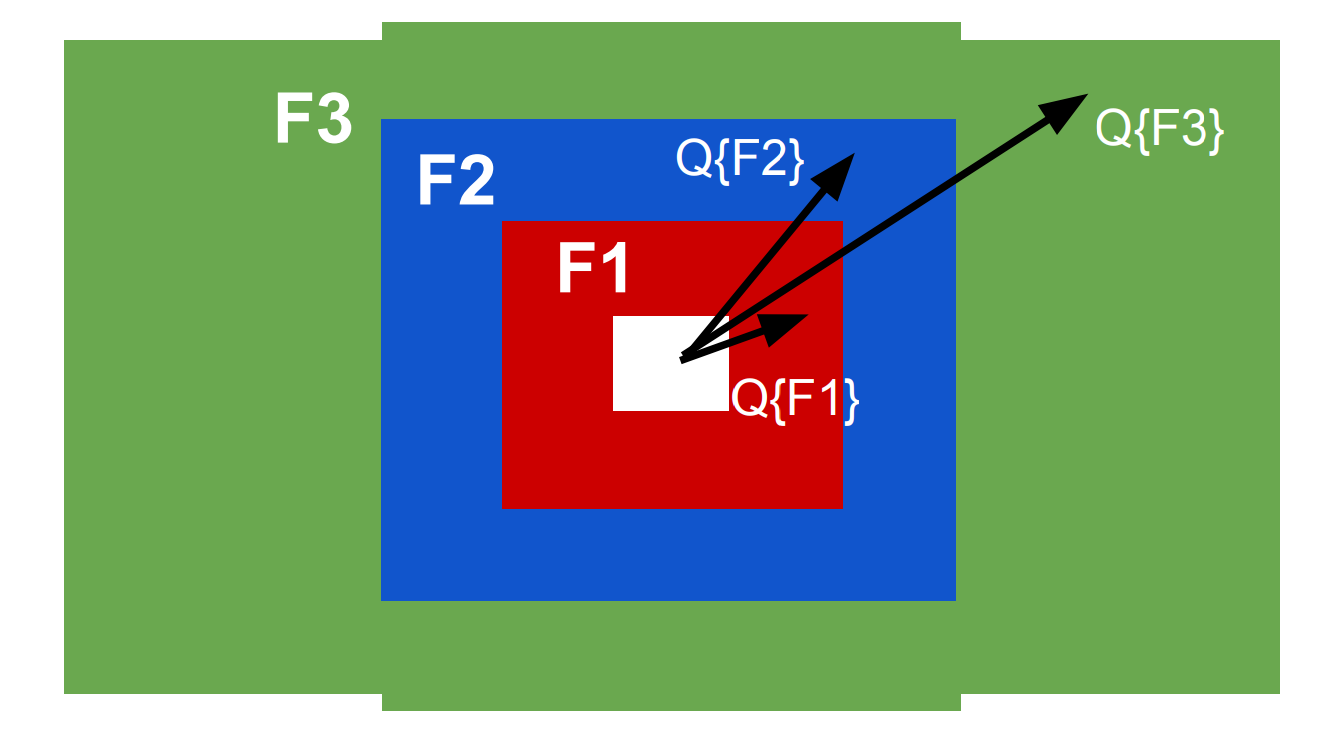}
    \includegraphics[width=0.33\textwidth]{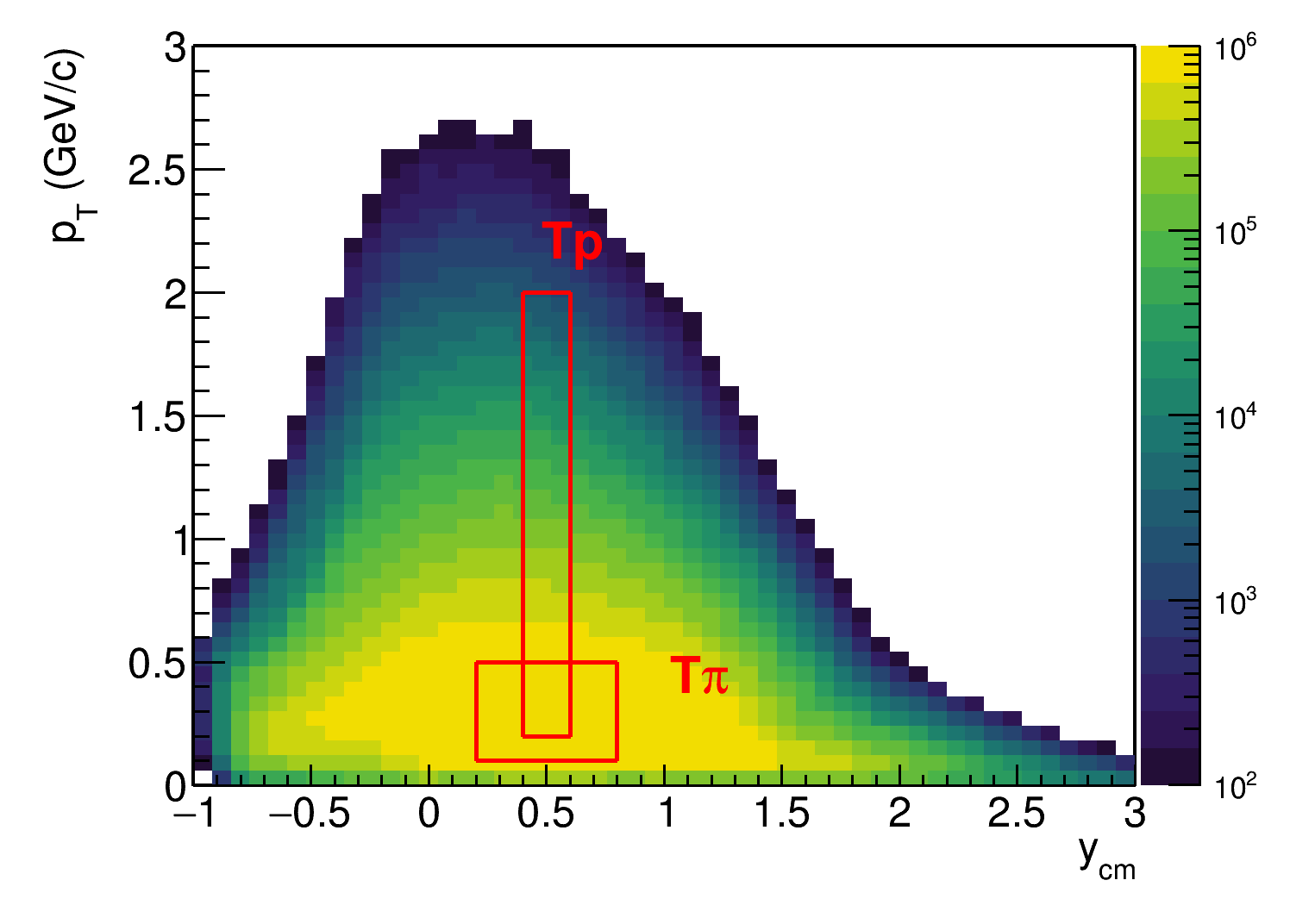}
    \caption{ Left part: schematic representation of modules of the Forwar Hadron Calorimeter divided in 3 groups. The corresponding sub-events are represented with different colors.
      Arrows denote the $Q_1$-vectors for  each sub-event in FHCaL (F1,F2,F3).
      Right part: schematic  representation of kinematic windows for  $Q_1$-vectors from tracks ($Tp$ and $T\pi$), see text for the details. }
    \label{fig:fhcal_subevents}
\end{figure}
Two additional sub-events were introduced from the  tracks of the charged  particles in the inner tracking system of BM@N. 
For the first group we used the  protons ($Tp$) in the kinematic window of $0.4<y_{cm}<0.6$ and $0.2 < p_{T} < 2.0$~$GeV/c$
and the negative charged pions ($T\pi$) for the second group with $0.2<y_{cm}<0.8$ and $0.1 < p_T < 0.5~GeV/c$.
The $Q_1$ vectors defined from the tracks of charged particles  ($Tp$ and $T\pi$) are calculated according to Eq.~\ref{eq:Qn_definition},
see the right panel of Figure~\ref{fig:fhcal_subevents}.

\begin{figure}[h]
    \centering
    \includegraphics[width=0.46\linewidth]{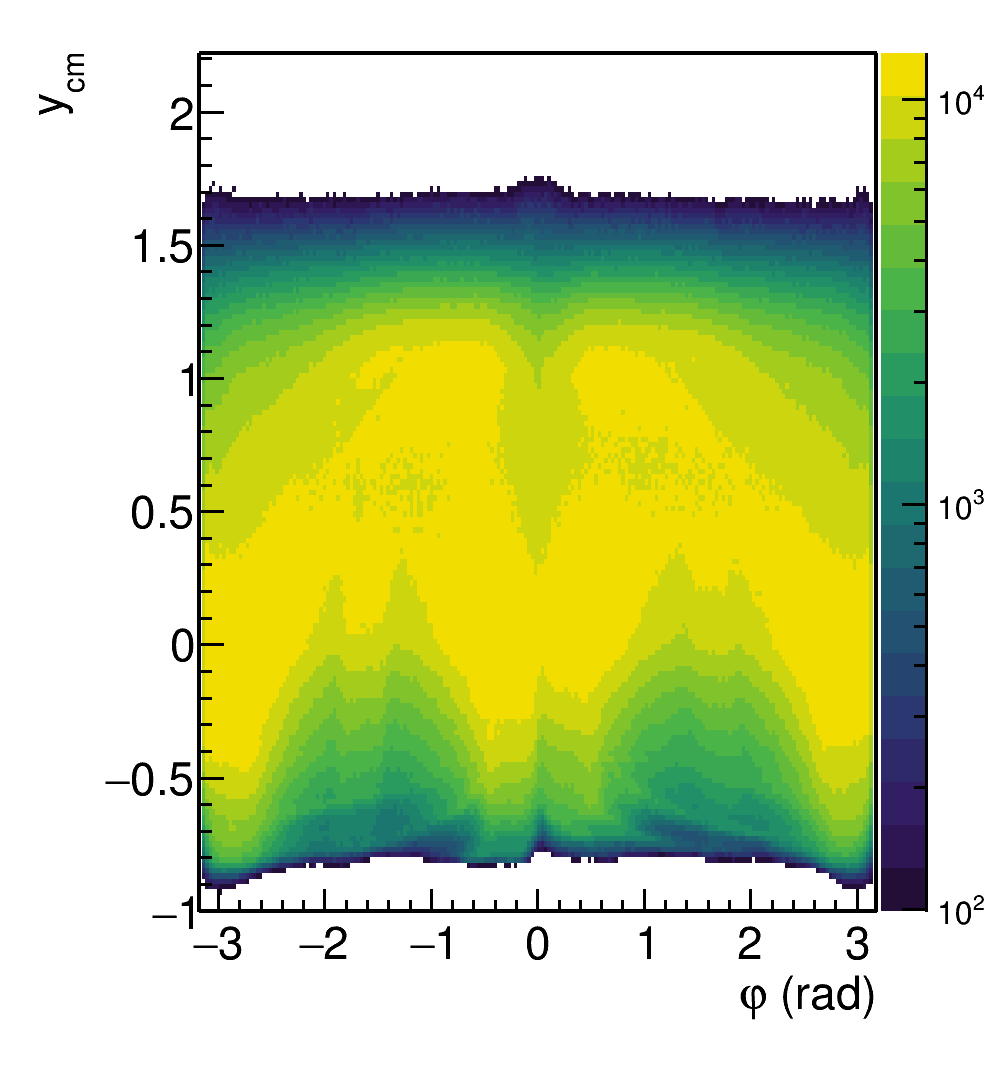}~%
    \includegraphics[width=0.46\linewidth]{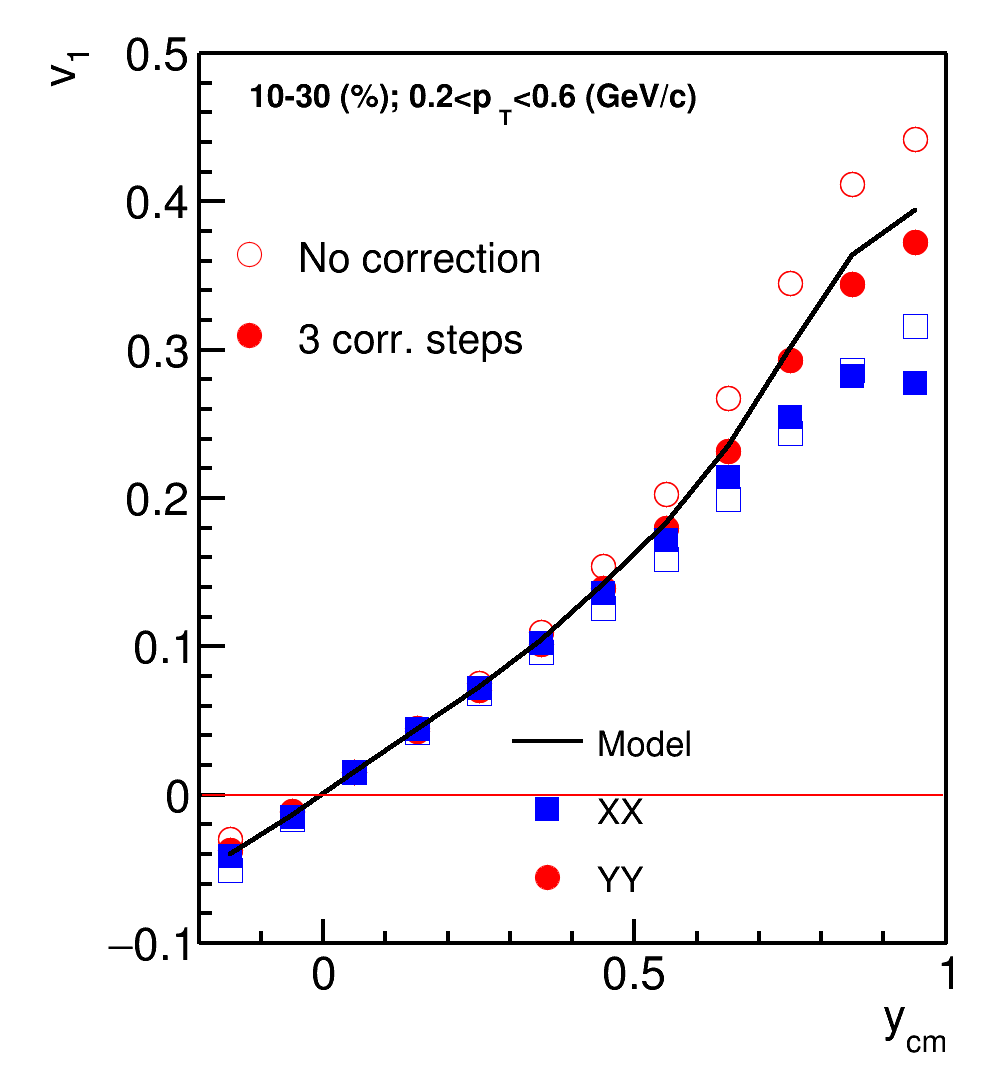}
    \caption{
    (Left) Raw yield of protons as a function of azimuthal angle $\varphi$ and center-of-mass rapidity $y_{cm}$.
      (Right) Comparison of the directed flow $v_1$ signal of protons before (open symbols) and after (closed symbols)
      corrections on the  non-uniformity of azimuthal acceptance, see the text for the details.
    }
    \label{fig:non_uniformity}
\end{figure}

The left part of Figure~\ref{fig:non_uniformity} shows the acceptance for selected  protons: azimuthal angle $\varphi$ vs center-of-mass rapidity $y_{cm}$.
The  azimuthal coverage of the tracking system in the BM@N is strongly non-uniform. 
QnTools framework \cite{qntool} with  recentering, twist and rescaling corrections has been applied for both $u_1$ and $Q_{1}$ vectors. The right part of
Figure~\ref{fig:non_uniformity} shows the $y_{cm}$ dependence of $v_1$ of protons with 0.2 $< p_T <$ 0.6 GeV/c  from 10-30\% central Xe+Cs(I) collisions.
The black colid line denotes the $v_1$ values of protons directly from the JAM model. The  symbols denote the $v_1(y_{cm})$ values of protons from the analysis
of the fully reconstructed model events: before   (open symbols) and after corrections for the non-uniform azimuthal acceptance  (closed symbols).
The application of  corrections yields to a better agreement between the reconstructed
(closed symbols) and  the model (line) $v_1$ signals in the  full range of rapidity. 
The agreement between reconstructed and model values of $v_1$ is better for the results obtained using the $YY$ correlation of vectors.
The magnetic field of BM@N is directed along the $y$ axis and it deflects the produced charged particles in $x$ direction.
This may introduce the additional correlation between the $XX$ components of the vectors and increase the
difference between the reconstructed  $v_1$ calculated from  the correlation of $XX$ components and the $v_1$ values from the JAM model. 
\begin{figure}[h]
    \centering
    \includegraphics[width=0.9\linewidth]{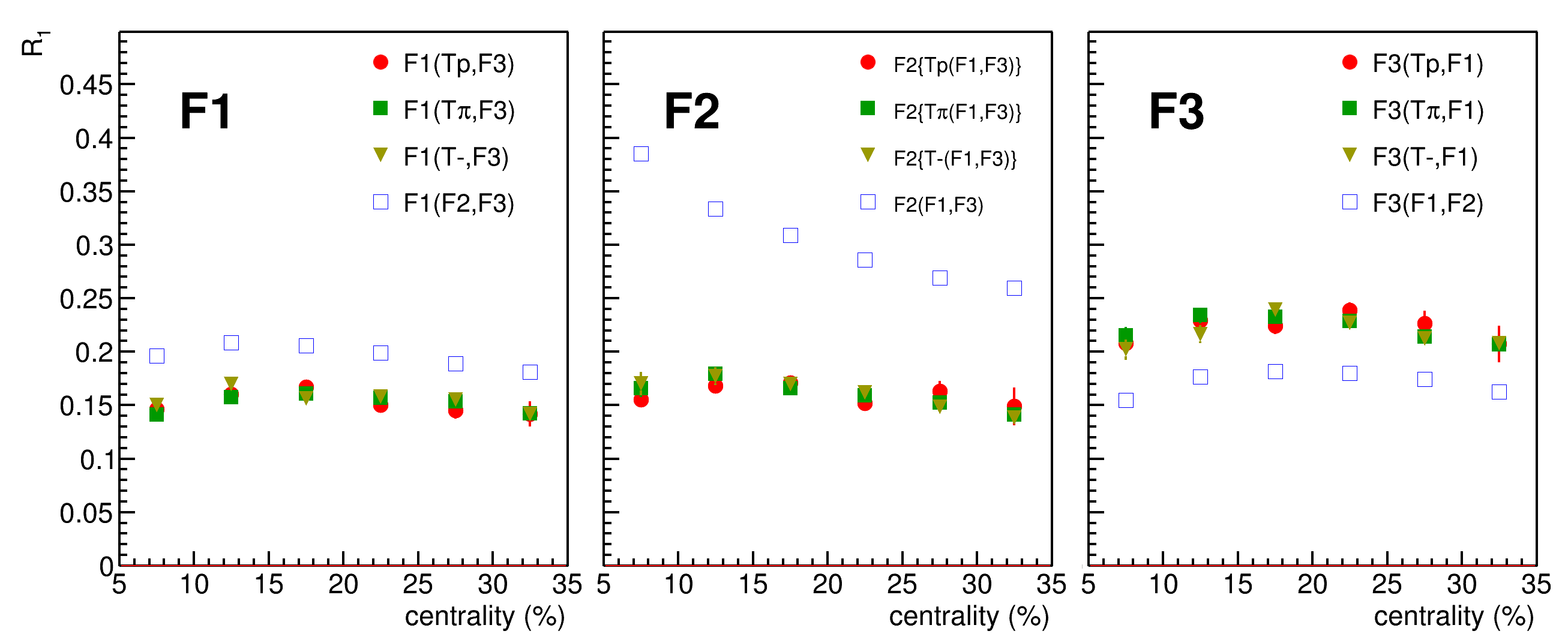}~%
    \caption{
      The centrality dependence of resolution correction factor $R_1$ for  different combinations
      of $Q_1$-vectors in the  3 and 4-subevents methods for F1, F2 and F3 symmetry planes from left to right. 
    }
    \label{fig:R1_combinations_centrality}
\end{figure}

\begin{figure}[h]
    \centering
    \includegraphics[width=0.9\linewidth]{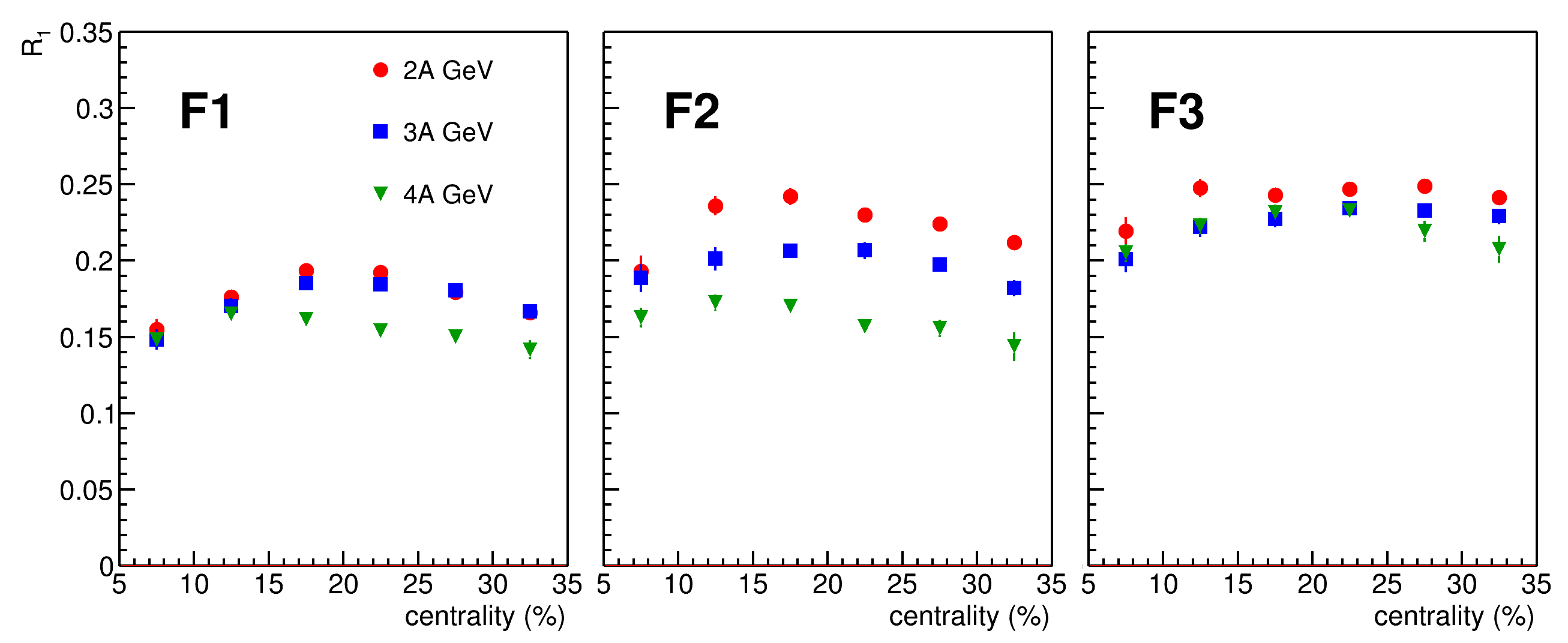}~%
    \caption{
    The centrality dependence of the resolution correction factor $R_1$ for spectator plane.
    The results are presented for sub-events F1, F2 and F3: panels  from left to right. Different symbols correspond
    to the results for Xe+Cs(I) collisions at  different beam energies: 2, 3 and 4A GeV.  
    }
    \label{fig:R1_centrality}
\end{figure}

Figure~\ref{fig:R1_combinations_centrality} shows the  centrality dependence of resolution correction factor $R_1$ for the  different combinations
      of $Q_1$-vectors in the  3 and 4-subevents methods for F1, F2 and F3 symmetry planes from left to right. 
      Due to the propagation of hadronic shower between the FHCal modules in the transverse direction,  the estimations for the $R_1$  resolution factor
      for  the combinations of
      neighboring sub-events such as F1 and F2 or F2 and F3 will be  strongly biased (blue markers).
In contrast, the $R_1$ values calculated using the combinations with significant rapidity separation (red, green and yellow markers)
are found to be in  agreement within the statistical errors.\\
Figure~\ref{fig:R1_centrality} shows the centrality dependence of the resolution correction factor for the spectator symmetry plane for different beam energies:
2 AGeV (left), 3 AGeV (middle) and 4 AGeV (right).
For all symmetry planes  F1, F2, F3  we observe a decrease of the resolution correction factor $R_1$ with increasing energy.
Shortening of the passage time of colliding nuclei at higher energies leaves less time for the interaction between
the matter produced within the overlap region and spectators, which leads to the smaller values of the spectators directed
flow and smaller magnitude of $Q_1$-vectors. As a consequence, one can expect smaller values for the resolution correction factor $R_1$.

\begin{figure}[h]
    \centering
    \includegraphics[width=0.35\textwidth]{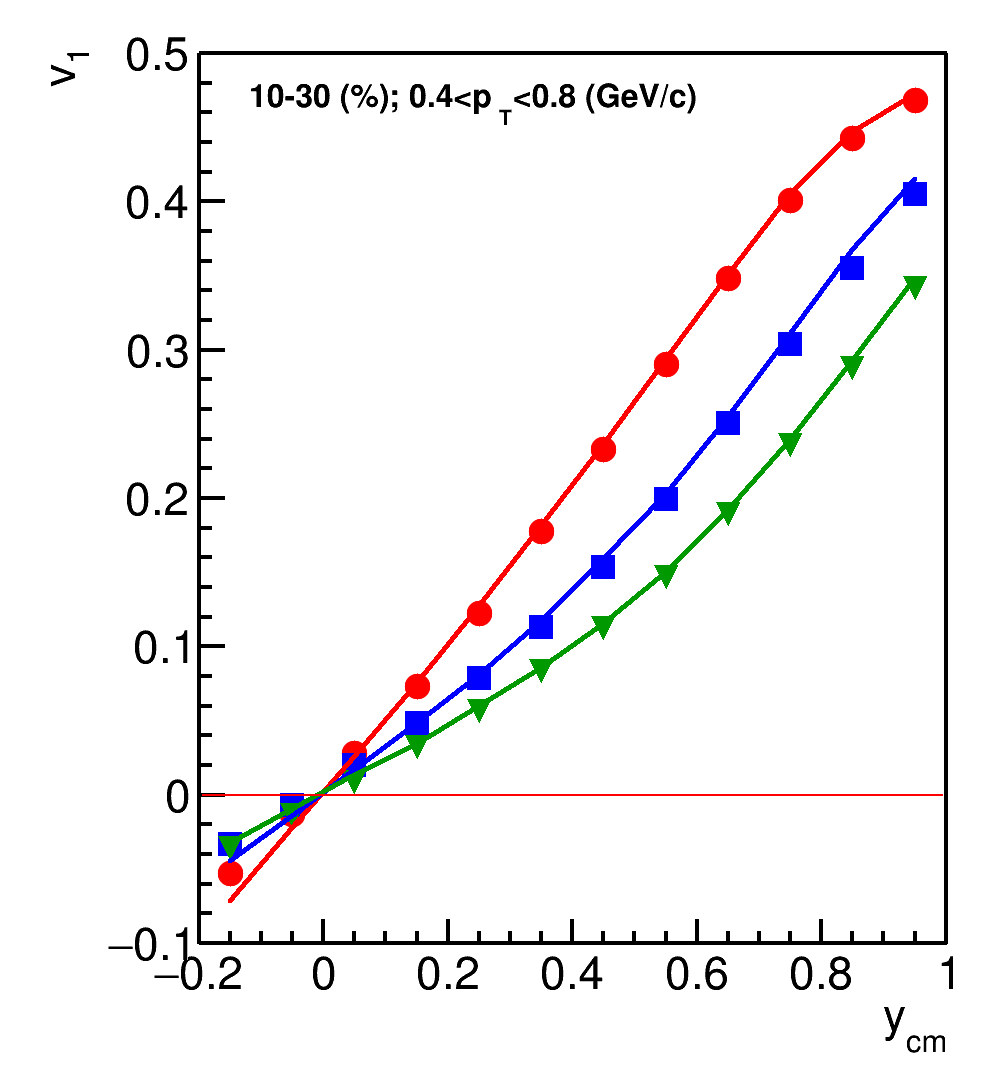}
    \includegraphics[width=0.35\textwidth]{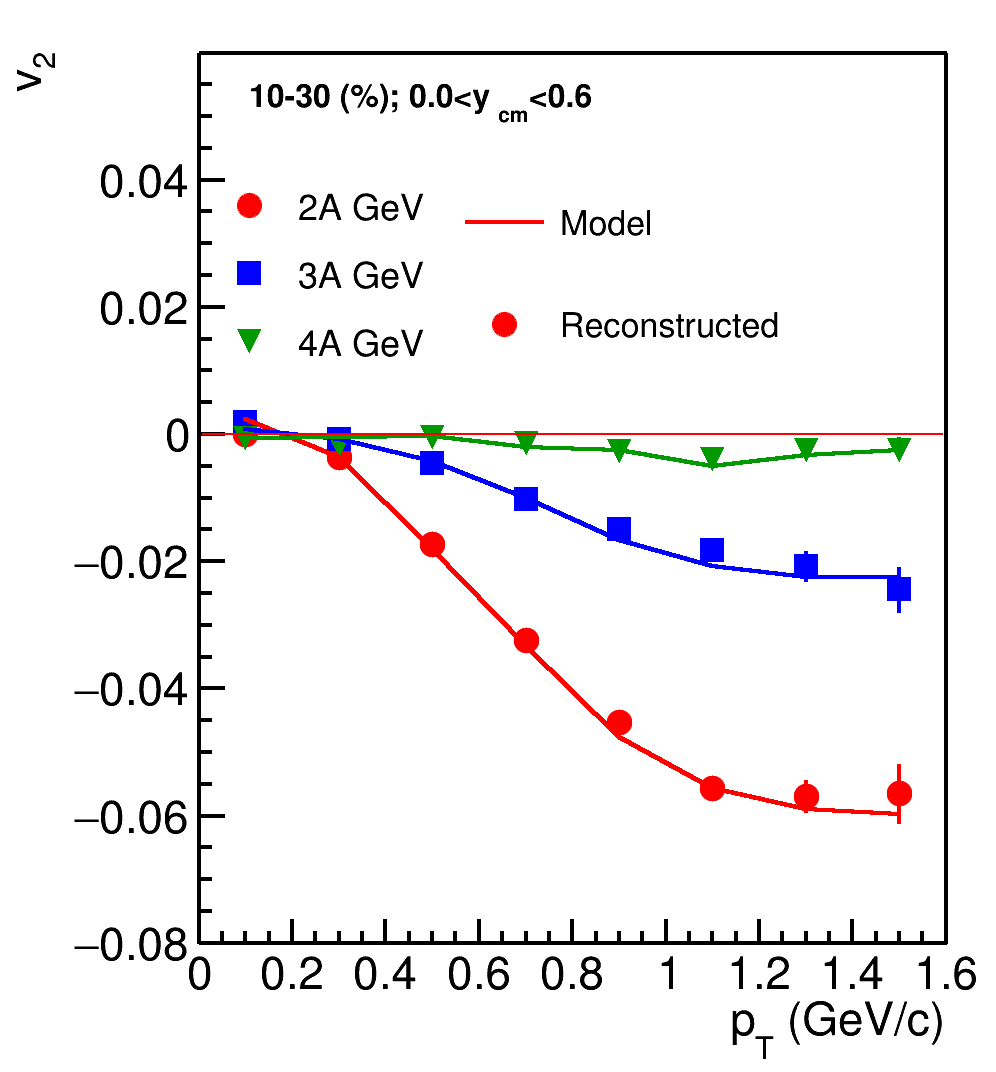}
    \caption{Left: directed flow  $v_1$ of protons as a function of center-of-mass rapidity $y_{cm}$ for 10-30\% central Xe+Cs(I) collisions
      at 2 AGeV (circles), 3 AGeV (boxes)  and 4 AGeV (triangles); Right:
      elliptic flow $v_2$ of protons as a function of transverse momentum $p_T$.
      Markers represent the results of the analysis of the fully reconstructed JAM model data  and lines the results obtained directly from the model.
      Figure is taken from \cite{mamaev}}
    \label{fig:v12_pT_y}
\end{figure}

Figure~\ref{fig:v12_pT_y} shows the directed $v_1$ (left part) and elliptic $v_2$ flow (right part) signals of protons from the analysis of JAM
model events for 10-30\% central Xe+Cs(I) collisions at 2 AGeV (circles), 3 AGeV (boxes)  and 4 AGeV (triangles). 
Markers represent the $v_n$ results from  the analysis of the fully reconstructed JAM model events  and lines the results obtained directly from the model
(output model particles without reconstruction were correlated with the RP).
A good agreement is observed between these two sets of $v_n$ results.

\subsection{The  analysis of $v_1$ of protons from  BM@N run8 data}
In this subsection, we discuss the details of analysis of directed flow $v_1$ of protons in Xe+Cs(I) collisions at 3.8 AGeV
using the BM@N run8 data. \\
{\bf 1)}  To address the effects of the non-uniform acceptance we applied  the corrections for both $u_1$ and $Q_1$ vectors
:recentering, twist and rescaling. The QnTools framework \cite{qntool}  was used for corrections of  $u_1$ and $Q_1$ vectors and flow analysis.
For $u_1$-vector corrections were employed multi-differentially on transverse $p_T$, rapidity $y$ and centrality.
For the $Q_1$-vectors corrections were applied only differentially on centrality.\\
{\bf 2)} The detailed performance study, persented in the previus subsection, shows that
due to magnetic field acting along the $y$-axis and deflecting charged particles along the $x$ axis, we can
measure the directed flow $v_1$ of protons  using only  the $y$ components of flow vectors:
\begin{equation}
    v_1 = 2\frac{ \langle y_1 Y_1^a* \rangle }{ R_1^y\{a\} },
\end{equation}
where the resolution correction factor is calculated using the method of three sub-events:
\begin{equation}
    R_1^y\{a(b,c)\} = \sqrt{ \frac{  \langle Y_1^a Y_1^b \rangle \langle Y_1^a Y_1^c \rangle }{  \langle Y_1^b Y_1^c \rangle } },
    \label{eq:three_subevent1}
\end{equation}
or by the four sub-event method:
\begin{equation}
    R_1^y\{a(d)(b,c)\} = \langle Y_1^a Y_1^d \rangle \sqrt{ \frac{  \langle Y_1^d Y_1^b \rangle \langle Y_1^d Y_1^c \rangle }{  \langle Y_1^b Y_1^c \rangle } },
    \label{eq:four_subevent1}
\end{equation}
    {\bf 3)}  The $Q_1$  vectors for symmetry planes in the FHCal have been obtained using the Eq.~\ref{eq:module_vector}.
Modules of the FHCal were divided into three groups (sub-events): F1, F2, F3  as it is shown in  the Figure.~\ref{fig:fhcal_subevents}.
According to the simulations, due to charge splitting in the dipole analyzing magnet SP-41, F1 sub-event primarily registers the spectator protons,
F2 --- spectator fragments and F3 --- neutrons, see the left part of Figure.~\ref{fig:fhcal_subevents}.
\begin{figure}[h]
  \centering
   
    \includegraphics[width=0.6\linewidth]{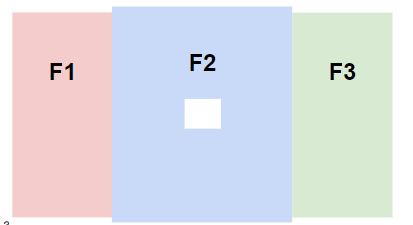}
    \caption{ Layout of the FHCal modules division into three groups (sub-events): F1, F2, F3.}
    \label{fig:fhcal_subevents}
\end{figure}

Two additional sub-events were introduced from the tracks of the charged particles in the inner tracking system of BM@N.
All negatively charged particles with pseudorapidity $1.5 < \eta < 3$ and transverse momentum $p_T > 0.2$~GeV/c comprise the T- sub-event.
The T+ sub-event consists of positively charged particles in following kinematic region: $2 < \eta < 3$ and $p_T > 0.2$~GeV/c.

\begin{figure}[h]
    \centering
    \includegraphics[width=0.95\linewidth]{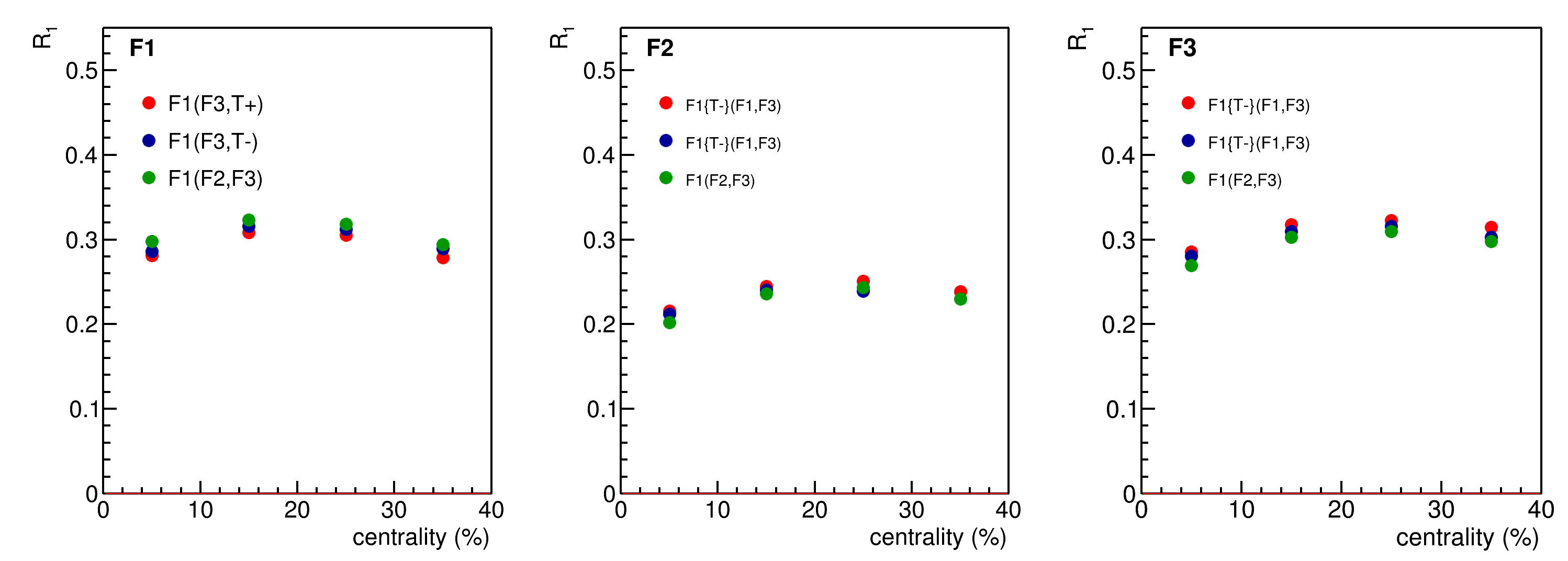}
    \caption{ Resolution correction factor $R_1$ calculated using different combinations as a function of centrality for
      sub-event  symmetry planes F1, F2 and F3 from left to right. }
    \label{fig:f123_centrality}
\end{figure}

\begin{figure}[h]
    \centering
    \includegraphics[width=0.45\linewidth]{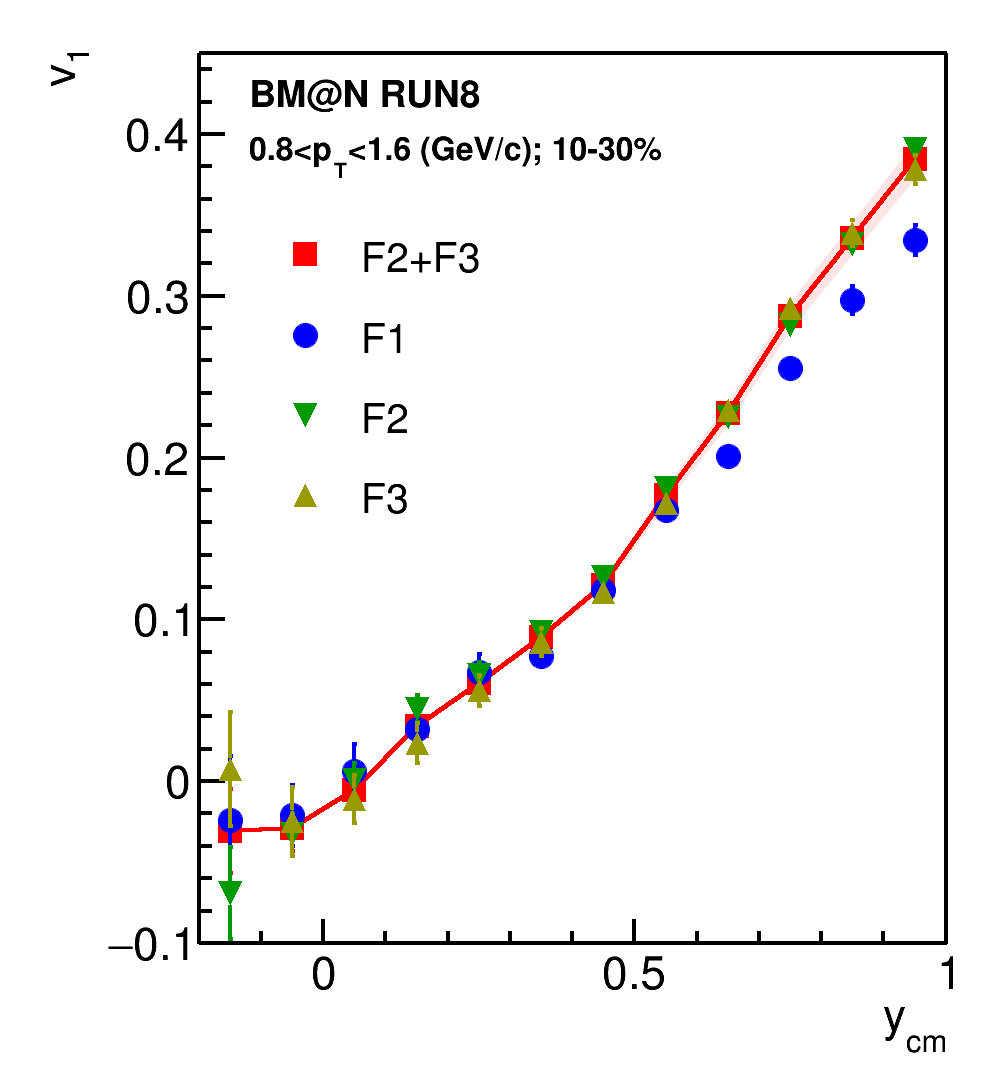}
    \caption{ Directed flow $v_1$ of protons as a function of rapidity $y_{cm}$ measured  with
      respect to different spectator symmetry planes:  F1, F2, F3 and combined (F2+F3), see text for the details. }
    \label{fig:sys_nonflow}
\end{figure}

Resolution correction factor was calculated for 3 spectator symmetry planes F1, F2, and F3 using the three sub-event
method (for F2 four sub-event technique was employed as well) using the Equation~\ref{eq:three_subevent1} (and for four sub-events
Equation~\ref{eq:four_subevent1}).
Figure~\ref{fig:f123_centrality} shows the centrality dependence of the resolution correction factors $R_1$
for sub-event symmetry planes F1, F2 and F3 from left to right.
For each symmetry plane $R_1$ was estimated using 3 combinations of sub-events (as indicated in the figure). One  can  observe that
all three estimations for each symmetry plane are in reasonable agreement.  may  suggests that the final values of $R_1$ resolution factors
This fact may suggest that the contribution of  non-flow correlations in the final values of $R_1$ is very small.\\
    {\bf 4)}  Figure~\ref{fig:sys_nonflow} shows the rapidity dependence $y_{cm}$ of directed flow $v_1$ of protons in 
    in 10-30\% central Xe+Cs(I) collisions at 3.8 A GeV. The measurements have been performed with respect to the F1, F2, F3 and combined
    (F2+F3)  symmetry planes. The  resulting $v_1$ values of protons are in a good agreement for the measurements with respect to F2, F3 and combined
    (F2+F3)  symmetry planes. The small difference in resulting $v_1$ values for the measurements with respect to the F1 plane, can be
    explained by the small contribution of non-flow effects.  In order to get the final results, the
    measurements of directed flow $v_1$ have been performed with respect to the combined (F2+F3)  symmetry plane, see Figure~\ref{fig:v1_y_pT}.

\begin{figure}[h]
    \centering
    \includegraphics[width=0.95\linewidth]{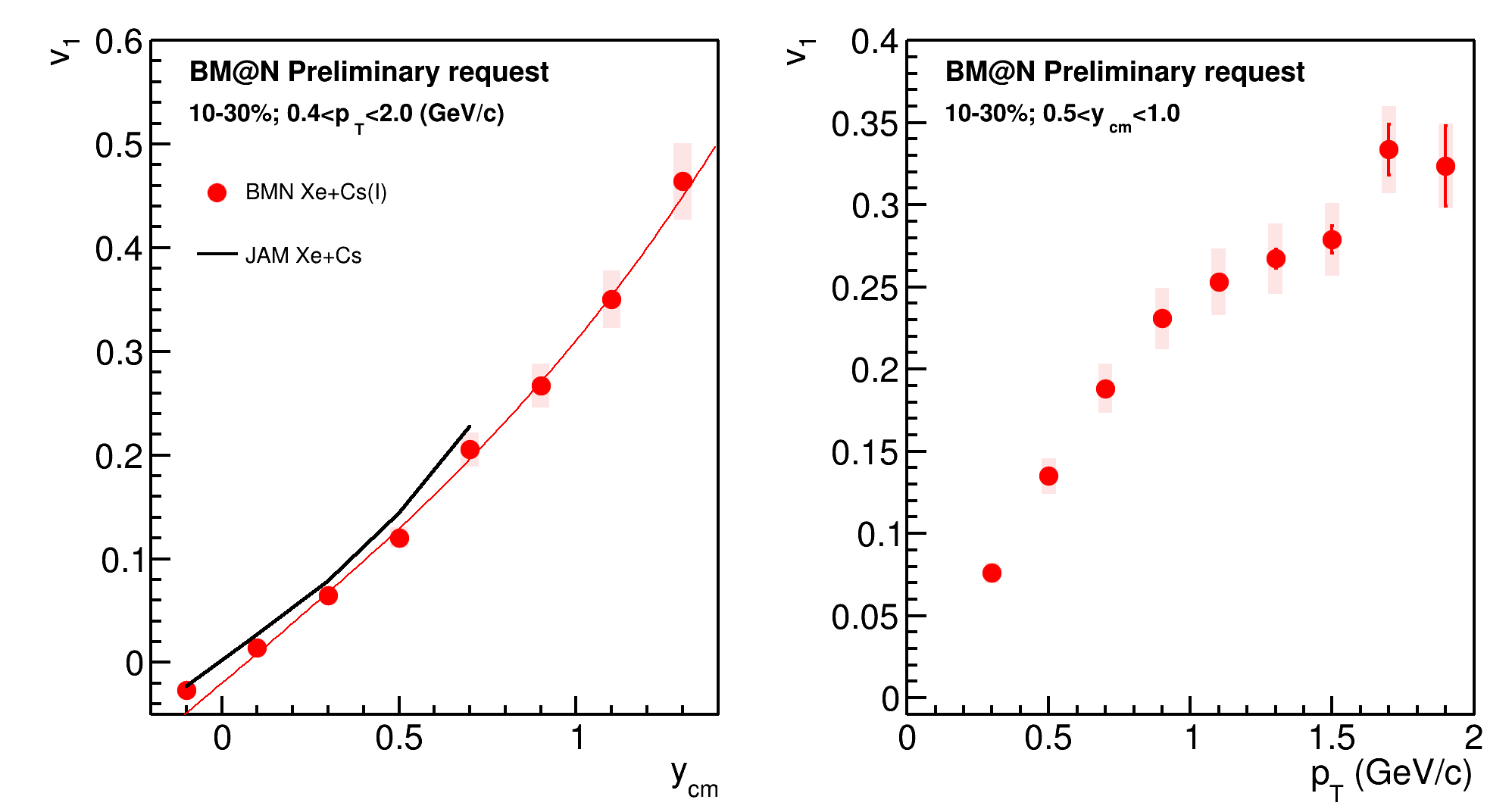}
    \caption{ Directed flow $v_1$ of protons in 10-30\% central Xe+Cs(I) collisions at 3.8 A GeV as a function of 
rapidity $y_{cm}$ (left panel) and transverse momentum $p_T$ (right panel).}
    \label{fig:v1_y_pT}
\end{figure}

\subsection{Systematic uncertainties of $v_1$ measurements}
\begin{figure}[h]
    \centering
    \includegraphics[width=0.45\linewidth]{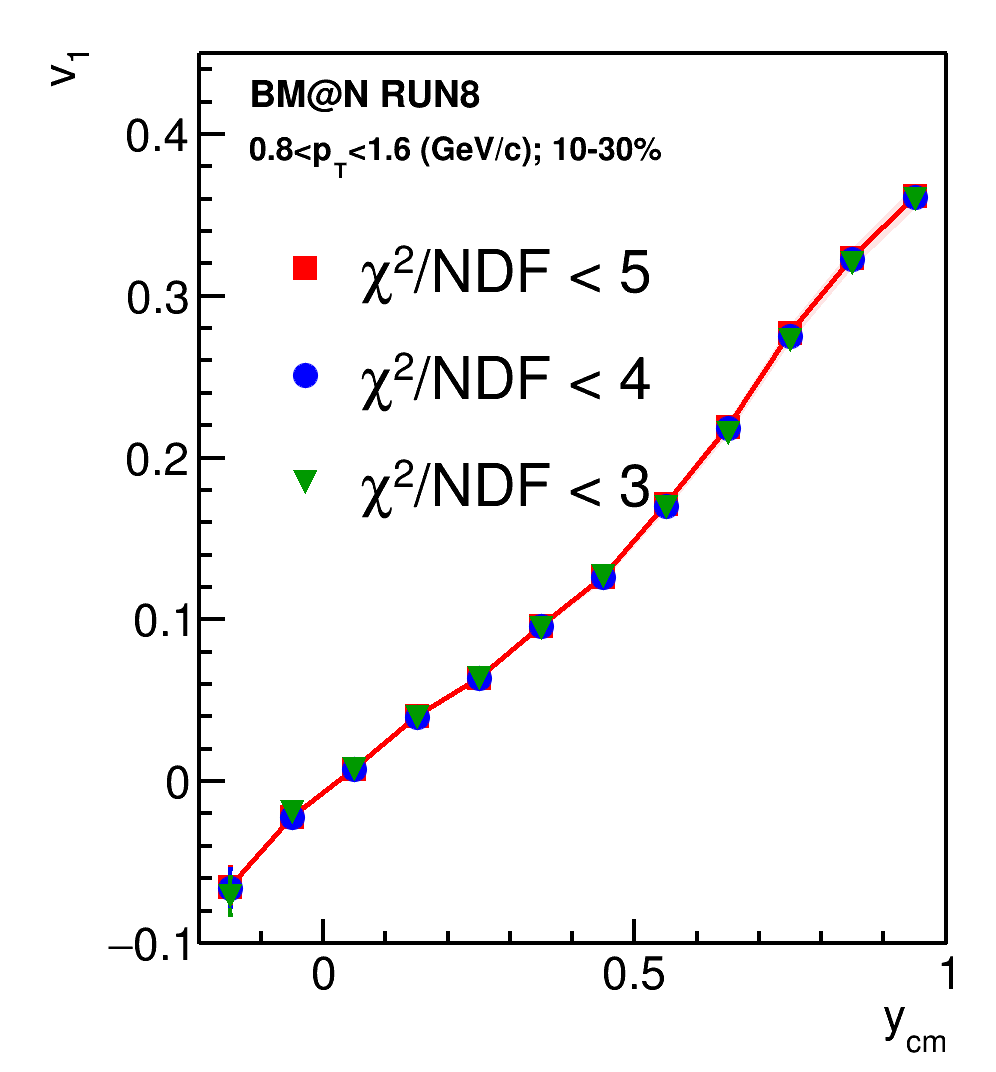}
    \includegraphics[width=0.45\linewidth]{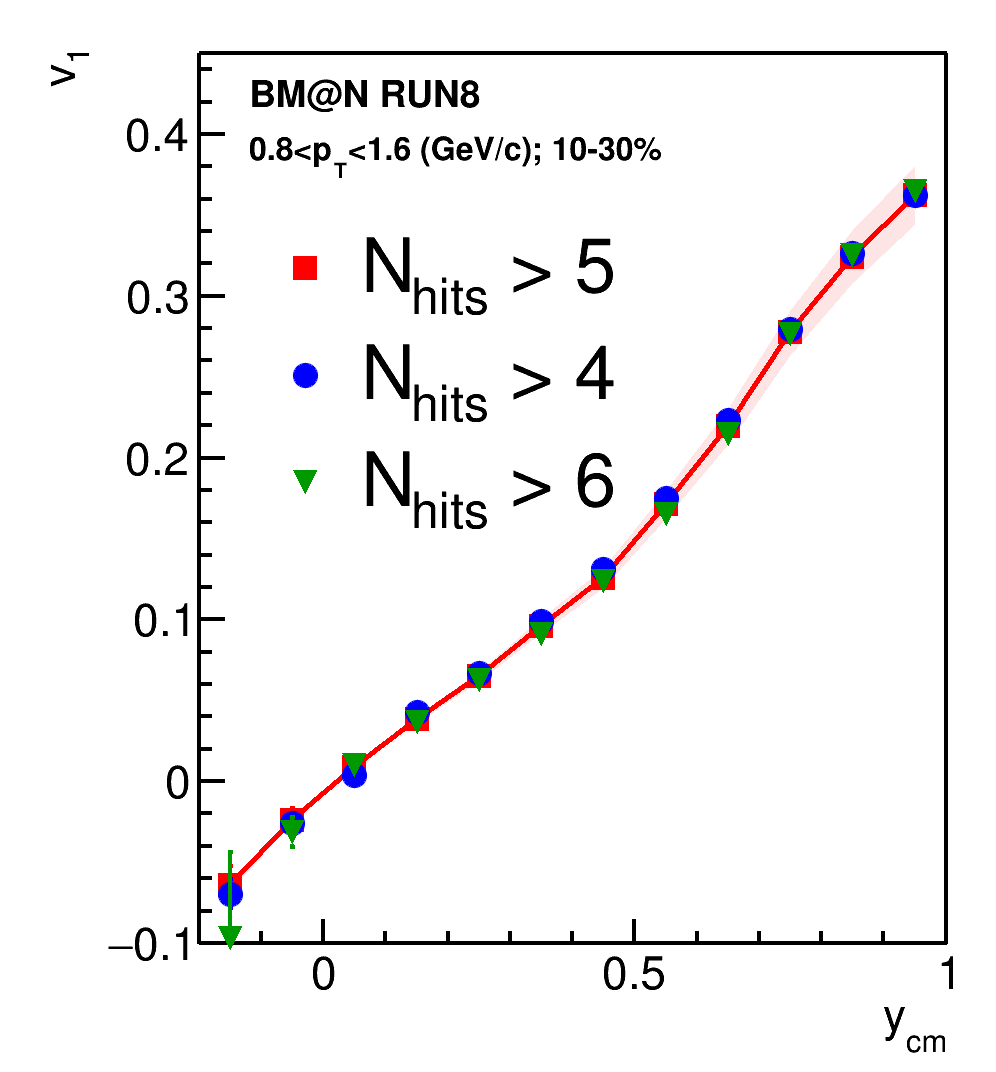}
    \caption{ Directed flow $v_1$ of protons as a function of rapidity $y_{cm}$ measured for different values of
      the track $\chi^2/NDF$ quality   (left) and the number of stations used for track reconstruction $N_{hits}$ (right). }
    \label{fig:sys_tracking}
\end{figure}

\begin{figure}[h]
    \centering
    \includegraphics[width=0.45\linewidth]{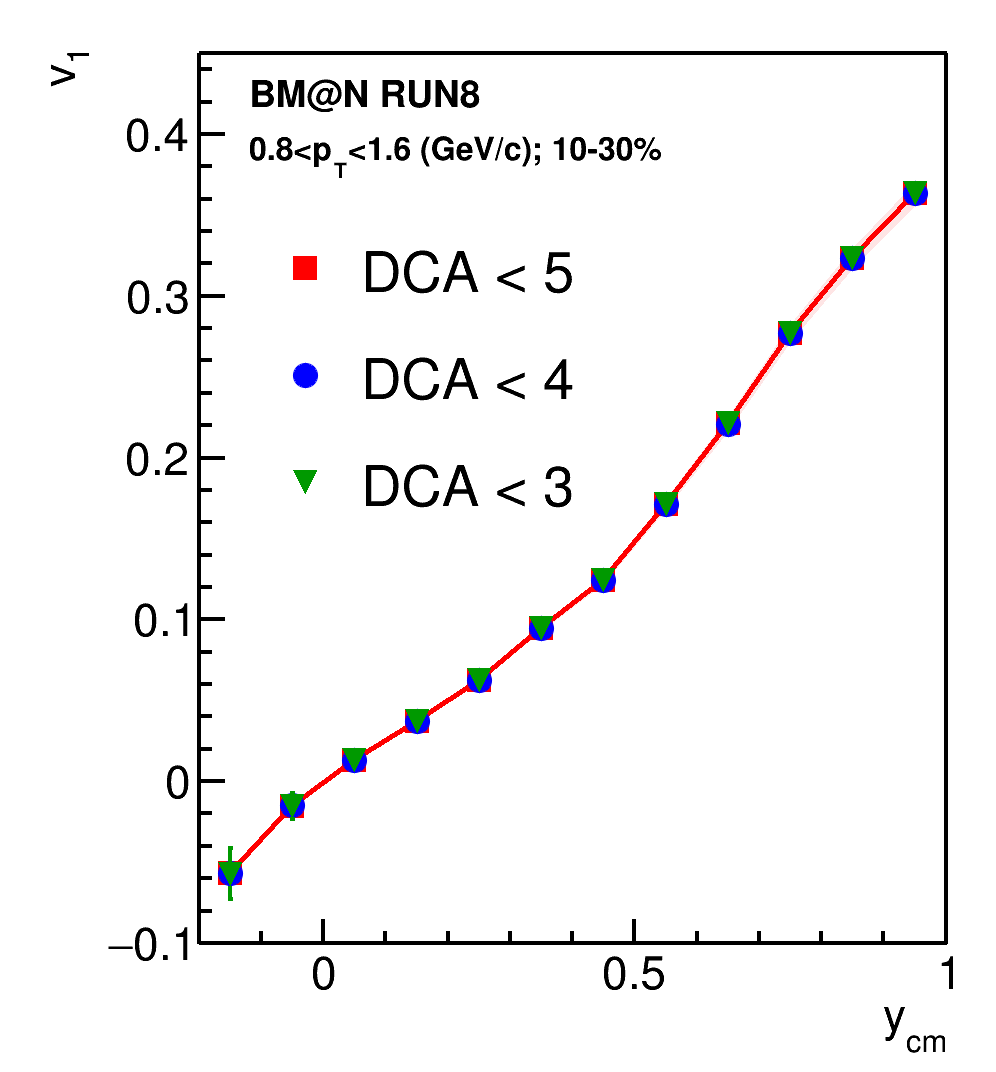}
    \includegraphics[width=0.45\linewidth]{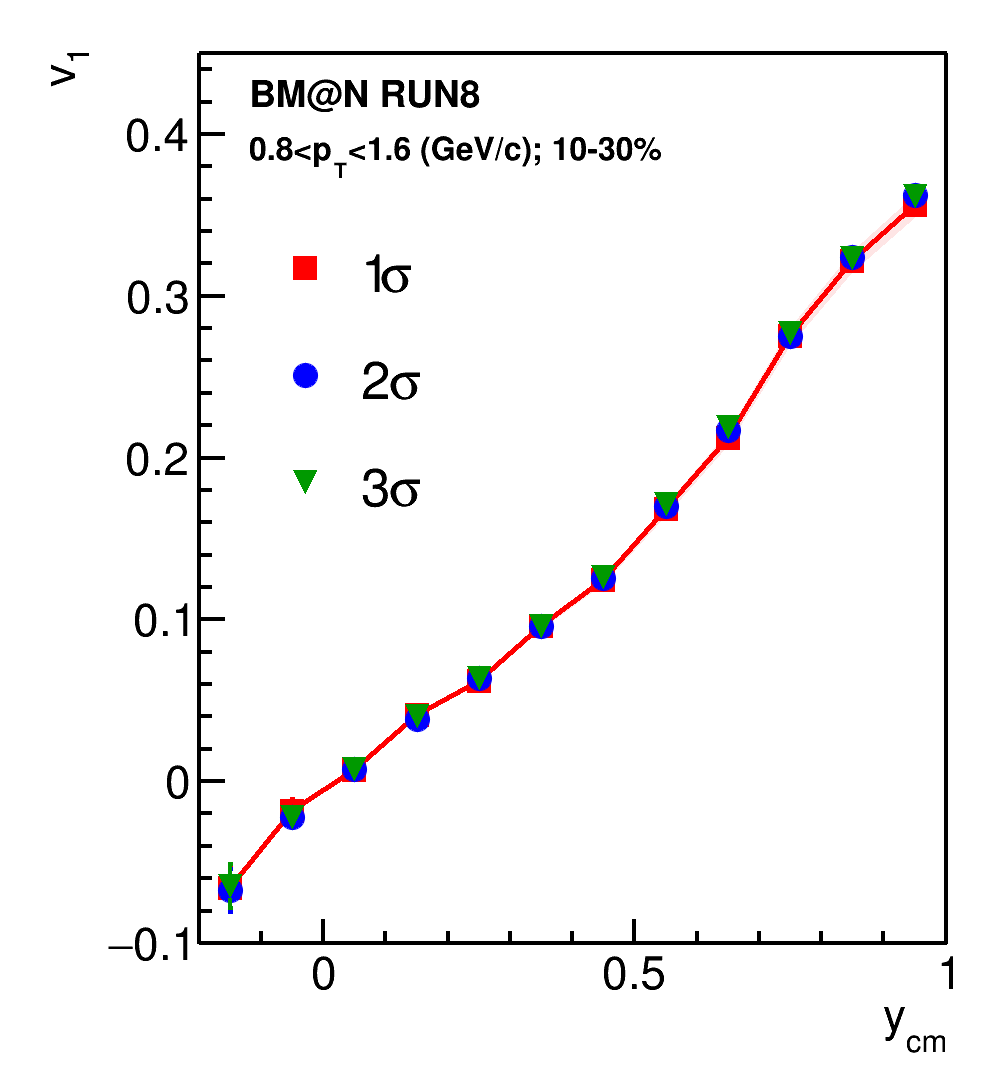}
    \caption{ Directed flow $v_1$ of protons as a function of rapidity $y_{cm}$ measured for different values of the $DCA$ cut
     and different n-$\sigma$ PID cuts for the proton identification:  ($m^2$-$\left\langle m_{p}^{2} \right\rangle$)  $< 1, 2, 3 ~\sigma_{m_{p}^2}$ cut (right). }
    \label{fig:sys_pid}
\end{figure}

\begin{figure}[h]
    \centering
    \includegraphics[width=0.45\linewidth]{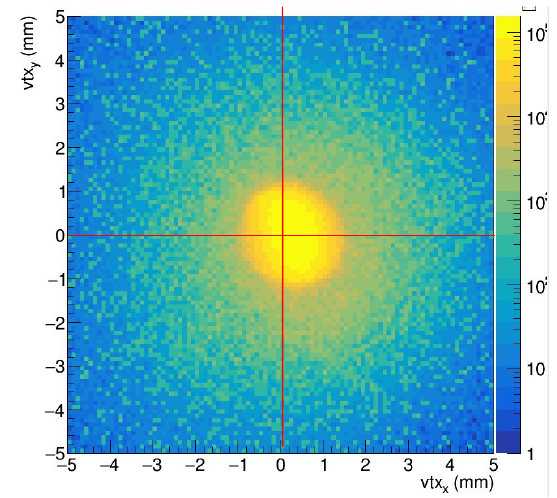}
    \includegraphics[width=0.45\linewidth]{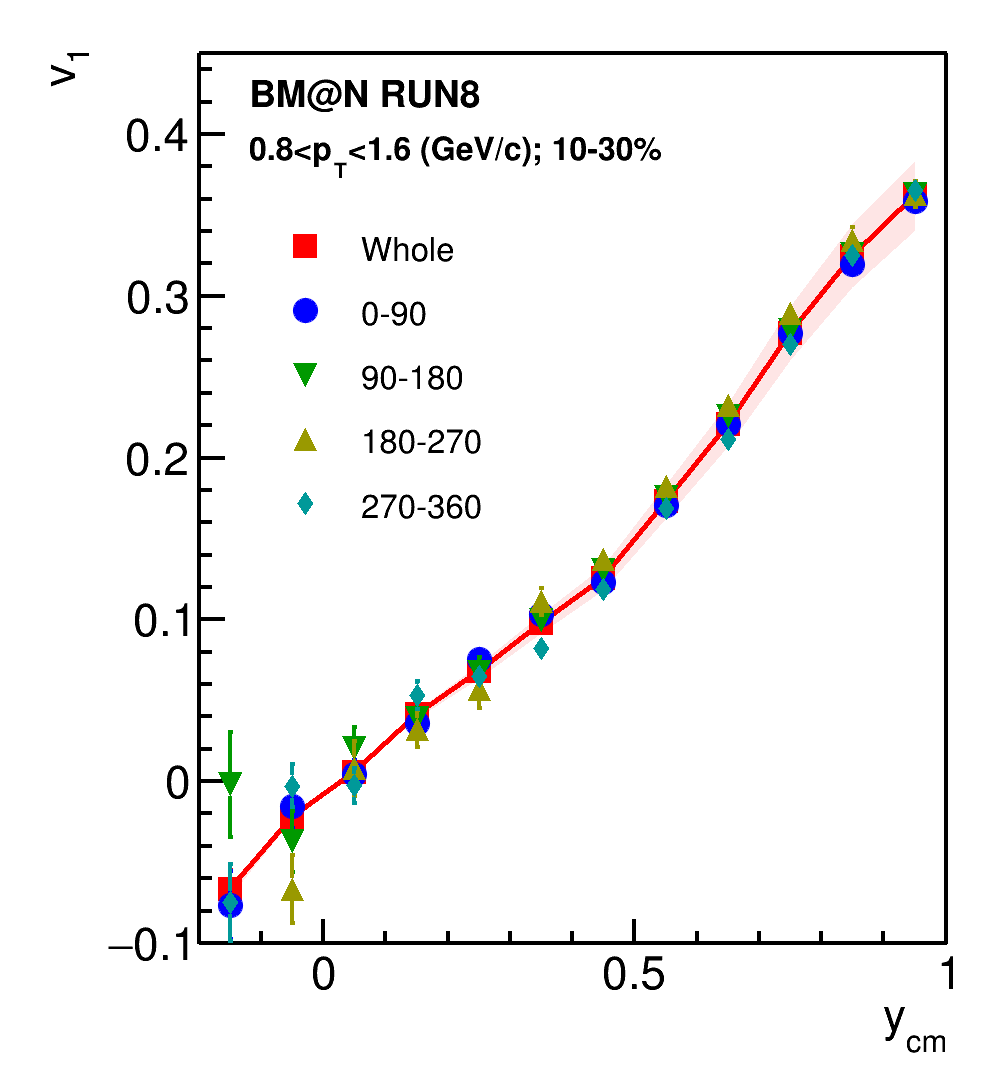}
    \caption{ Left: the distribution of the primary vertex in X-Y plane. Right:
      Directed flow $v_1$ of protons as a function of rapidity $y_{cm}$ calculated with varying
      the reconstructed primary vertex position of the collision.}
    \label{fig:sys_event}
\end{figure}

\begin{figure}[h]
    \centering
    \includegraphics[width=0.45\linewidth]{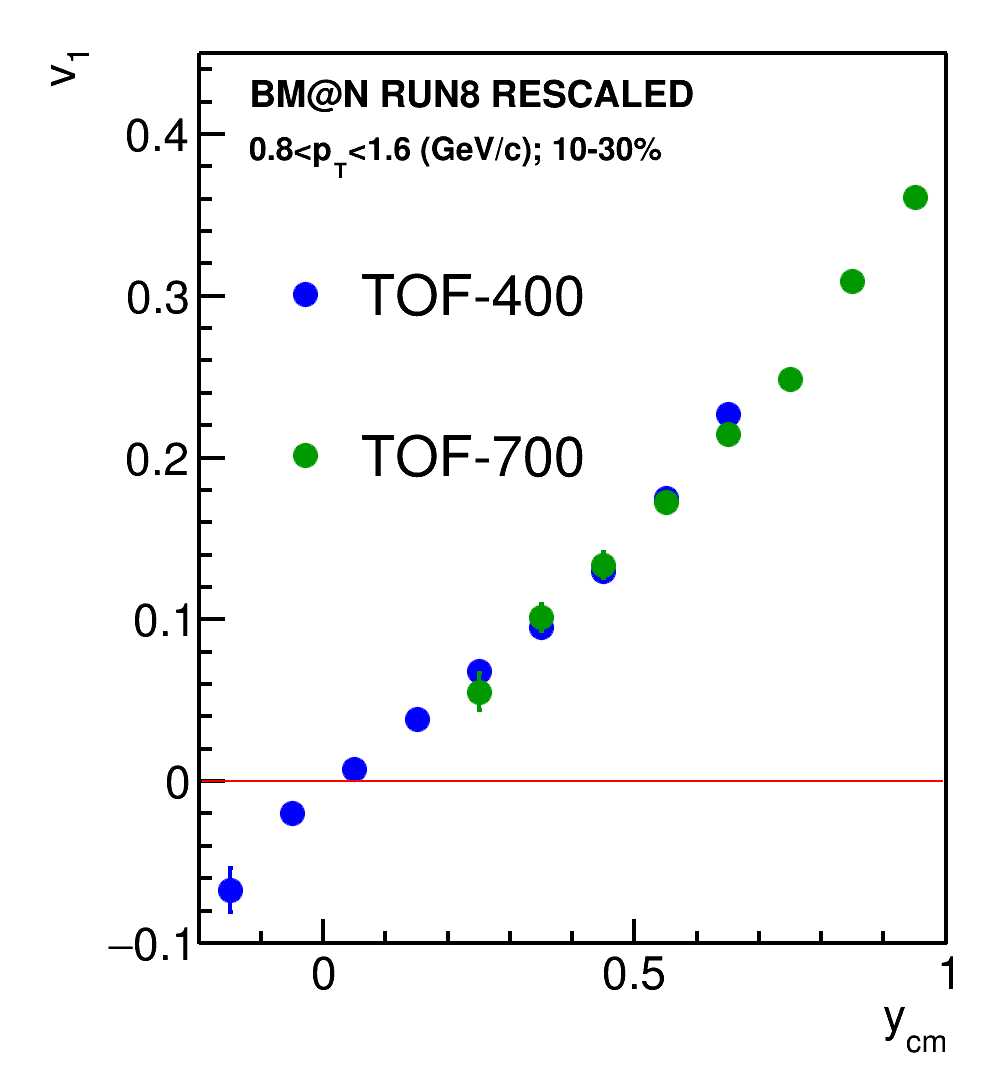}
    \includegraphics[width=0.45\linewidth]{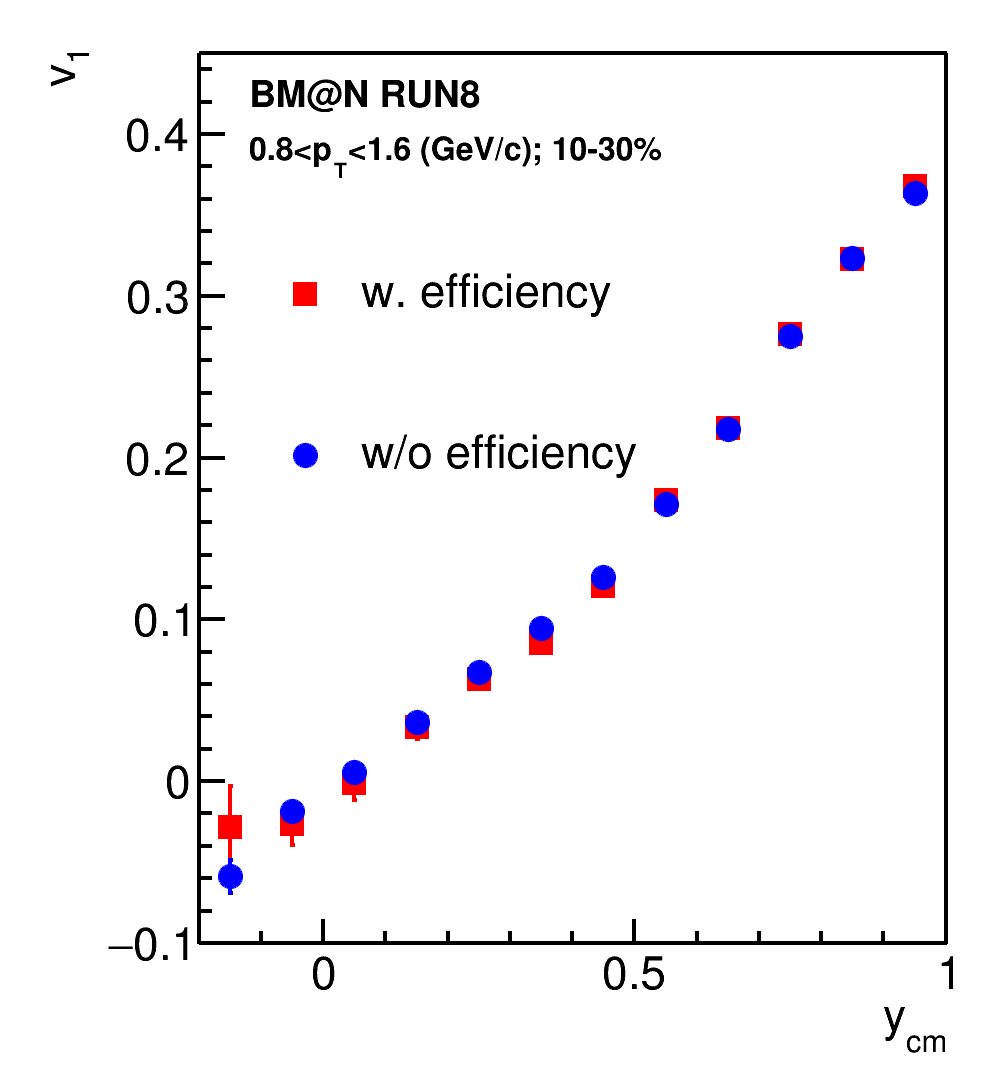}
    \caption{ Directed flow $v_1$ of protons as a function of rapidity $y_{cm}$ measured
      for protons identified using different TOF-systems (left) and protons weighted and not weighted with efficiency based on MC simulations for run8 (right). }
    \label{fig:tof_efficiency}
\end{figure}

\begin{figure}[h]
    \centering
    \includegraphics[width=0.45\linewidth]{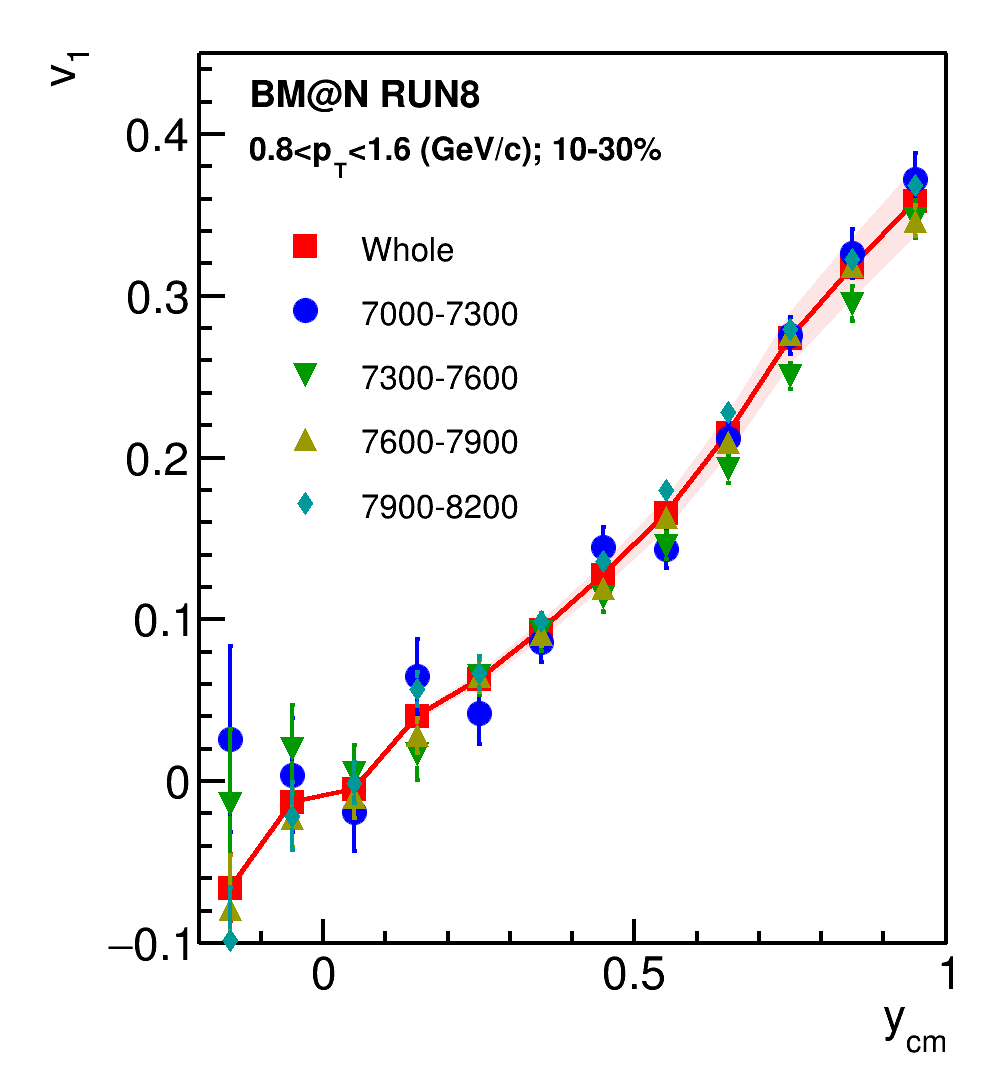}
    \includegraphics[width=0.45\linewidth]{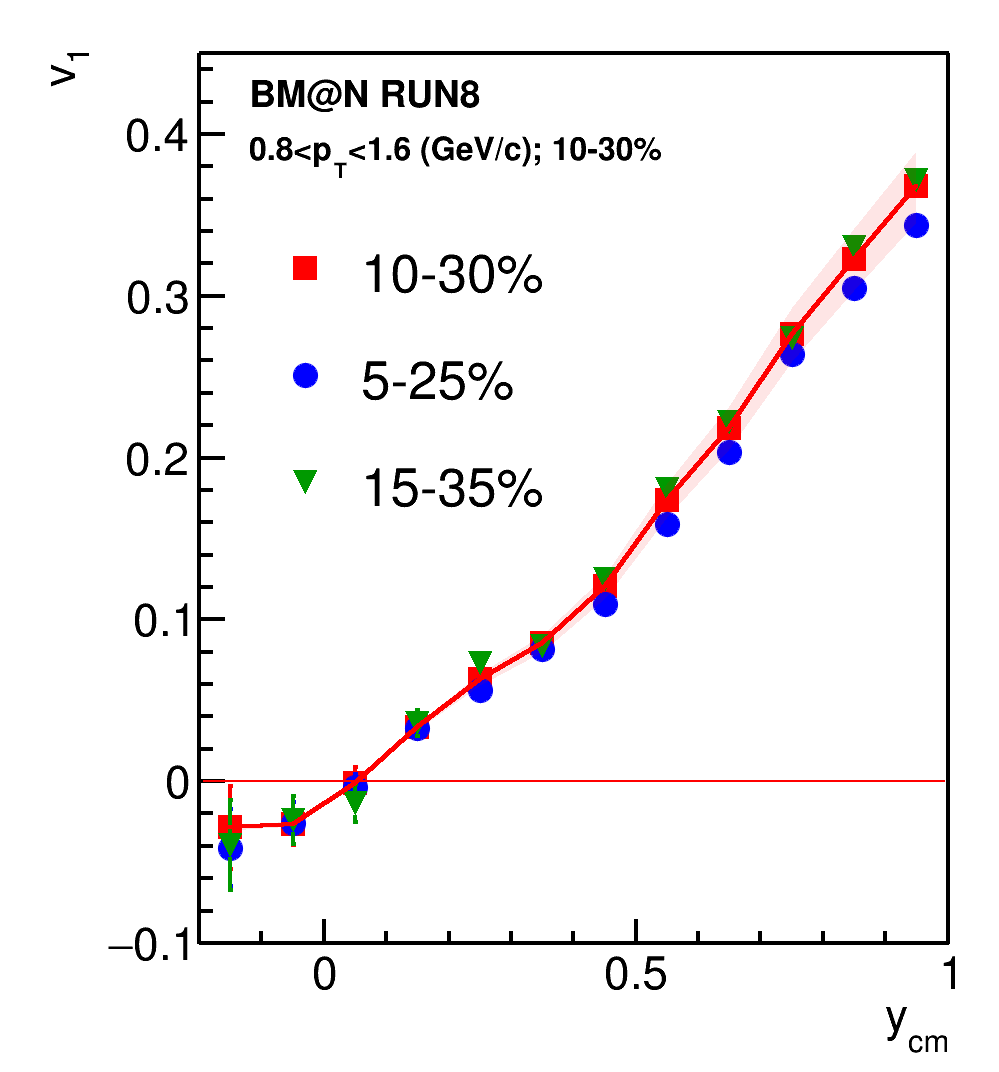}
    \caption{ Directed flow $v_1$ of protons as a function of rapidity $y_{cm}$ measured in the different run periods (left) and for different bins in collision centrality (right).}
    \label{fig:sys_run}
\end{figure}

In order to estimate systematic uncertainties of $v_1$ measurements ,
the following sources were considered:

\begin{itemize}
\item Uncertainty in proton momentum  reconstruction. We varied the number of stations $N_{hits}$ in
  inner tracking system used for track reconstruction as well as the values of track $\chi^2/NDF$ quality, see results in
  Figure~\ref{fig:sys_tracking}. The overall systematic uncertainty is found to be bellow 2-5\%.
\item Contribution from the secondary particles. We studied the difference in $v_1$  results
  for tracks with different Distance of the Closes Approach (DCA) to the primary vertex, see the left panel of Figure~\ref{fig:sys_pid} for results.
  It is found that proton $v_1$ values obtained with different DCA cut are in agreement within  1-2\%.
\item Contamination from the different particle species. We varied the identification selection criteria for protons, see the right
  panel of Figure~\ref{fig:sys_pid} for results. Observed is the systematic uncertainty is bellow 2-4\%
    \item Contribution due to off-target collisions. We divided the events based on
      the azimuthal angle of the vertex position  and compared the $v_1$ of protons in each group of events, see results in Figure~\ref{fig:sys_event}.
      Systematic variation stays bellow 5\%.
    \item Acceptance and efficiency. We preform the $v_1$ flow measurements for protons detected in TOF-400 and TOF-700 separately. We perform the measurements
      with and without the applying the efficiency correction for protons based on MC simulations for run8, see Figure~\ref{fig:tof_efficiency} for results.
      The results are in a good agreement and we can conclude that the mean value of transverse momentum $p_T$ is not shifted in this rapidity range.
    \item Run-by-run systematics was estimated dividing the events into several run periods and comparing the results in each group, see the left panel of
      Figure~\ref{fig:sys_run} for results.  The
systematic uncertainty is less than 5\% and found to be less than statistical. 
\end{itemize}

Systematic uncertainties were calculatad by the square root of quadratic sum of uncertainties from each source.

\section{Results of the directed flow measurements}

Directed flow $v_1$ of protons was measured in 10-30\% central Xe+Cs(I) collisions at 3.8 AGeV as a function of  rapidity $y_{cm}$
and transverse momentum $p_T$, see  Figure.~\ref{fig:v1_y_pT}. 
Rapidity-dependence of  $v_1$ of protons from the experimental data has been
compared with predictions from the  model JAM transport model \cite{jam1,jam2}  with momentum dependent mean field\cite{parfenov1,mamaev}.
JAM model roughly captures the overall magnitude and trend of the measured  $v_1(y_{cm})$ signal of protons, see black solid line in Figure.~\ref{fig:v1_y_pT}.
The slope of the directed flow $v_1$  at midrapidity $dv_1/dy_{cm}|_{y_{cm}=0}$ is extracted by fitting the $v_1(y_{cm})$ with polynomial
function $v_1=a + by_{cm} + cy_{cm}^3$ as it was done in other experiments \cite{hadesvn,starv1,star3gev,fopi}.
\begin{figure}[h]
    \centering
    \includegraphics[width=0.95\linewidth]{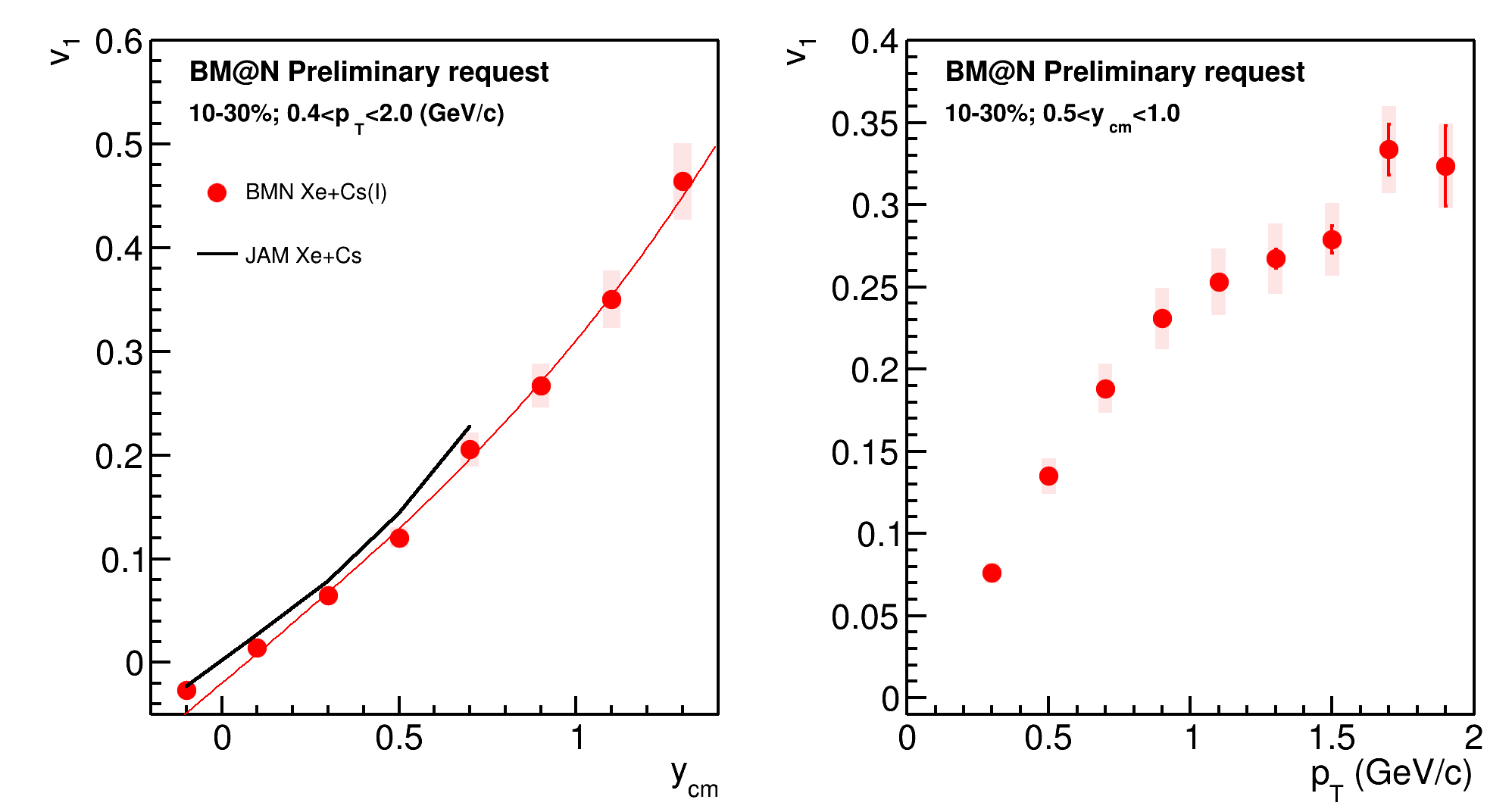}
    \caption{Directed flow $v_1$ of protons in 10-30\% central Xe+Cs(I) collisions at 3.8 A GeV as a function of 
rapidity $y_{cm}$ (left panel) and transverse momentum $p_T$ (right panel).}
    \label{fig:v1_y_pT}
\end{figure}

\begin{figure}[h]
    \centering
    \includegraphics[width=0.95\linewidth]{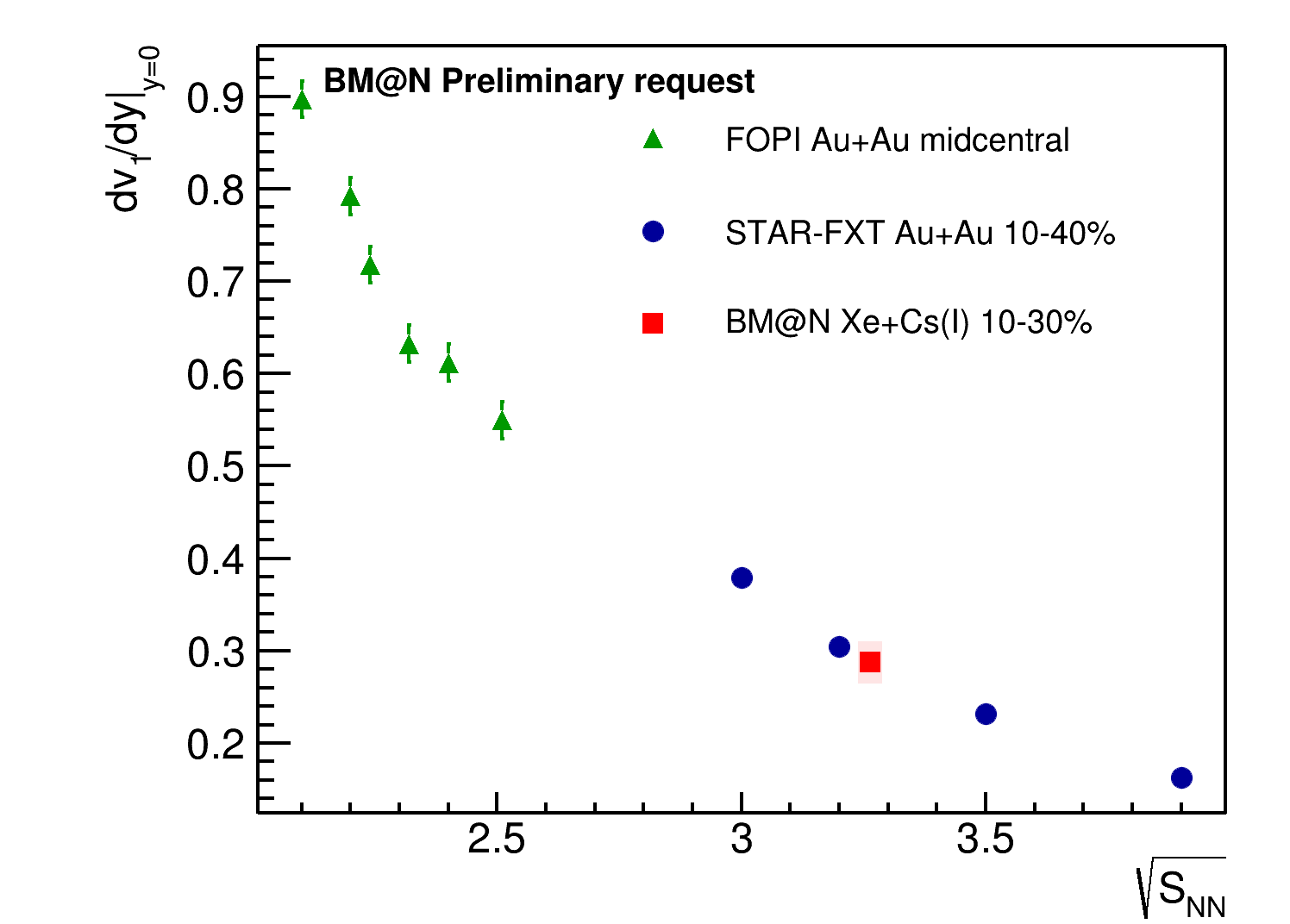}
    \caption{The  slope of $v_1$ of protons at midrapidity $dv_1/dy_{cm}|_{y_{cm}=0}$ as a function of collision energy. The obtained BM@N results were
      compared with existing  data from other experiments  \cite{fopi,star3gev,starv1}. }
    \label{fig:dv1dy_sqrt_snn}
\end{figure}

The  slope of $v_1$ of protons at midrapidity $dv_1/dy_{cm}|_{y_{cm}=0}$ as a function of collision energy is presented in the fig.~\ref{fig:dv1dy_sqrt_snn}.
The results for the BM@N experiment are compared with existing  data from other experiments  \cite{fopi,star3gev,starv1}.
Directed flow slope at midrapidity $dv_1/dy_{cm}|_{y_{cm}=0}$ are found to be in a reasonable agreement with the existing measurements.

\printbibliography
    
\end{document}